\documentclass[twocolumn]{aastex62}

\newcommand{\arcs}{$^{\prime\prime}$} 
\newcommand{\beq}{\begin{equation}\begin{aligned}}
\newcommand{\eeq}{\end{aligned}\end{equation}}

\newcommand{\msun}{M$_\odot$}

\usepackage{scalerel,amssymb}
\usepackage{amsmath}
\usepackage{enumitem}
\usepackage[]{url}
\usepackage{lineno}
\usepackage[caption=false]{subfig}
\defcitealias{karachentsev2013}{K13}
\defcitealias{kourkchi2017}{KT17}

\accepted{\today}
\submitjournal{ApJ}

\shorttitle{Dwarf Satellites of MW-like Galaxies}
\shortauthors{Carlsten et al.}

\begin{document}

\title{The Exploration of Local VolumE Satellites (ELVES) Survey: A Nearly Volume-Limited Sample of Nearby Dwarf Satellite Systems}

\correspondingauthor{Scott G. Carlsten}
\email{scottgc@princeton.edu}

\author[0000-0002-5382-2898]{Scott G. Carlsten}
\affil{Department of Astrophysical Sciences, 4 Ivy Lane, Princeton University, Princeton, NJ 08544}

\author[0000-0002-5612-3427]{Jenny E. Greene}
\affil{Department of Astrophysical Sciences, 4 Ivy Lane, Princeton University, Princeton, NJ 08544}

\author[0000-0002-1691-8217]{Rachael L. Beaton}
\altaffiliation{Hubble Fellow}
\affiliation{Department of Astrophysical Sciences, 4 Ivy Lane, Princeton University, Princeton, NJ 08544}
\affiliation{The Observatories of the Carnegie Institution for Science, 813 Santa Barbara St., Pasadena, CA~91101\\}

\author[0000-0002-1841-2252]{Shany Danieli}
\altaffiliation{Hubble Fellow}
\affil{Department of Astrophysical Sciences, 4 Ivy Lane, Princeton University, Princeton, NJ 08544}

\author[0000-0003-4970-2874]{Johnny P. Greco}
\affiliation{Center for Cosmology and AstroParticle Physics (CCAPP), The Ohio State University, Columbus, OH 43210, USA}

\begin{abstract}
We present the final results of the Exploration of Local VolumE Satellites (ELVES) Survey, a survey of the dwarf satellites of a nearly volume-limited sample of Milky Way (MW)-like hosts in the Local Volume. Hosts are selected simply via a cut in luminosity ($M_{K_s}<-22.1$ mag) and distance ($D<12$ Mpc). We have cataloged the satellites of 25 of the 31 such hosts, with another five taken from the literature. All hosts are surveyed out to at least 150 projected kpc ($\sim R_\mathrm{vir}/2$) with the majority surveyed to 300 kpc ($\sim R_\mathrm{vir}$). Satellites are detected using a consistent semi-automated algorithm specialized for low surface brightness dwarfs. As shown through extensive tests with injected galaxies, the catalogs are complete to $M_V\sim-9$ mag and $\mu_{0,V}\sim26.5$ mag arcsec$^{-2}$. Candidates are confirmed to be real satellites through distance measurements including redshift, tip of the red giant branch, and surface brightness fluctuations. Across all 30 surveyed hosts, there are 338 confirmed satellites with a further 105 candidates awaiting distance measurement. For the vast majority of these, we provide consistent multi-band S\'{e}rsic photometry. We show that satellite abundance correlates with host mass, with the MW being quite typical amongst comparable systems, and that satellite quenched fraction rises steeply with decreasing satellite mass, mirroring the quenched fraction for the MW and M31. The ELVES survey represents a massive increase in the statistics of surveyed systems with known completeness, and the provided catalogs are a unique dataset to explore various aspects of small-scale structure and dwarf galaxy evolution.

\end{abstract}

\section{Introduction}
The Exploration of Local VolumE Satellites (ELVES) Survey is a survey to detect and characterize low-mass, dwarf satellite galaxies around nearby, massive hosts in the Local Volume (LV; $D<12$ Mpc). The explicit survey goal is to fully map the ``classical''-mass satellites ($M_V<-9$ mag, $M_\star\gtrsim 5\times10^5$ \msun) of all 31 LV hosts with $M_{K_s}<-22.1$ mag throughout most of their virial volumes. Initial results of the survey have been presented in \citet{carlsten2020a, carlsten2020b, carlsten2021a}. This work represents the final results of the photometric part of the survey, and we present the satellite systems of nearly all (30 out of 31) hosts in this volume-limited sample.

Broadly, the motivation of the ELVES Survey is to push our understanding of dwarf satellite galaxies beyond just the satellite system of the Milky Way (MW). The satellites of the MW have been richly characterized over the years \citep[e.g. see][and references therein]{mateo1998, koposov2008, simon2019, drlica2020}. For decades the MW dwarfs (and to a lesser extent the satellites of M31 and other dwarfs in the Local Group) have acted as the \textit{de-facto} benchmark for models of small-scale structure formation and dwarf galaxy evolution in the $\Lambda$CDM paradigm \citep[e.g.][]{klypin1999, moore1999, mayer2006, bk2011, bk2012, brooks2013, brooks2014, elvis, sawala2016, wetzel2016, bullock2017,  pawlowski2018, nadler2020}. Yet, there is no current consensus on whether the MW and its dwarf satellites are ``typical'', making these comparisons with models difficult to interpret.  On the theoretical front, multiple groups have now produced large samples of high-resolution simulated MW-mass systems \citep{simpson2018, gk2019, applebaum2020, libeskind2020, font2020, font2021, engler2021} or semi-analytic models \citep{jiang2020}, which are ready to be compared to similarly large samples of observed systems. Thus it is critical to characterize a statistical sample of satellite systems around galaxies other than the MW.

Some of the significant open questions that can be addressed with such a sample of satellite systems include: the galaxy-(sub)halo connection in low-mass ($M_\star\lesssim 10^9$ \msun) galaxies \citep{kim2018, nadler2019, munshi2021}, the role of the massive host on disrupting and/or quenching star formation in the satellites \citep{wetzel2015, fillingham2015, hausammann2019, samuel2020, akins2021, font2021b}, and the prevalence of coherent, kinematic structures within satellite systems \citep{pawlowskiVPOS, pawlowski2020, ibataGPOA, muller_plane}.

In recent years, numerous groups have started work on the formidable observational task of surveying the very low-mass satellites of nearby $(D\lesssim 50$ Mpc) massive hosts. Many groups have used deep, wide-field imaging or spectroscopic surveys to catalog candidate satellites around various hosts in the LV, including MW-mass hosts \citep[e.g][]{irwin2009, kim2011, sales2013, spencer2014, merritt2014, karachentsev2015, bennet2017, bennet2019, bennet2020, tanaka2017, danieli2017, danieli2020, smercina2018, kondapally2018, park2017, park2019, byun2020, davis2021, garling2021, mutlu2021} and hosts of somewhat lower mass \citep{madcash, carlin2021, muller2020_low_mass, drlica2021} and higher mass \citep{stierwalt2009, trentham2009, chiboucas2009, chiboucas2013, crnojevic2014, crnojevic2016, crnojevic2019, muller2015, muller2017, muller2018, muller2019, muller2021, smercina2017, cohen2018} mass. Pushing to larger distances, deep surveys are starting to map the dwarf content of various galaxy groups \citep[e.g.][]{greco2018, zaritsky2019, habas2020, tanoglidis2021, prole2021}. At the highest host mass end, nearby galaxy clusters are also being surveyed for dwarf satellites \citep{ferrarese2012, ferrarese2016, ferrarese2020, munoz2015, eigenthaler2018, venhola2018, venhola2021, su2021, lamarca2021}.

The process of surveying nearby satellite systems generally consists of two primary steps. The first is to detect candidate satellites. While satellites of the MW and the nearest hosts ($D\lesssim4$ Mpc; i.e., M31, M81, CenA, and M94) can be found via resolved stars, satellites of more distant hosts in the LV can currently only be found via integrated light\footnote{Space-based observatories with wide fields of view, like the \emph{Roman Space Telescope} \citep{spergel2015} and \emph{Euclid} \citep{euclid}, will change this.}. Due to the fact that dwarfs, on average, get lower in surface brightness at lower masses \citep{mateo1998, kormendy2009, danieli_field, carlsten2021a}, detecting very low-mass satellites in integrated light requires deep imaging and specialized detection methods \citep[e.g.][]{bennet2017, greco2018, danieli2019, carlsten2020a}. An important part of this first step is also quantifying the completeness of the dwarf search. 

The second step is to measure distances to the dwarfs to confirm their environment (in other words, their status as satellites). While contamination from background galaxies is not much of a concern for rich groups or clusters, contamination can account for a majority of candidate satellites \citep[$>80$\%;][]{sbf_m101, bennet2019} in sparse, MW-mass groups. As discussed in \citet{carlsten2020b}, the number of satellite systems that have been surveyed \citep[prior to ELVES and the SAGA Survey,][see below]{mao2020} with quantified completeness  and distance confirmation for the candidate satellites is actually quite small ($\sim5$). We note that it is possible to treat contamination statistically, without individual satellite distance measurements, by using a background subtraction technique \citep[e.g.][]{wang2012, wang2021, nierenberg2016, tanaka2018, xi2018, roberts2021, wu2021}. However, such an approach does not allow for detailed follow-up of individual dwarfs (e.g., for gas or star cluster properties) and, thus, is not the approach that ELVES takes. With that said, the statistics afforded by these works make them valuable comparisons for ELVES.

Over the years, \emph{HST} has been employed significantly to measure precise tip of the red giant branch (TRGB) distances to LV dwarfs \citep[e.g.][]{karachentsev2006, karachentsev2007, karachentsev2013, karachentsev2020, jacobs2009, anand2021}. However, for the scale ELVES aims to address, it would be infeasible to procure a statistical sample of hundreds of satellites relying solely on \emph{HST}. Many low-mass satellites are quenched \citep[i.e., without nebular emission lines or HI reservoirs;][]{karunakaran2020, putman2021} and are of low surface brightness (LSB) which makes acquiring a redshift extremely difficult. In earlier papers, we have shown that surface brightness fluctuations \citep[SBF;][]{tonry1988, jerjen_field, jerjen_field2, cantiello2018} offer an efficient way to get distances to LSB dwarfs, often from the same deep ground-based imaging used to detect them in the first place \citep{sbf_calib, sbf_m101, greco2020}. A combination of novel image detection algorithms and use of SBF facilitates the ELVES Survey in rapidly establishing a large sample of satellite systems.

A particularly relevant contemporary project is the Satellites Around Galactic Analogs Survey \citep[SAGA,][]{geha2017, mao2020}. The SAGA Survey is an ongoing spectroscopic survey of the bright classical satellites ($M_r<-12.3$ mag) of 100 MW-analogs in the distance range $20 < D < 40$ Mpc. SAGA is complementary to ELVES in several important ways. First, while SAGA is not sensitive to as faint of satellites as ELVES (ELVES goes $\sim3$ mag fainter), it will achieve much better statistics at the bright end of the satellite luminosity function due to the larger number of hosts. Second, while SAGA selects ``MW-analogs'' via multiple criteria, ELVES simply selects all hosts in the LV above a certain $K$-band luminosity (see Section \ref{sec:hosts}). Thus, SAGA will better probe what a ``typical'' MW-like system is, while ELVES will be more sensitive to the effect that differing host properties have on satellite systems. Finally, SAGA, being a spectroscopic survey, catalogs and characterizes satellites in a substantially different way than ELVES meaning that the projects offer an important check on each other.

As mentioned above, this paper extends the surveys presented in \citet{carlsten2020a, carlsten2020c, carlsten2020b}. Since those papers, we have been able to nearly triple the number of surveyed satellite systems, essentially completing the full volume-limited sample. These new dwarf findings have been used in \citet{carlsten2021a} and \citet{carlsten2021b} to explore different aspects of dwarf galaxy evolution but have not been fully described until the current paper. This paper is structured as follows: in Section \ref{sec:hosts} we describe the host selection process and final host list, in Section \ref{sec:data} we outline the different sources of data used, in Section \ref{sec:detection} we describe the detection of candidate satellites, in Section \ref{sec:distances} we detail how we confirm the distance of the satellites, in Section \ref{sec:sats} we describe how we characterize the satellites,   in Section \ref{sec:sat_systems} we give some overview of the properties of the satellite systems, including satellite abundance, spatial distribution, and the fraction of star-forming satellites, and, finally, in Section \ref{sec:conclusions} we give an overview of the key results of the ELVES Survey to this point.

\section{Host Selection}
\label{sec:hosts}
The primary goal of the ELVES Survey is to obtain a volume limited sample of well-surveyed satellite systems around massive, roughly MW-like host galaxies. In this section, we describe the selection of the host sample and give some details on the host properties.

To make the initial host list, we use both the group catalog of \citet{kourkchi2017} (hereafter KT17) and the Updated Nearby Galaxy Catalog (UNGC) of \citet{karachentsev2004, karachentsev2013} (hereafter K13). The group catalog of \citetalias{kourkchi2017} is based on the Cosmic-Flows \citep{tully2016} distance database.  For massive galaxies, the catalog relies heavily on the 2MASS Redshift Survey \citep{huchra2012}, which is complete down to $K_s=11.75$ mag. At the edge of the LV (12 Mpc), this corresponds to a luminosity of $M_{K_s}=-18.6$ mag and thus will include all potential massive host galaxies. The catalog of \citetalias{karachentsev2013} includes many more up-to-date (e.g., very recent TRGB distances) than \citetalias{kourkchi2017} but is missing several massive hosts that are known to be within the LV. Using these two catalogs, almost all of the potential hosts have direct (i.e. not redshift-based) distance measurements with the majority being from TRGB observations.

The cuts that we make on the catalogs are as follows:
\beq
    \mathrm{Distance}:& \;\; D < 12 \; \textrm{Mpc} \\ 
    \mathrm{Host\;Luminosity}:& \;\; M_{K_s} < -22.1 \; \textrm{mag} \\
    \mathrm{Galactic\;Latitude}:& \;\; |b| > 17.4^\circ
\eeq

In making the host list, we merge the two catalogs but always use the $K_s$ magnitude reported by \citetalias{kourkchi2017}. To these catalogs, we add the host NGC 4565 which has a TRGB distance from \citet{rs11} of within 12 Mpc but whose distance in \citetalias{kourkchi2017} is greater than 12 Mpc\footnote{With the larger distance, \citetalias{kourkchi2017} group NGC 4565 with NGC 4494, which has an SBF distance of 16.9 Mpc. With the TRGB distance of 11.9 Mpc we take for NGC 4565, it seems likely that these two galaxies are not physically associated. Therefore, for NGC 4565, we do not take the group properties calculated by \citetalias{kourkchi2017} because they will include NGC 4494. Instead, we calculate them ourselves using the members of NGC 4565's group cataloged in ELVES.}. 

The chosen $K_s$ luminosity cut corresponds to a stellar mass of roughly $M_\star\approx 10^{10}$ \msun, assuming $M_\star/L_{K_s}=0.6$ \citep{mcgaugh2014}. For each host, we search NED and SIMBAD for more up-to-date distances (particularly looking for precise distances, like those from TRGB, SNe Ia, or Cepheid observations). Where appropriate, we update the $K_s$ luminosity from \citetalias{kourkchi2017} using the newer distance. The cut on Galactic latitude, $b$, is chosen to be as restrictive as possible to eliminate highly extincted hosts while still including NGC 891 and  Cen A, both major focuses of recent dwarf satellite research \citep[e.g.][]{trentham2009, muller2019_n891, crnojevic2019, muller2021}.

With these cuts, we often select multiple members within the same galaxy group. In these cases, we choose the member with the greatest $K_s$ luminosity as the `host' and remove the the fainter objects from consideration as a `host'. The specific galaxies that obey our other cuts but are likely affiliated with a more massive host are given in Appendix \ref{app:host_rejects}.

\begin{deluxetable*}{ccccccccccc}
\tablecaption{ELVES Host List\label{tab:hosts}}
\tablewidth{\textwidth}
\tablehead{
\colhead{Name} & \colhead{Dist} & \colhead{$v_\mathrm{rec}$} & \colhead{$M_{K_s}$} & \colhead{$M_{K_s}^{\mathrm{group}}$} &  \colhead{$M_V$}  &  \colhead{$B-V$}  &  \colhead{$\log(M_\star/M_\odot)$}  & \colhead{$r_\mathrm{cover}$} & \colhead{Data Source} & \colhead{References}  \\ 
\colhead{} & \colhead{Mpc} & \colhead{km/s}  &  \colhead{mag}  & \colhead{mag}  & \colhead{mag} & \colhead{mag} & \colhead{} & \colhead{kpc} & \colhead{}  & \colhead{} } 
\startdata
M31 & 0.78 & -300 & -24.81 & -24.89  &  -21.19 & 0.87  &  11.01  &  300  &  M12  &  CF, RC3, Sick15 \\
NGC253 & 3.56 & 259 & -23.95 & -23.96  &  -20.7 & 0.94  &  10.77  &  300  &  D- -D  &  CF, Cook14a, Cook14b \\
NGC628 & 9.77 & 656 & -22.79 & -22.81  &  -20.68 & 0.48  &  10.45  &  300  &  D-G-D  &  CF, Cook14a, Cook14b \\
NGC891 & 9.12 & 528 & -23.83 & -23.83  &  -20.05 & 0.82  &  10.84  &  200  &  C-C-C  &  CF, GALEX, $M_{K_s}$ \\
NGC1023 & 10.4 & 638 & -23.78 & -23.98  &  -20.91 & 0.95  &  10.6  &  200  &  C-C-C  &  NED, GALEX, $M_{K_s}$ \\
NGC1291 & 9.08 & 838 & -23.94 & -23.97  &  -21.01 & 1.0  &  10.78  &  300  &  D-D,M-D  &  CF, Cook14a, Ler19 \\
NGC1808 & 9.29 & 1002 & -23.12 & -23.78  &  -19.98 & 0.79  &  10.01  &  300  &  D-D,G,H,M-D  &  CF, RC3, Ler19 \\
NGC2683 & 9.4 & 409 & -23.49 & -23.5  &  -20.17 & 0.86  &  10.5  &  300  &  D-H-D  &  K15, Cook14a, Ler19 \\
NGC2903 & 9.0 & 556 & -23.68 & -23.69  &  -20.78 & 0.65  &  10.67  &  300  &  D-G,C-D  &  K13\&T19, Cook14a, Cook14b \\
M81 & 3.61 & -42 & -23.89 & -25.27  &  -21.07 & 0.88  &  10.66  &  300  &  C13-C13-C13,D  &  CF, GALEX, Ler19 \\
NGC3115 & 10.2 & 666 & -24.12 & -24.14  &  -21.27 & 0.93  &  10.76  &  300  &  D-G-D  &  Pea15, RC3, Ler19 \\
NGC3344 & 9.82 & 585 & -22.27 & -22.27  &  -20.19 & 0.65  &  10.27  &  300  &  D-H-D  &  CF, Cook14a, Ler19 \\
NGC3379 & 10.7 & 876 & -23.8 & -25.37  &  -20.46 & 0.78  &  10.63  &  370  &  D-H-D  &  M18, HT11, Ler19 \\
NGC3521 & 11.2 & 798 & -24.4 & -24.41  &  -21.39 & 0.76  &  10.83  &  330  &  D-H-D  &  This Work, GALEX, Ler19 \\
NGC3556 & 9.55 & 698 & -22.76 & -22.76  &  -19.96 & 0.69  &  9.94  &  300  &  D- -D  &  CF, RC3, Ler19 \\
NGC3621 & 6.7 & 731 & -22.37 & -22.37  &  -19.79 & 0.55  &  9.9  &  0  &   - -   &  CF, RC3, Ler19 \\
NGC3627 & 10.5 & 789 & -24.17 & -25.33  &  -21.28 & 0.7  &  10.66  &  300  &  D-C,H-D  &  Lee13, GALEX, Ler19 \\
NGC4258 & 7.2 & 462 & -23.7 & -23.94  &  -20.92 & 0.71  &  10.62  &  150  &  C,D-C,H-C,D  &  H13, Cook14a, Ler19 \\
NGC4517 & 8.34 & 1136 & -22.13 & -22.16  &  -19.27 & 0.67  &  9.93  &  300  &  D-H-D  &  K14, RC3, $M_{K_s}$ \\
NGC4565 & 11.9 & 1261 & -24.26 & -24.27  &  -20.84 & 0.82  &  10.88  &  150  &  C-C-C  &  RS11, RC3, Ler19 \\
M104 & 9.55 & 1092 & -24.91 & -25.01  &  -22.04 & 0.94  &  11.09  &  150  &  C-C-C  &  McQ16b, RC3, Ler19 \\
NGC4631 & 7.4 & 606 & -22.73 & -22.9  &  -20.21 & 0.55  &  10.05  &  200  &  C,D-C,G-C,D  &  RS11, RC3, Ler19 \\
NGC4736 & 4.2 & 287 & -22.96 & -23.1  &  -19.93 & 0.74  &  10.29  &  300  &  D-S18-D  &  RS11, RC3, Ler19 \\
NGC4826 & 5.3 & 402 & -23.24 & -23.24  &  -20.21 & 0.8  &  10.36  &  300  &  D-H,C-D  &  CF, RC3, Ler19 \\
NGC5055 & 8.87 & 503 & -24.02 & -24.04  &  -21.2 & 0.71  &  10.72  &  300  &  D-H-D  &  McQ17, RC3, Ler19 \\
CENA & 3.66 & 548 & -23.87 & -24.41  &  -21.29 & 0.89  &  10.92  &  200  &  M/C19-M/C19-Dc,M17  &  CF, RC3, Ler19 \\
NGC5194 & 8.58 & 465 & -24.08 & -24.51  &  -21.4 & 0.56  &  10.73  &  150  &  C-C-C  &  McQ16a, GALEX, Ler19 \\
NGC5236 & 4.7 & 520 & -23.64 & -23.67  &  -21.02 & 0.53  &  10.37  &  300  &  M15-M18,D-D  &  CF, Cook14a, Ler19 \\
NGC5457 & 6.5 & 243 & -23.12 & -23.16  &  -21.22 & 0.44  &  10.33  &  300  &  D-B19,C,H-C,D  &  Beat19, RC3, Ler19 \\
NGC6744 & 8.95 & 839 & -23.58 & -24.0  &  -21.64 & 0.86  &  10.64  &  200  &  Dc-Dc-Dc  &  CF, RC3/HyperLeda, Ler19 \\
MW & 0 & 0 & -24.0 & -24.13  &  -20.74 & 0.67  &  10.78  &  300  &  M12  &  Lic15, BG16 \\
\enddata
\tablecomments{Volume limited sample of massive hosts, selected with $M_{K_s}<-22.1$ mag, $|b|>17.4^\circ$, and $D<12$ Mpc. $M_{K_s}^{\mathrm{group}}$ denotes the total luminosity for the group, as determined by \citet{kourkchi2017}. $r_\mathrm{cover}$ denotes the projected radial extent of the satellite survey. The data source column lists the data sets used for satellite detection, confirmation (i.e. measuring candidate satellite distances), and photometry. The letters stand for: D-DECaLS, Dc-DECam (but not from DECaLS), C-CFHT/MegaCam, H-Subaru/HSC, G-Gemini/GMOS, M-Magellan/IMACS. The literature references used in this column are: S18-\citet{smercina2018}, B19-\citet{bennet2019}, M12-\citet{mcconnachie2012}, C13-\citet{chiboucas2013}, M15-\citet{muller2015}, M17-\citet{muller2017}, M18-\citet{muller2018_trgb}, M19-\citet{muller2019}, C19-\citet{crnojevic2019}. The reference column lists references for the host distance, $BV$ photometry, and stellar mass, respectively.  The sources are:  Beat19-\citet{beaton2019}, CF-\citet{kourkchi2017} and \citet{tully2016}, RS11-\citet{rs11}, H13-\citet{humphreys2013}, K13-\citet{karachentsev2013}, K14-\citet{karachentsev2014}, K15-\citet{karachentsev2015b}, T19-\citet{tully2019}, Pea15-\citet{peacock2015},  McQ16a-\citet{mcquinn2016a}, McQ16b-\citet{mcquinn2016b}, McQ17-\citet{mcquinn2017}, Lee13-\citet{lee2013}, M18-\citet{muller2018}, Ler19-\citet{leroy2019}, Cook14a-\citet{cook2014a}, Cook14b-\citet{cook2014b}, GALEX-\citet{galex2007}, RC3-\citet{rc3}, HT11-\citet{ht11}, Sick15-\citet{sick2015}, BG16-\citet{bland2016}, Lic15-\citet{licquia2015}. A stellar mass reference of $M_{K_s}$ denotes that the stellar mass came from the $K_s$ luminosity and a $M_\star/L_{K_s}=0.6$, assuming $M^\odot_{K_s}=3.27$ mag in the Vega system \citep{willmer2018}. }
\end{deluxetable*}

The host list is given in Table \ref{tab:hosts}. There are 31 primary, massive hosts that pass our selection criteria. 30 of these are surveyed for satellites in this work or by previous work in the literature, leaving only one host (NGC 3621) unanalyzed. We list names, distances, redshifts\footnote{The velocity for NGC 3627 is actually the average of the three centrals in the Leo Triplet: NGC 3627, NGC 3623, and NGC 3628.}, and various photometry for each of the hosts. The stellar masses generally come from multi-wavelength photometry (including IR band-passes) and have been corrected for the updated distances that we use. In Table \ref{tab:hosts}, the $r_\mathrm{cover}$ column lists the approximate radial coverage (in projected kpc from the host) of the survey imaging used to find candidate satellites.

\begin{figure*}
\includegraphics[width=\textwidth]{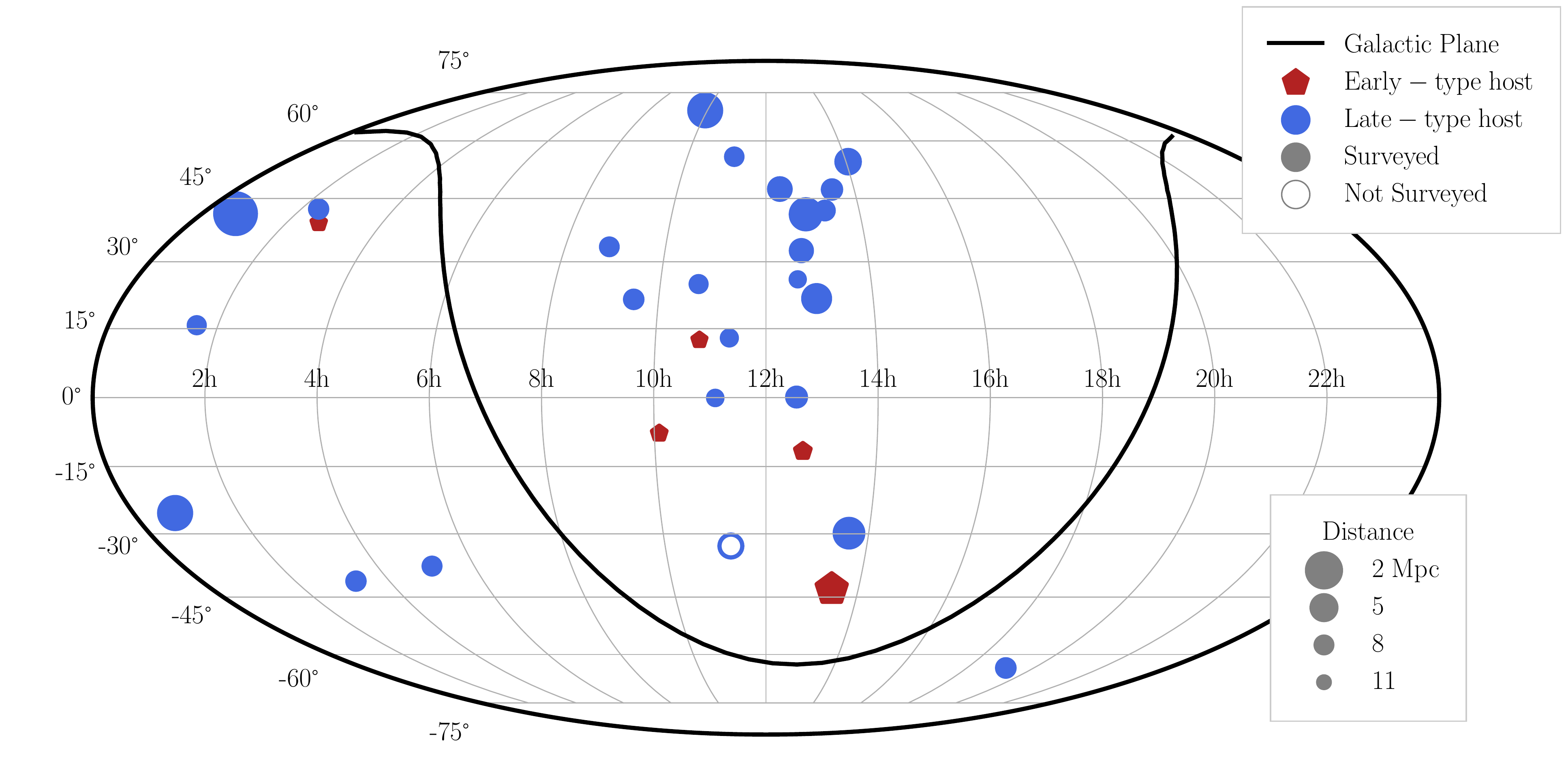}
\caption{Position on the sky of the ELVES host sample. Hosts are colored by their morphology: late-type spirals vs. early-type ellipticals and lenticulars. Point size indicates the host distance. The unfilled point labelled `Not Surveyed' indicates a host where we have no survey coverage (NGC 3621). All other hosts have candidate satellite lists.}
\label{fig:host_posn}
\end{figure*}

\begin{figure*}
\includegraphics[width=\textwidth]{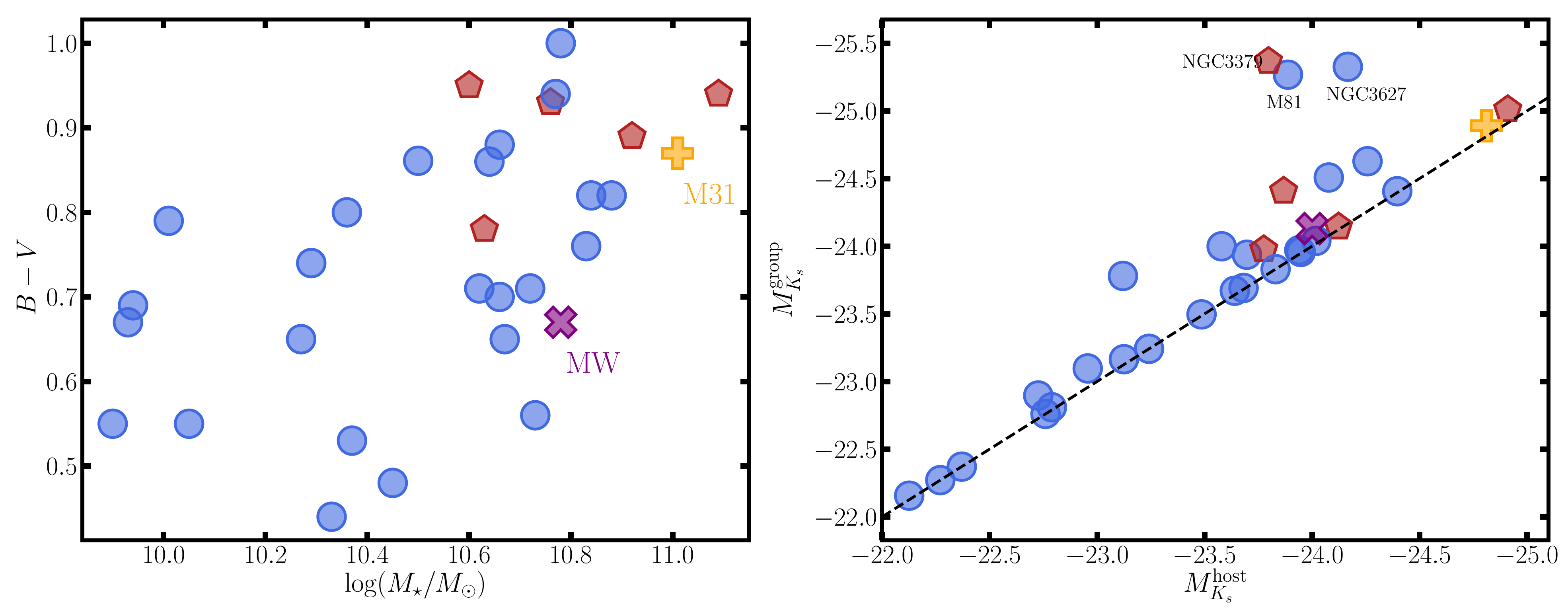}
\caption{Properties of the full ELVES host sample. \textit{Left:} The host color vs. stellar mass. Points are the same as in Figure \ref{fig:host_posn}. \textit{Right}: Host $K_s$ luminosity vs. group $K_s$ luminosity. The three hosts with the largest difference between the two luminosities are labelled. NGC 3379 and NGC 3627 are the primaries of the Leo I and Leo Triplet groups, respectively.}
\label{fig:host_prop}
\end{figure*}

The on-sky locations of these hosts are shown in Figure \ref{fig:host_posn}. In Figure \ref{fig:host_prop}, we show various properties of the host sample. The hosts are fairly evenly distributed between $10^{10}$ and $10^{11}$\msun~ in stellar mass. The right panel shows the group $K_s$ luminosity (the sum of the luminosity for all group members) reported in the group catalog of \citet{kourkchi2017} versus the luminosity of just the primary host. In most cases, these values are quite close, indicating the host completely dominates the luminosity of the group. For a few cases, the group luminosity is significantly larger than that of the primary host. These include the M81/M82 group, the Leo I group, and the Leo Triplet (labelled in the Figure), each of which has several massive galaxies. In previous papers on this survey, we have referred to these hosts as `small-groups' to distinguish that they are notably higher mass than bona fide MW-analogs.

\section{Observational Data}
\label{sec:data}
The full ELVES survey is an extension of our earlier survey presented in \citet{carlsten2020a, carlsten2020b}. That work presented a set of LV hosts surveyed with archival CFHT/MegaCam data\footnote{Specifically 10 hosts were included, but only 8 made it into the ELVES host sample, 2 of which are redone here with DECaLS data due to the limited radial coverage of the CFHT/MegaCam data.}. The ELVES sample is considerably expanded by using the extremely wide-field DECaLS imaging data sets \citep{decals, bass1, bass2}. 

As stated above, the process of cataloging nearby satellite systems is two-fold. First, candidate satellites need to be detected either visually or with an automated algorithm. Second, the candidate satellites need to have their distances measured to confirm that they are actually physically associated with a purported host. In \citet{carlsten2020a, carlsten2020b}, the CFHT/MegaCam data was of high-enough quality to do both steps, with the second step being accomplished via measurements of surface brightness fluctuations (SBF). However, the DECaLS data is almost always too shallow and with too poor of seeing to do the distance measurements, and thus we generally only use it for the object detection. The distance measurements come from SBF using various follow-up data, including archival Hyper Suprime-Cam (HSC) and new Gemini and Magellan imaging. 

Table \ref{tab:hosts} lists the data sources used for each host, including for candidate detection and distance confirmation. The table also lists data sources for the satellite photometry used throughout the paper, which is almost always the same as the candidate detection data, except for where we use a literature reference for the satellite detection. Overall, we use literature sources for the satellite lists for 5 hosts (MW, M31, CenA, NGC 5236, and M81) and do our own object detection for 25 hosts. 17 of these use DECaLS imaging, 6 (NGC 1023, NGC 4258, NGC 4565, NGC 4631, NGC 5194, and M104) use the CFHT/MegaCam search of \citet{carlsten2020a}\footnote{Two of these (NGC 4631 and NGC 4258) additionally use DECaLS imaging to extend the radial coverage slightly, as explained more below.}, one (NGC 891) uses separate CFHT/MegaCam data, and one (NGC 6744) uses (non-DECaLS) DECam data.  

Regardless of the source for the imaging data, we always use two bands for candidate detection: either $g/r$ or $g/i$. The one exception is NGC 891, where for about half of the surveyed area, only $r$-band data are available. Deeper follow-up data for distance confirmation are often only a single band (either $r$ or $i$) since those are the ideal SBF bands, and a color measurement already exists from the candidate detection data. 

The reduced DECaLS imaging data are taken directly from the web server\footnote{\url{https://www.legacysurvey.org/}.} where $g$ and $r$ band data are available for download. We use DECaLS DR8 for the candidate detection but have redone the satellite photometry using DR9 as the DR9 reduction includes a more optimized sky-subtraction for LSB galaxies\footnote{See \url{https://www.legacysurvey.org/dr9/sky/}.}. Overall, we find minimal difference between DR8 and DR9.

CFHT/MegaCam data are downloaded raw from the CFHT archive server\footnote{\url{http://www.cadc-ccda.hia-iha.nrc-cnrc.gc.ca/}} and are reduced in the fashion described in \citet{sbf_calib} and \citet{carlsten2020a}. DECam data that are not part of DECaLS (NGC 6744 is the only host that uses this) are downloaded raw from the NOIRLab archive\footnote{\url{https://astroarchive.noirlab.edu/}} and reduced in a very similar way to the CFHT data. Generally, SDSS \citep{sdss_df14} or Pan-STARRS \citep{panstarrs, panstarrs2} are used for the photometric calibration. For the few southern hemisphere hosts outside those footprints, we use \emph{Gaia} \citep{gaia_dr2} or APASS \citep{apass} for the photometric calibration. 

Gemini/GMOS data are reduced using the \texttt{DRAGONS} software package\footnote{\url{https://dragons.readthedocs.io/en/v3.0.0/}}. Gemini/GMOS data from programs FT-2020A-060, US-2020B-037, and US-2021B-0018 (PI: S. Carlsten) are used. Magellan/IMACS data are reduced using a custom pipeline that does all the main steps in a similar way to the MegaCam and DECam pipelines. Finally, the HSC data are downloaded from the Subaru archive\footnote{\url{https://smoka.nao.ac.jp/fssearch}} and reduced using the pipeline written by the HSC team \citep{bosch2018}. All HSC data are archival and include (but are not limited to) data from the HSC-SSP survey \citep{aihara2018, aihara2019}. 

All of these telescopes and camera systems use Sloan-like filters. However, there will be some slight differences between the different filter sets. Following \citet{carlsten2021a}, we do not make any attempt to bring them all onto the same filter system. That work (Figure 16 in that paper) showed that the differences between the filters is expected to be $\lesssim0.1$ mag based on synthetic photometry from stellar evolution models \citep{mist_models}. One instance where systematic offsets of this magnitude might matter is integrated color. However, the effect of this systematic uncertainty in inferring stellar mass (the primary use of color) is generally less than the statistical uncertainty coming from photon noise and/or sky subtraction issues and thus we do not account for it. 

The actual areal coverage of each host is shown in Appendix \ref{app:host_areas}. For the DECaLS hosts, we almost always are able to extend fully to 300 kpc, an approximate value for the virial radius of halos of the typical mass expected for most ELVES hosts ($M_\mathrm{halo}\sim10^{12}$\msun). We note that the halo masses of the most massive hosts in the ELVES sample (e.g. M104, NGC 3627, and NGC 3379) are likely closer to $\sim10^{13}$ \citep{karachentsev2014_masses, kourkchi2017} and will have much larger virial radii. On the other hand, the least massive ELVES hosts (e.g. NGC 3344 and NGC 4517) will have smaller virial radii. The hosts from the non-DECaLS data sources (CFHT/MegaCam or DECam) all have less coverage. For a couple of these (NGC 4631 and NGC 4258), we extend the CFHT/MegaCam data with some DECaLS data and additional SBF follow-up, with the specific footprints shown in Appendix \ref{app:host_areas}. The approximate radial extent of the search footprints for each host are listed as $r_\mathrm{cover}$ in Table \ref{tab:hosts}.

\section{Candidate Satellite Detection}
\label{sec:detection}

For the 25 hosts where we conduct our own candidate satellite detection, we apply the semi-automated detection algorithm optimized for LSB dwarf galaxies detailed in \citet{carlsten2020a}. This algorithm draws heavily on the LSB galaxy detection pipelines of \citet{bennet2017} and \citet{greco2018}. 

We use a custom search algorithm as opposed to simply using common object detection algorithms \citep[e.g. \texttt{SExtractor},][]{sextractor}, because those algorithms are not optimized for large, diffuse, LSB dwarf galaxies. Such sources are likely to get ``shredded'' (i.e. over-deblended). More advanced algorithms \citep[e.g.][]{noise_chisel, lang2016, profound, melchior2018, haigh2021} are being developed that demonstrate promise for LSB dwarf surveys \citep{venhola2021}. While our search algorithm is more complicated than simply using \texttt{SExtractor}, it still does make use of \texttt{SExtractor} in multiple `hot-and-cold' runs (i.e. runs with high and low thresholds, see below) intermixed with aggressive masking. 

Like many other object detection algorithms, the one used here explicitly separates source detection from source extraction. In this section, we detail the first step which is simply to find the dwarf galaxies. A later section (\S\ref{sec:sats}) describes how we measure the photometric properties of the candidate satellites.

\subsection{Overview of Detection Algorithm}
\label{sec:det_algo}
The search algorithm is detailed in \citet{carlsten2020a} and is applied essentially unchanged to the DECaLS data, but we provide a brief overview of the important steps in this section. The main steps of the process are:

\begin{enumerate}
    \item Create bright star masks from \emph{Gaia} star catalogs for stars brighter than $20^{th}$ magnitude. Custom shaped masks that cover the scattered light halos and saturation spikes of bright stars are applied to all of the survey fields. 
    
    \item Initial detection step with a large size and moderate significance threshold to find large, relatively high surface brightness candidate satellites. In particular, \texttt{SExtractor} is run with a $5\sigma$ threshold and $\sim2000$ pixel minimum area (corresponding to a radius of $\sim 7$\arcs~ for a circular source). To remove some massive background galaxies, an effective surface brightness cut of $\langle \mu \rangle_{r_e} \gtrsim 22$ mag arcsec$^{-2}$ (calculated from \texttt{SExtractor} fluxes and sizes) is applied to the detections. Note that this is not the only stage where relatively massive satellites can be detected as their diffuse/LSB outskirts will also be detected in the LSB detection step below \citep[and dwarfs universally have diffuse outskirts, see][]{kadofong2020}. 
    
    \item Mask bright sources and their associated diffuse emission as in \citet{greco2018}. Sources are detected with both a high, $15\sigma$, and low, $\sim1\sigma$, detection threshold. The LSB detections are associated with a high surface brightness (HSB) detection if they overlap more than a certain fraction ($5\%$) of their pixels with the HSB detection. The detected LSB pixels that do get associated with an HSB detection get masked. These represent, for example, the extended envelopes of background galaxies, intracluster light from galaxy clusters, and/or scattered light around stars not bright enough to be covered by the \emph{Gaia} star mask. After this step, there is a another simple masking of all sources over $\sim4\sigma$ above the background.
  
    \item Filter the image to emphasize diffuse, LSB emission. The masked image is convolved with a Gaussian with FWHM $\sim2\times$ the field's PSF size. 
    
    \item Second detection step with a low threshold to find faint candidate satellites. Sources in the masked and filtered image are detected with a significance threshold of $\sim2\sigma$ and size greater than $\sim800$ pixels (corresponding to a radius of $\sim 4$\arcs~ for a circular source). 
    
    \item Visual inspection to remove artifacts and other false positives. The vast majority ($>95\%$) of the things removed in this step are clear artifacts such as saturation spikes or emission from bright stars or the outskirts of massive elliptical galaxies that leaked through the masks. Additionally, detected galaxies that are clear background contaminants are removed. For instance, small ($r_e \lesssim 5$\arcs) galaxies that have well-defined spiral arms or a bulge$+$disk morphology are clearly not in the Local Volume. Low-mass dwarfs will necessarily be fairly featureless and diffuse. \citet{carlsten2020a} gives more details on this step and examples of visually rejected galaxies. Often during this step, particularly ambiguous detections are queried in Simbad and any galaxies with known redshifts more than $275$\footnote{We discuss this velocity criterion later in \S\ref{sec:distances}.} km/s greater than the host are background objects and not further considered. 
    
\end{enumerate}

The first five steps are done independently in the $g$ and $r$ (or $i$) bands and then the two detection catalogs are merged before the visual inspection step. Objects need to be detected in both bands to pass to the visual inspection step.

Many of the algorithm's parameters needed to be slightly tweaked for each field to account for the differing depths of the different datasets and host distances. Each parameter preceded by a $\sim$ in the description above was often tuned for each host. Hosts with shallower imaging, for instance, needed a larger smoothing kernel (i.e. more blurring), and more nearby hosts had larger size thresholds for detected objects. The parameters were tuned in an initial step where a subset of the survey coverage for each host was visually searched for candidate satellites against which the algorithm was tested. The specific parameters used for each host are not particularly important to the results as the search algorithm's completeness for each host is quantified using those same parameters. Thus, all affects they have on the search completeness are accounted for.

\begin{figure*}
\includegraphics[width=\textwidth]{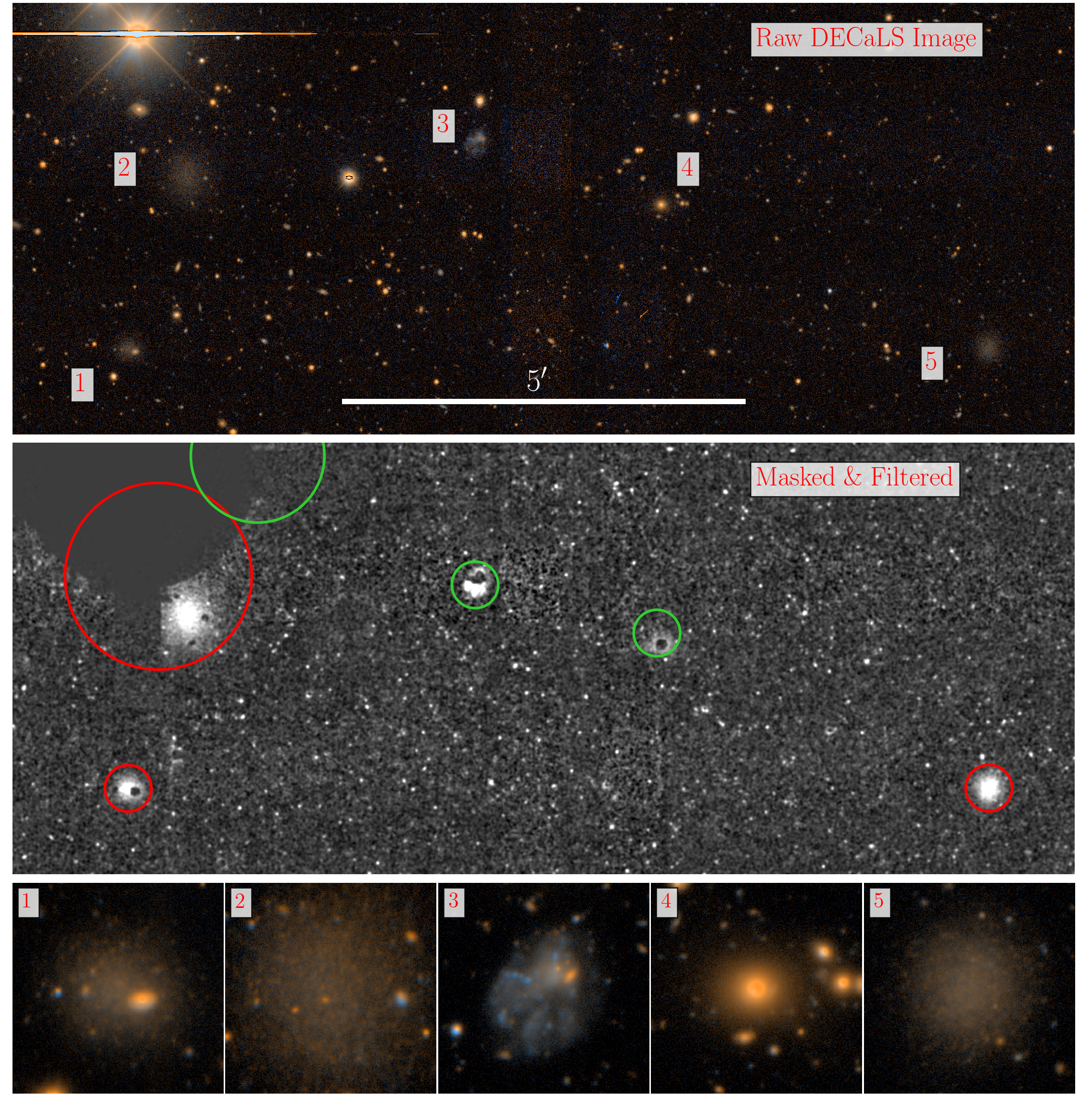}
\caption{Demonstration of the detection algorithm on a surveyed area near NGC 3379. Top panel shows the raw DECaLS detection images ($g+r$ color composite). The middle panel shows the detection image after it has been masked for bright stars and galaxies and has been filtered to bring out LSB emission. Detections from the pipeline are shown as circles. The detections that pass visual inspection are in red while those that do not are shown in green. The rejected detection in the top left is from scattered light from the bright star leaking through the mask while the other two detections are clear background galaxies (see text for more details). The five labeled detections are shown in the bottom row in deep HSC images. These images are not used in the detection process but are used to confirm the candidate satellites via SBF distance measurements. The images on the bottom row are 40\arcs~ (2 kpc at the distance of NGC 3379) to a side.}
\label{fig:algo}
\end{figure*}

Figure \ref{fig:algo} demonstrates the main steps through an example part of the field around NGC 3379, the main host of the rich Leo I group. The top panel shows the raw DECaLS image ($g+r$ color composite). The middle panel shows the image with bright stars and background galaxies masked and filtering applied. The masking in Step 3 above is clearly quite aggressive, but we show in the next section on completeness that it is not overly aggressive because it does not completely mask candidates. The detections from Step 5 are shown as the superimposed circles. Those that pass visual inspection are in red while those that are rejected are in green. The five detections that are labeled in the top image are shown in close-up in the bottom row using extremely deep ($>2$ hour integration) HSC images\footnote{The HSC images are just being shown for clarity, the visual inspection step is done just on the detection images (DECaLS in this case).}. The sixth rejected detection is simply emission that leaked around the star mask and is not shown in close-up. Detections 1, 2, and 5 are high-quality candidate satellites. In fact, these candidates are all confirmed satellites via SBF distance measurements. The two rejected detections (3 and 4) are background galaxies. Object 3 has a complicated morphology and small size ($r_e\sim5$\arcs), indicating it is not in the Local Volume. In this case, its HI redshift from ALFALFA \citep{haynes2018} of several thousand km/s greater than that of NGC 3379 confirms this, although not all visually rejected galaxies have redshifts. Object 4 is a massive background elliptical whose outskirts leaked through the masking. While this visual inspection step might appear to introduce ambiguity, visual selection of dwarfs in the Local Volume is a powerful tool due to their proximity \citep[e.g.][]{karachentseva1968, karachentsev2007} and many contemporary searches are primarily visual \citep[e.g.][]{byun2020, habas2020}.

The median false positive rate as determined by the visual inspection step is $\sim30$ rejected detections (e.g. artifacts, halos around bright stars, or background galaxies) for each candidate satellite\footnote{Note that this `false positive rate' describes the ratio of the number of total detections to those that pass as candidate satellites. There is another `false positive rate' to describe the ratio of total number of candidate satellite to those that actually get confirmed as physical satellites via a distance measurement which will be discussed in \S\ref{sec:distances}.}. As stated above, this is mostly a result of insufficient masking of bright stars or bright background galaxies that allows too much emission to pass through. This step did require a fairly significant amount of human involvement, meaning that this algorithm is not scalable for the extremely wide-field surveys of the Roman Space Telescope or Vera Rubin Observatory. For the relatively small areas surveyed in ELVES ($\sim400$ square degrees total), removing these contaminants by visual inspection was tractable.

\subsection{Completeness of Surveys}
\label{sec:completeness}
In order to facilitate comparison to other satellite dwarf searches and theoretical models, it is critical to quantify the completeness of the survey. We do this by injecting artificial galaxies into the survey images and quantifying the fraction of injected galaxies recovered as a function of input luminosity and size. The fact that our search process is semi-automated (as opposed to purely visual) greatly facilitates this process. We do not apply any visual inspection step to the artificial galaxies and assume that any injected galaxy detected by the search algorithm would pass the visual inspection. In this section, we describe the artificial galaxy injection and show the overall average completeness of the survey.

The process for injecting artificial galaxies is detailed in \citet{carlsten2020a}. The DECaLS and non-DECaLS hosts are treated a little differently, primarily in that the non-DECaLS hosts have their artificial galaxies injected at the chip level before sky subtraction and coaddition. On the other hand, the DECaLS hosts have their artificial galaxies injected directly in the coadds due to the significantly larger volume of imaging data, both in number of hosts and area covered per host.  In either case, exponential profiles (i.e. $n=1$ S\'{e}rsic profiles) are injected at various $r$ or $i$ luminosities and central surface brightness levels, with colors in the range $g-r\in[0.3,0.6]$ or $g-i\in[0.4,0.8]$ depending on which filters are used. The data are then run through the detection pipeline, and the fraction of galaxies in each luminosity and surface brightness bin that are recovered is recorded. Simulated dwarfs that fall on top of areas lost to the star-masking are not treated differently, thus, we do not expect the recovery fraction to ever be $100\%$. The area lost to star-masking is generally around $\sim5\%$.

As shown in \citet{carlsten2021a}, the input galaxy parameters roughly resemble the S\'{e}rsic indices and colors of the real satellites, both quenched and star-forming. However, true dwarf galaxies generally have slightly lower average S\'{e}rsic index of $n\sim0.7$, a result also seen in other surveys \citep[e.g.][]{eigenthaler2018, ferrarese2020}. At fixed luminosity and central surface brightness, a higher S\'{e}rsic index dwarf will be more difficult to find since more of the emission will be be at a very low surface brightness level in its outskirts, and more likely to be buried in the sky noise. Thus, we expect these completeness estimates to be slightly conservative.

\begin{figure*}
\includegraphics[width=\textwidth]{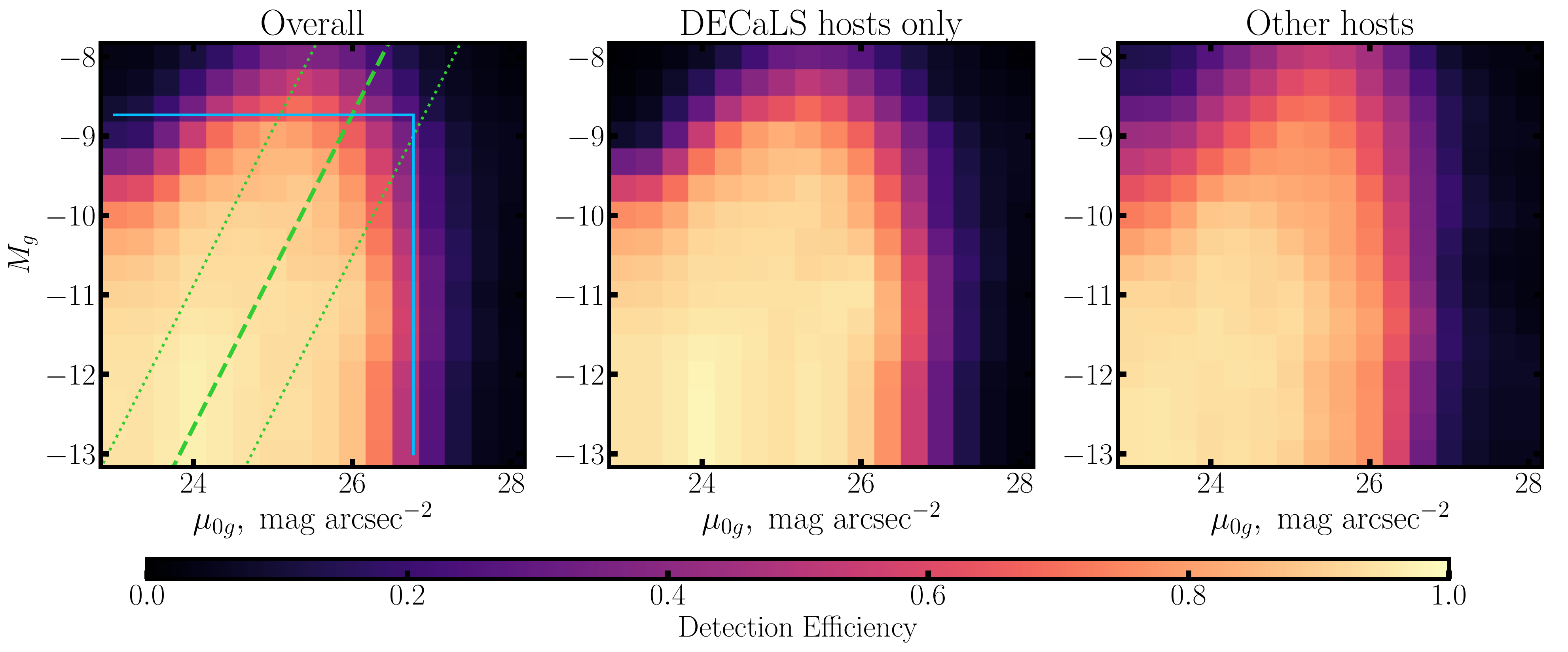}
\caption{Average completeness of the dwarf searches as quantified by image simulations with injected artificial galaxies. Left panel shows the average of all hosts, while center and right panels split the hosts by the source of the imaging data. The blue region in the left panel demarcates the completeness quoted in prior ELVES papers ($M_V<-9$ mag and $\mu_{0,V}<26.5$ mag arcsec$^{-2}$). The green dashed line shows the mass-size relation of \citet{carlsten2021a} converted into this plane, and the green dotted lines show the $1\sigma$ intrinsic scatter measured in the mass-size relation.}
\label{fig:completeness_avg}
\end{figure*}

Figure \ref{fig:completeness_avg} shows the results of the injection simulations. The left panel shows the average recovery fraction across all 25 hosts. In doing the average, the hosts are weighted by the number of confirmed and possible satellites (see \S\ref{sec:distances} below). The center and right panels show the average recovery fraction of the DECaLS hosts separately from the other hosts (primarily CFHT/MegaCam hosts). 

Results for individual hosts are shown in Appendix \ref{app:host_completeness}. The recovery fractions are generally similar across the different hosts and data sources. We note that the BASS part of DECaLS (which 4 DECaLS hosts use) and the CFHT/Megacam data used for NGC 891 are significantly shallower than the rest of the hosts with the surface brightness completeness being $\sim0.5$ mag lower. However, this is a relatively small number of hosts, and we do not further distinguish these from the others.

The blue region in the left panel of Figure \ref{fig:completeness_avg} shows the completeness limits quoted in previous ELVES papers: $M_V<-9$ mag and $\mu_{0,V}<26.5$ mag arcsec$^{-2}$, assuming $g-V\approx0.25$ mag. The green dashed line shows the mass-size relation of \citet{carlsten2021a} converted into this plane assuming $M_\star/L_g=1.24$ which comes from the color-$M/L$ relation of \citet{into2013} and the average color of $g-i=0.74$. An $n=1$ S\'{e}rsic profile is assumed in determining the central surface brightness from luminosity and effective radius. 

Overall, the recovery fraction is $\sim90-95\%$ at bright magnitudes and surface brightness, reflecting the area lost from the search to star masks. There is a clear drop-off in completeness between $\mu_{0,g}\sim26-27$ mag arcsec$^{-2}$. The $\mu_{0,V}<26.5$ mag arcsec$^{-2}$ we have quoted in the past is roughly the 50\% completeness limit. 

In addition to the artificial galaxy injection tests, our completeness is proven through comparison with previous satellite searches in the literature for several of the ELVES hosts. In Appendix \ref{app:prev_work}, we go through the hosts in more detail, and provide detailed comparison with previous searches from the literature for hosts that have them. The comparisons are quite favorable for ELVES with no candidate satellites missed above the asserted completeness level across the 8 ELVES hosts that have been significantly surveyed in the past.

For the five hosts that we do not perform our own object detection (MW, M31, CenA, M81, M83), we assume they are complete down to the fiducial level of $M_V<-9$ mag and $\mu_{0,V}<26.5$ mag arcsec$^{-2}$. The actual completeness limits of these literature surveys are generally quite a bit deeper due to the closer proximity of these hosts relative to most ELVES hosts. Details can be found in the sources for those satellite lists (see Table \ref{tab:hosts}).

While we do detect numerous candidate satellites fainter than $M_V=-9$ mag, \textbf{all ELVES satellite lists and figures in this paper have a $M_V<-9$ mag cut applied to them.}

\section{Candidate Satellite Confirmation}
\label{sec:distances}
In this section, we describe the process of confirming candidate satellites as actual physical satellites of a LV host. Many of the candidate satellites have prior distance information, either \emph{HST}-based TRGB distances or redshift, but the majority do not. Thus, we rely heavily on efficient surface brightness fluctuation (SBF) measurements to provide a large number of distance constraints with modest ground-based data. We try to constrain the distance to as many candidate satellites as possible, but several remain with no distance constraints. We keep careful track of these `unconfirmed/candidate' satellites throughout this paper as they require special treatment in any analysis using ELVES satellite lists.

\subsection{SBF Measurement}
Details of the SBF measurement are given elsewhere \citep{sbf_calib, carlsten2020b}, so we only provide a brief summary here. SBF entails measuring how `patchy' a galaxy is in the semi-resolved regime. Distance smooths out Poisson fluctuations in the number of bright stars per resolution element such that a galaxy will look smoother further away. In measuring a galaxy's SBF, this `patchiness' is quantified as the `SBF magnitude', $\bar{M}$. 

To quantify this, we start with a smooth model of the galaxy's surface brightness profile. The reddest band we have ($r$ or $i$) is modeled, as SBF is brighter and seeing is generally better for redder pass-bands \citep[e.g][]{jensen2003, scott_psf}. Due to the small sizes and low luminosities of most of the dwarfs, we always use a parametric S\'{e}rsic profile as a model for the smooth profile (see \S\ref{sec:sats} for more information on the S\'{e}rsic profile photometry). This model is subtracted out from the raw image and the result is normalized by dividing it by the square root of this model. We then calculate the azimuthally averaged power spectrum of this result via a fast Fourier transform. The fluctuation power due to SBF is the component of the power spectrum on the scale of the PSF. For details on how we mask contaminants, we refer the reader to \citep{sbf_calib, carlsten2020b}.

\subsection{SBF Calibration}
Due to the fact that the average brightness of stars in a galaxy will depend on the stellar population, the level of SBF also will depend on the stellar population present in a galaxy. Thus a calibration is used to relate the absolute SBF magnitude to galaxy color. For $i$-band measurements, we use the calibration provided in \citet{sbf_calib}, while for $r$-band measurements, we use the calibration in \citet{carlsten2020b}. 

These calibrations are specifically for CFHT/MegaCam. In this work we use five different telescope and instrument combinations (CFHT/MegaCam, Subaru/HSC, Blanco/DECam, Gemini/GMOS, and Magellan/IMACS) that all have Sloan-like $r/i$ filters and that all will be slightly different. However, from experiments calculating theoretical SBF magnitudes using the MIST isochrones and synthetic photometry \citep{mist_models} for MegaCam, HSC, and DECam, we find that the SBF magnitudes are always the same within $\sim0.1$ mag, which is small compared to the statistical error of the SBF measurement. With that said, we do incorporate a small $-0.1$ mag adjustment to the CFHT-based calibration when using DECam filters. We do not perform any adjustments when using HSC, GMOS, or IMACS data, however. \citet{kim2021} show that their empirical SBF calibration in the HSC filter system is quite close to that of \citet{sbf_calib} with any difference more likely attributable to different masking strategies than filter differences.

\subsection{SBF Distances}
\label{sec:sbf_criteria}
With the SBF measurements and calibration in hand, we can constrain the distance to each dwarf. Following \citet{carlsten2020b}, there are three outcomes to this measurement. First, if the SBF measurement is of high S/N ($>5$) with a distance that is consistent with that of the host within $2\sigma$, we consider the candidate satellite to be a `confirmed' satellite.  Second, if the $2\sigma$ lower bound of the SBF distance constraint is beyond the distance of the host, we consider the candidate to be rejected as a background contaminant\footnote{Alternatively, the $2\sigma$ upper bound to the distance could be nearer than the distance of the host (i.e. the candidate is foreground). However, this only happens once.}. The actual distances of these objects are generally not known as the SBF measurement only sets a distance lower bound from the absence of measurable SBF. Third, if the SBF measurement is of too low S/N and the distance lower bound is within the distance of the host, the candidate is kept in the `unconfirmed/candidate' satellite category. These objects may or may not be satellites and need deeper data or \emph{HST} observations for their satellite status to be decided. 

We use this cutoff of S/N $>5$ in the SBF measurement as a safeguard against false positive satellite confirmation. In our experience, this S/N level is roughly where SBF becomes clearly distinguishable from other sources of fluctuation power, such as scattered star-forming clumps in the galaxy or background galaxies. \citet{sbf_m101} used SBF to confirm two satellites around NGC 5457 (M101). These satellites had SBF S/N $\gtrsim7$, above the threshold used here. Both of these have since been confirmed by \emph{HST} TRGB distances in \citet{bennet2019}. \citet{sbf_m101} also discussed 2 other candidates as promising follow-up targets that had moderate signal with S/N$\sim2-3$. The \emph{HST} imaging of \citet{bennet2019} showed that the galaxies were background contaminants. The small fluctuation signal appeared to be coming from unmasked background galaxies, not SBF. Using the threshold adopted here, these two candidates would be conservatively included in the `unconfirmed/candidate' satellite category.


In \citet{sbf_calib}, we showed that the ground-based SBF distances agreed with \emph{HST} TRGB distances with an RMS of $15\%$.  While this is lower precision than the $\sim5-10\%$ precision that \emph{HST} TRGB distances can attain \citep{beaton2018}, it is still quite useful in confirming or rejecting candidate satellites as actual, physical satellites of a certain host. Thus, by including satellites with SBF distances within $2\sigma$ of the host,  we do expect that some near-field dwarfs (e.g. dwarfs analogous to the Local Group field dwarfs nearby to the MW) will be included as `satellites'. Experimenting with halos from the IllustrisTNG simulations \citep{tng1, tng2}, in \citet{carlsten2020b}, we found that this distance constraint led to a contamination of field interlopers of around $15-20\%$. Until more precise distances are possible, nothing can be done regarding this contamination, and it must be kept in mind when analyzing ELVES satellite lists. We note that this is a similar contamination fraction to what occurs when selecting satellites with a recessional velocity cut, as in the SAGA Survey \citep{mao2020}.

\subsection{Other Distance Information}
In addition to SBF measurements, we make use of TRGB distances and redshift information, when available. We search the Updated Nearby Galaxy Catalog of \cite{karachentsev2004, karachentsev2013} for TRGB distances for all of the candidate satellites. Similarly, we query SIMBAD for any redshift information. Dwarfs with TRGB distances within $1\sigma$ (or 1 Mpc, if no distance uncertainty is available) of the distance to their host are considered `confirmed' with those outside of this range rejected as contaminants. For dwarfs with redshifts, those with velocities within 275 km/s of their host are confirmed satellites \citep[e.g.][]{mao2020} while those outside this range are rejected. We note that the search for redshift information often occurs very early on in the candidate detection pipeline (in particular, in the visual inspection step) with background contaminants being removed at this point, even before the candidate satellite stage. Thus, many otherwise passable candidate satellites are not in the final candidate satellite lists since they had redshift information indicating they are background.

\begin{deluxetable*}{cccccccc}
\tablecaption{Overview of Distance Constraints.\label{tab:dists_overview}}
\tablehead{
\colhead{Name} & \colhead{$M_{K_s}^\mathrm{group}$} & \colhead{\# candidates}  & \colhead{\# SBF conf.}  & \colhead{\# other conf.}  & \colhead{\# SBF rejected}   & \colhead{\# other rejected}   & \colhead{\# unconfirmed}   \\ 
\colhead{} & \colhead{(mag)}  & \colhead{}   & \colhead{}  & \colhead{} & \colhead{} & \colhead{} & \colhead{}   }  
\startdata
NGC253 & -23.96 & 8 & 0 & 5 & 0 & 2 & 1\\
NGC628 & -22.81 & 17 & 7 & 6 & 3 & 0 & 1\\
NGC891 & -23.83 & 8 & 1 & 2 & 1 & 0 & 4\\
NGC1023 & -23.98 & 22 & 9 & 5 & 5 & 0 & 3\\
NGC1291 & -23.97 & 21 & 12 & 2 & 1 & 2 & 4\\
NGC1808 & -23.78 & 18 & 7 & 3 & 4 & 0 & 4\\
NGC2683 & -23.50 & 12 & 3 & 3 & 2 & 0 & 4\\
NGC2903 & -23.69 & 16 & 4 & 3 & 9 & 0 & 0\\
NGC3115 & -24.14 & 27 & 13 & 4 & 8 & 0 & 2\\
NGC3344 & -22.27 & 9 & 3 & 0 & 2 & 0 & 4\\
NGC3379 & -25.37 & 85 & 16 & 28 & 17 & 2 & 22\\
NGC3521 & -24.41 & 18 & 9 & 2 & 6 & 0 & 1\\
NGC3556 & -22.76 & 17 & 0 & 1 & 2 & 1 & 13\\
NGC3627 & -25.33 & 33 & 6 & 14 & 1 & 0 & 12\\
NGC4258 & -23.94 & 21 & 3 & 4 & 12 & 1 & 1\\
NGC4517 & -22.16 & 56 & 7 & 2 & 39 & 8 & 0\\
NGC4565 & -24.27 & 11 & 2 & 2 & 2 & 0 & 5\\
M104 & -25.01 & 18 & 11 & 0 & 3 & 0 & 4\\
NGC4631 & -22.90 & 23 & 7 & 5 & 8 & 2 & 1\\
NGC4736 & -23.10 & 17 & 0 & 7 & 0 & 3 & 7\\
NGC4826 & -23.24 & 12 & 3 & 4 & 2 & 1 & 2\\
NGC5055 & -24.04 & 13 & 5 & 4 & 0 & 0 & 4\\
NGC5194 & -24.51 & 13 & 2 & 1 & 8 & 2 & 0\\
NGC5457 & -23.16 & 42 & 2 & 7 & 33 & 0 & 0\\
NGC6744 & -24.00 & 15 & 4 & 1 & 4 & 0 & 6\\
\hline\\
M31 & -24.89 & -- & -- & 20 & -- & -- & --\\
M81 & -25.27 & -- & -- & 24 & -- & -- & --\\
CENA & -24.41 & -- & -- & 22 & -- & -- & --\\
NGC5236 & -23.67 & -- & -- & 11 & -- & -- & --\\
MW & -24.13 & -- & -- & 10 & -- & -- & --\\
\hline\\
\hline\\
Total & -- & 552 & 136 & 202 & 172 & 24 & 105\\
\enddata
\tablecomments{Overview of the distance confirmation results. For each host the total number of candidates, number of confirmed satellites, number of rejected background contaminants, and number of remaining unconfirmed/possible satellites are given. The numbers of confirmed and rejected dwarfs are split by whether the constraint is from SBF or otherwise (TRGB and/or velocity). Many of the satellites confirmed via redshift or TRGB also have an SBF distance with overall close agreement between the methods. The five previously surveyed hosts are given at the bottom with the bottom-most row listing the totals in each category.}
\end{deluxetable*}

\subsection{Distance Results}
\label{sec:distance_results}

The distance results for all the satellites for which we found or measured a distance are provided in Appendix \ref{app:distances}.  There are three tables: one for dwarfs which get confirmed as satellites, one for dwarfs that get rejected as contaminants, and one for dwarfs that remain as possible/candidate satellites. We do not provide any distance constraints for these latter dwarfs since any distance constraints for them are, by definition, not informative. However, we do list what data source was used, if any, in attempting an SBF distance.

Table \ref{tab:dists_overview} gives an overview of the distance results, including number of confirmed and rejected objects and number of lingering unconfirmed/possible candidates that did not have conclusive distance results. All together, including the five previously surveyed satellite systems (MW, M31, M81, CenA, NGC 5236), there are 338 confirmed satellites and 105 remaining candidates. 136 of these confirmed satellites are confirmed via SBF measurements. Only 31 ($23\%$) of these SBF-based confirmations came from our previous work in \citet{carlsten2020b}. In total, there are 182 confirmed satellites with SBF distances, with 53 also having other distance confirmation (either TRGB or redshift)\footnote{There are seven confirmed satellites with SBF distances and another distance measurement which we count as being confirmed by the SBF. This is because either the SBF distance was published before the other distance or the other distance measurement is quite ambiguous.}. In all but four cases ($7.5\%$), the SBF result agrees with the TRGB or redshift result. In these few exceptions, there is some issue (e.g. nearby bright star halo or irregular morphology) that clearly biases the SBF measurement, causing it to be inaccurate. These are discussed more in Appendix \ref{app:distances}.

\subsection{Candidate Satellite Contamination Estimate}
\label{sec:cont_mod}
In this section, we provide a simple estimate for the probability that each of the remaining unconfirmed/possible candidates is a real satellite of its host. To do this, we consider the fraction of similar candidate satellites that did have conclusive distance results (either from SBF or otherwise) that ended up being confirmed as a satellite. 

We do this as a function of luminosity and surface brightness as the likelihood a candidate is an actual satellite will depend on these. This is similar in principle to the $\mathcal{R}_\mathrm{sat}$ estimate in the SAGA Survey \citep[see \S 5.3 of][]{mao2020} that accounts for photometric candidates in that work that did not have robust spectroscopic measurements. 

\begin{figure*}
\includegraphics[width=\textwidth]{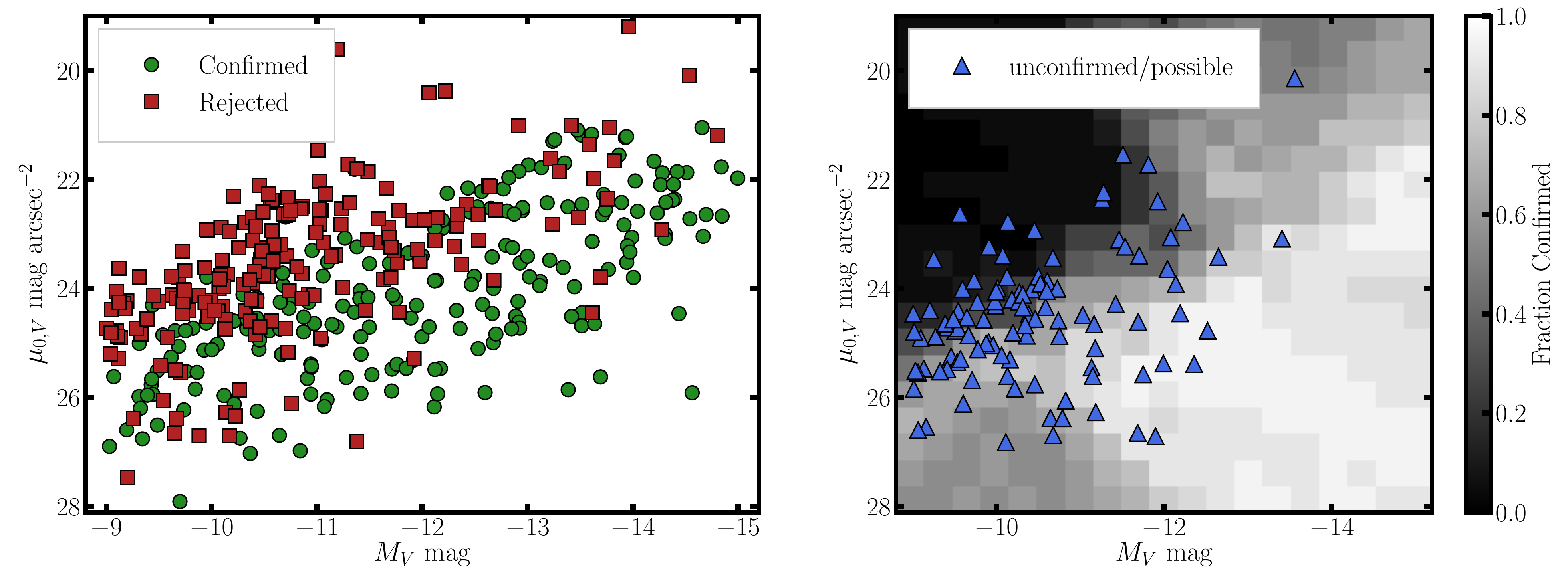}
\caption{A simple estimate for the probability that the remaining unconfirmed/possible candidates are actual physical satellites. The left panel shows the luminosity and surface brightness of both the confirmed satellites and the rejected contaminants (if the latter were to be at the distance of their purported host) that had conclusive distance constraints. The confirmed satellites show a pronounced luminosity-surface brightness relation while background contaminants tend to be smaller with higher surface brightness. The right panel shows an estimate for the likelihood that a candidate will turn out to be a real satellite as a function of luminosity and surface brightness. This likelihood is estimated from the fraction of confirmed satellites out of the closest 20 confirmed or rejected candidates for each point on the $M_V$-$\mu_{0,V}$ plane. The locations of the actual remaining candidates are over-plotted.}
\label{fig:contam}
\end{figure*}

Figure \ref{fig:contam} shows the surface brightness versus luminosity relation for candidates that had conclusive distance constraints (either confirmed as a satellite or rejected). The luminosity of a background contaminant is calculated as if it was at the distance of the host. In most cases, the actual distance of these contaminants are not known; the SBF results only show that they must be beyond the distance of the host. The confirmed satellites show a clear relation between surface brightness and luminosity \citep[this is explored in more detail in][]{carlsten2021a} whereas the rejected contaminants are generally higher surface brightness with smaller sizes. This is suggestive that most of the background contaminants are dwarf galaxies that are $\sim2-3\times$ further in distance than the LV hosts. The confirmed and rejected dwarfs are clearly segregated in the $M_V-\mu_{0,V}$ plane such that the location of an unconfirmed/possible candidate in this plane contains information on how likely it is to be a real satellite. 

To quantify the likelihood of each unconfirmed/possible candidate being real, we simply consider the 20 nearest candidates that had conclusive distance results in the $M_V-\mu_{0,V}$ plane and calculate the fraction of those that were real satellites. We find very similar results if we had instead used anywhere between the 10 to 50 nearest candidates. The right panel of Figure \ref{fig:contam} shows the value of the confirmed fraction at different locations in the $M_V-\mu_{0,V}$ plane. 

There are several simplifying assumptions being made in constructing this likelihood estimate. The first is that this estimate will certainly not take into account the host-to-host scatter in the contamination fraction due to prominent background groups along the line of sight. For instance, NGC 5457 and NGC 4517 have significant contamination due to rich background groups at $D\sim15-30$ Mpc. 

Second, it implicitly assumes that the threshold in data quality required to confirm a satellite via SBF at a given luminosity and surface brightness is the same as that required to reject a background contaminant at that luminosity and surface brightness. It is easy to understand why this assumption is important if we consider an alternate scenario where, for instance, rejecting a background contaminant is possible with much worse data than that required to confirm a satellite (for instance if we required a S/N $>100$ measurement of the SBF to `confirm' a dwarf). In this case, the rejected contaminants would be over-represented in the set of candidates with conclusive distance results, and Figure \ref{fig:contam} would underestimate the likelihood that an unconfirmed candidate is a real satellite. We test this assumption in Appendix \ref{app:sbf_sims} where we show the results of applying the SBF confirmation and rejection criteria (see \S\ref{sec:sbf_criteria}) to simulated dwarf galaxies across a range in luminosity and surface brightness. Taking two different hosts (NGC 4258 and NGC 3379) with high-resolution, deep data as examples, we simulate both dwarfs with SBF at the level appropriate for the distance of the host and dwarfs without SBF, representing background contaminants. Given a level of data quality, it is possible to reject candidates that are fainter by about a magnitude (in luminosity or surface brightness) than the faintest candidates that can be confirmed with that data. This discrepancy would be decreased if we did not require a S/N $>5$ measurement of the SBF to consider a dwarf `confirmed', but we explained the rationale behind that criterion in \S\ref{sec:sbf_criteria}. This $\sim1$ mag difference is smaller than the dynamic range of the remaining candidates, and we expect this estimate of the confirmed fraction is fairly accurate. With that said, we recommend any analysis that uses ELVES satellite lists to test their results considering the cases where all or none of the remaining candidates are real satellites.

In total, summing these satellite probability estimates for all 105 remaining unconfirmed objects yields 49.8 real satellites, meaning that, most likely, there are an additional $\sim50$ actual satellites amongst the unconfirmed candidates. The SAGA Survey \citep[\S 5.3 of][]{mao2020} estimates a total of 24 actual satellites amongst the candidates that did not have successful spectroscopic follow-up. As a fraction of the total number of confirmed satellites in each survey, this is a similar rate of failure in distance follow-up between the two surveys.

\section{Properties of the Satellites}
\label{sec:sats}
In this section, we turn to how we measure various properties of the satellites, including optical and UV photometry. 

All satellite photometry is in the AB system and corrected for MW dust extinction using the $E(B-V)$ maps of \citet{sfd} recalibrated by \citet{sfd2}. We take the solar $g$-band magnitude to be $M^\odot_g=5.03$ mag \citep{willmer2018}.

\subsection{Optical S\'{e}rsic Photometry}
\label{sec:sersic_fitting}
The primary photometric results come from fitting the optical images with S\'{e}rsic profiles. We refer the reader to \citet{carlsten2021a} for an in depth description of how the S\'{e}rsic profiles are fit to the galaxies. In brief, the $g$-band images are fit first since they are generally deeper, and then the $r$- or $i$-band image is fit using the shape parameters fixed, varying only the intensity\footnote{For the few dwarf candidates in the NGC 891 field with only $r$-band coverage, we assume $g-r=0.4$. Note that no SBF measurements (which require color) are attempted for these dwarfs.}. The masking threshold and image cutout size are adjusted in an iterative fashion to minimize the dependence that the final photometry results have on those choices.

As detailed in Table \ref{tab:hosts}, the photometry always uses DECaLS/DECam or CFHT data. Even though we have deeper Gemini, Magellan, and Subaru data for many dwarfs for SBF, we do not present the photometry from that data in an attempt to limit the number of filter systems used. For three of the five previously surveyed systems (M81, CenA, and NGC 5236), we are able to provide photometry using our consistent methodology for almost all of the satellites. There were a few M81 satellites out of the DECaLS footprint and a few CenA satellites not covered in DECam archival data.

Following \citet{carlsten2021a}, we estimate uncertainties in the photometric parameters from image simulations where we inject S\'{e}rsic profile galaxies into CFHT or DECaLS data and quantify how well the input values are recovered. \citet{carlsten2021a} also gives the equations we use to convert between $g-r$ and $g-i$ and convert $g$-band photometry into $V$-band.

For each dwarf, we use the color and $g$-band luminosity to estimate its stellar mass using the color-mass-to-light ratio relations in \citet{into2013}. For several of the brightest ($M_V\lesssim-18$ mag) satellites, we did not attempt S\'{e}rsic photometry due to the clear inadequacy of a single profile in fitting these large, complicated galaxies. Instead, for these we simply provide stellar mass estimates calculated from 2MASS \citep{2mass} $M_{K_s}$ values from \citet{kourkchi2017}. For the MW and M31 satellites, we do not have colors but use average color-luminosity trends from the dwarfs that do to separately define luminosity-mass to light ratio relations for early- and late-type dwarfs \citep[see Equation 4 of][]{carlsten2021a}.

Appendix \ref{app:sat_lists} presents the main table for the photometry of the satellites. The confirmed and possible satellites are all given in one table, along with a flag to indicate which satellites have robust distance confirmation.

\subsection{Galaxy Morphology}
\label{sec:morph}
In addition to the optical photometric measurements, we visually classify each dwarf to have either late- or early-type morphology. We use all available optical imaging, including the deep data often available for SBF measurements, to make this distinction. Dwarfs with smooth, regular, and generally low surface brightness morphology are classed as early-types. On the other hand, dwarfs with clear star-forming regions, blue clumps, dust-lanes, or any other complications in their surface brightness profile are classed as late-type. Given the data available, we believe this is the most robust way to split the dwarfs. More details of this separation and color images of example dwarfs classified as early- and late-type can be found in \citet{carlsten2021a}.

\subsection{GALEX Data}
\label{sec:galex}
To complement the optical photometry, particularly to provide more information on recent star formation, we use archival GALEX \citep{galex} data to measure the UV photometry of the dwarfs, where available. We largely follow the methods of \citet{greco2018_two} and \citet{karunakaran2021}. For each confirmed dwarf, we search the MAST archive for GALEX coverage. We find at least some coverage in NUV and FUV for a majority of the confirmed dwarfs, 271 in total. About a third of these are only covered by the shallow GALEX All-sky Imaging Survey with the remainder having deeper data, often from the Nearby Galaxy Survey \citep{galex2007} or the 11 Mpc H$\alpha$ and Ultraviolet Galaxy Survey \citep[11HUGS;][]{kennicutt2008, lee2011} both of which targeted many LV galaxies. 

From the MAST archive, we download two files for each GALEX frame and filter: the intensity maps (\texttt{-int.fits} files; $I$) and the high resolution relative response maps (\texttt{-rrhr.fits} files; $R$). We use these frames to construct the variance image as: $V = I/R$.

Due to the fact that many of the dwarfs are not actually detected in the UV, we do not do S\'{e}rsic photometry like with the optical data. Instead, we perform aperture photometry using elliptical apertures with radii of twice the effective radii found from the optical S\'{e}rsic fits. Contaminating point sources are masked before performing the photometry with an initial \texttt{sextractor} \citep{sextractor} run. The sky contribution to the flux is estimated from the median of 50 apertures placed in the vicinity of the galaxy. 

The measured flux and uncertainty for the apertures are: 

\beq
F &= \sum_i I_i, \\
\sigma^2_F &= \sum_i V_i  + \sigma_{\rm sky}^2/N,
\eeq

with $i$ ranging over the pixels covered by the aperture and $\sigma_{\rm sky}$ being the standard deviation of sky values measured in the $N$ sky apertures. We use the zeropoints given in \citet{morrissey2007} to convert the measured fluxes to AB magnitudes and correct for galactic extinction using $R_{\rm NUV}=8.2$ and $R_{\rm FUV}=8.24$ \citep{wyder2007}. 

We do not attempt to correct for emission outside of the $2r_e$ aperture. Even without this, we find good agreement when applying this methodology to the SAGA satellites \citep{mao2020} compared to the GALEX measurements of \citet{karunakaran2021} who performed a curve-of-growth analysis. The aperture NUV magnitudes are biased high by $\lesssim0.1$ mag. We also find good agreement with the FUV fluxes in the UNGC \citep{karachentsev2013} for ELVES satellites that are in that catalog. 

Appendix \ref{app:sat_lists} presents a table with the UV photometry for the confirmed satellites that have it. For dwarfs that had a S/N$<2$ result, we report $2\sigma$ upper limits to the UV flux.

\section{Properties of the Satellite Systems}
\label{sec:sat_systems}

In this section, we briefly explore features and trends amongst the satellite systems, including the luminosity functions and abundance, spatial distribution, and star formation properties. In-depth comparisons with galaxy formation models are out of the scope of this survey overview paper but will be pursued in future papers.

\begin{figure*}
\includegraphics[width=\textwidth]{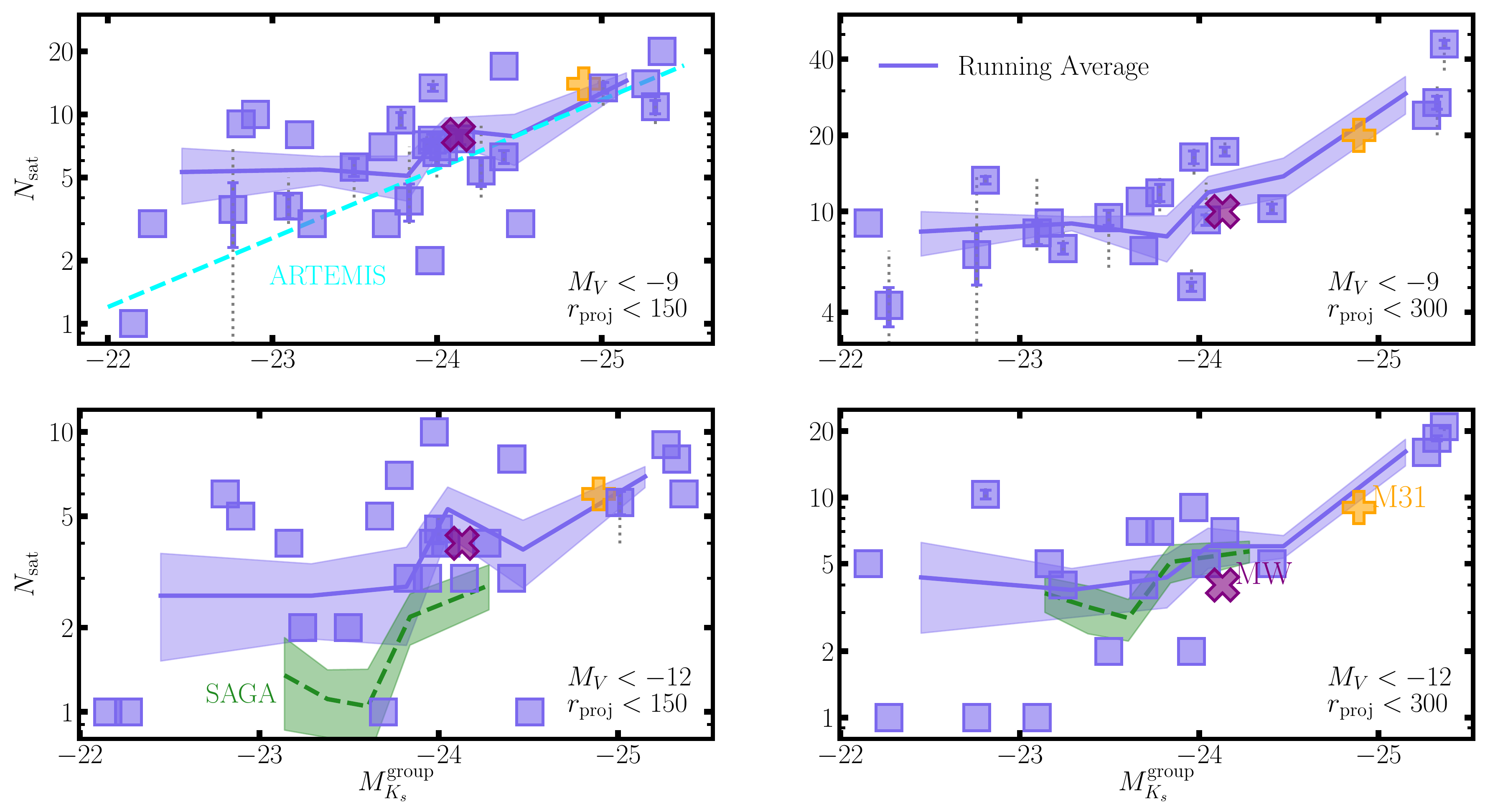}
\caption{The satellite abundance versus the $K$-band luminosity of the host. Top row shows all satellites $M_V<-9$ mag, while the bottom row is restricted to the bright, $M_V<-12$ mag, satellites. The left panels show all hosts, but are restricted to satellites within $r_\mathrm{proj}<150$ kpc. Right panels show the subsample of hosts that are complete out to 300 kpc. Solid errorbars show the uncertainty in satellite abundance due to the unconfirmed satellites, using the satellite probabilities from \S\ref{sec:cont_mod}.  Dotted errorbars show the upper and lower limits of abundance if all, or none, of the unconfirmed candidates are actual satellites. The MW and M31 are plotted individually in the crosses. The solid lines show running averages. The abundance-$M_K$ relation from the ARTEMIS simulations \citep{font2021} is shown in the upper left. The average trends from the SAGA Survey \citep{mao2020} are shown in the bottom panels.}
\label{fig:nsat}
\end{figure*}

\subsection{Satellite Abundance}
\label{sec:sat_abund}

The question of satellite abundance around MW-like galaxies has been a central component of the debate around `small-scale problems in $\Lambda$CDM' for decades \citep[e.g.][]{moore1999, klypin1999, bullock2017}. The abundance and luminosity function of satellites are sensitive probes to the underlying stellar-to-halo-mass relation \citep[SHMR; e.g.][]{gk_2017, nadler2020}. Satellite abundance is also an important probe of recent accretion events experienced by a host \citep{smercina2021}.  Using early results from ELVES, in \citet{carlsten2020b}, we showed that the MW's abundance of satellites was quite typical compared to LV systems of similar mass. We also showed that this abundance and the host-to-host scatter in satellite abundance were well-reproduced by recent galaxy formation models. In this section, we briefly highlight some abundance-related results and provide a comparison with recent SAGA Survey results \citep{mao2020}.

Figure \ref{fig:nsat} shows the relation between satellite abundance and host mass, as proxied by the group $K$-band luminosity. The group $K$-band luminosity is used instead of just the host's in order to account for the groups with multiple massive primaries, like M81, NGC3627, and NGC3379. The top panels show all satellites down to the ELVES limit of $M_V<-9$ mag while the bottom panels show just the bright satellites, $M_V<-12$ mag. The left panels show the inner satellite systems, $r_\mathrm{proj}<150$ kpc, while the right panels show abundances out to $300$ kpc for the systems that are surveyed out that far. Running averages are shown in the blue bands. The unconfirmed satellites are included in the satellite counts using their satellite probability from \S\ref{sec:cont_mod} (Figure \ref{fig:contam}). The solid errorbars show the $1\sigma$ spread in abundance when stochastically including, or not, the unconfirmed candidates over many trials. The dotted errorbars show the upper and lower limits of abundance in the cases where all, or none, of the unconfirmed candidates turn out to be physical satellites. The lower panels show few errorbars because there are few unconfirmed candidates more luminous than $M_V=-12$ mag.

In all panels, a strong trend is seen between satellite abundance and host mass, as expected. Both the MW and M31 seem quite typical in terms of satellite abundance amongst similar mass systems. We found this result in \citet{carlsten2020b} and show it here with a much larger statistical sample of hosts and satellites. This result is one of the primary results, if not the main result, from the ELVES survey.

In the top left panel, we show the scaling relation fit from the ARTEMIS simulation suite in \citet{font2021}. This particular fit was using the `LV-selection' which was chosen to mimic roughly the ELVES luminosity and surface brightness sensitivity (and restricted to the inner $r_\mathrm{proj}<150$ kpc regions). The fit reproduces the average abundance well, particularly at high host mass, but seems to underpredict the satellite abundance somewhat at the lowest masses. In fact, in all panels, the observed satellite abundance seems to flatten out at the lowest masses. Note that several of these low-mass hosts have complete, or nearly complete, distance confirmation for all candidates so this is not simply a floor caused by background contaminants. Including satellite abundances for even lower-mass, LMC-like, hosts will be an important extension of this work \citep[see][for initial work in this direction]{madcash, carlin2021, muller2020_low_mass}.

In the bottom panels we show a comparison to the average trends from the SAGA Survey \citep{mao2020}. The $M_V<-12$ mag cut used closely matches the luminosity limit of SAGA. The redshift follow-up incompleteness in SAGA is accounted for using the satellite probability model calculated in \S5.3 of \citet{mao2020}. This is treated in an analogous way to the unconfirmed/candidate satellite correction used in ELVES. SAGA also shows a relation with higher $K$-luminosity hosts having higher satellite abundance. ELVES satellite abundance is slightly higher than SAGA, with the difference being quite noticeable for the inner, $r_\mathrm{proj}<150$ kpc, satellites.

\begin{figure}
\includegraphics[width=0.48\textwidth]{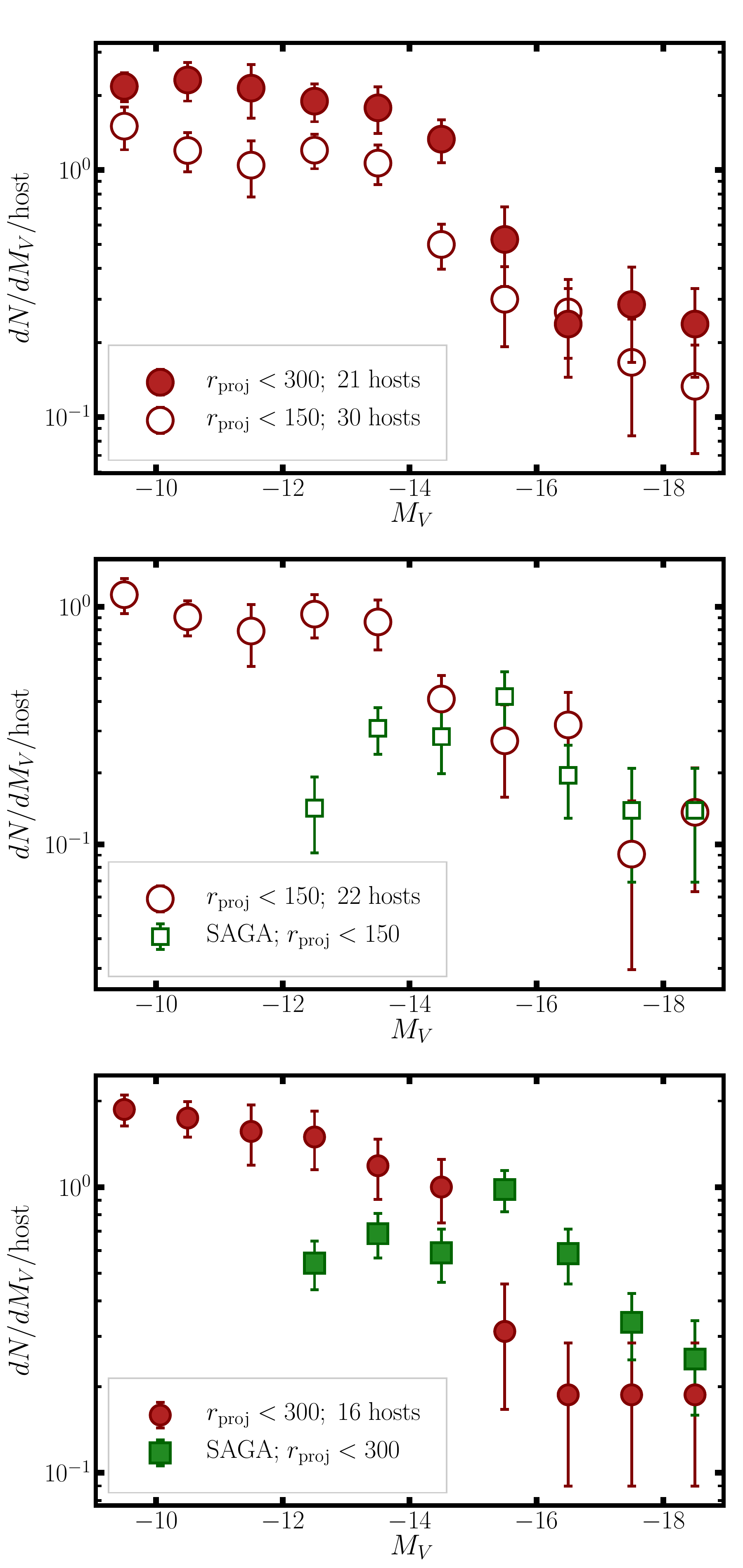}
\caption{The average satellite abundance per host in 1 mag wide luminosity bins within both $r_\mathrm{proj}<150$ and 300 kpc. The legends indicate the number of hosts contributing to each stacked LF. The middle and bottom panels show a comparison to the SAGA Survey \citep{mao2020}. Only ELVES hosts that would satisfy the SAGA selection criteria are included. The SAGA Survey LF shows a noticeable drop below $M_V>-15$ mag compared to ELVES, even when including a correction for incompleteness in redshift follow-up in that survey.}
\label{fig:lf}
\end{figure}

Figure \ref{fig:lf} shows a more detailed look at the ELVES satellite luminosity function (LF). The average differential luminosity function is shown for both $r_\mathrm{proj}<150$ and $r_\mathrm{proj}<300$ kpc. The unconfirmed candidate satellites are included using their satellite probability values from \S\ref{sec:cont_mod}. The errorbars show the error in the mean number of satellites per host within each 1 magnitude wide bin. This is calculated as the standard deviation of satellite abundance across the hosts divided by $\sqrt{N_\mathrm{host}}$. Thus, these errorbars account for uncertainty coming from the intrinsic host-to-host scatter in satellite abundance. The errorbars are essentially the same as would be calculated from bootstrapping the host sample. Within the errorbars, the average LFs appear to be simple power laws. The $r_\mathrm{proj}<300$ kpc LF flattens somewhat at the lowest luminosities, likely due to ELVES starting to lose the lowest surface brightness satellites (cf. Figure \ref{fig:completeness_avg}). The satellite abundance within 150 kpc is generally $\sim2\times$ lower than that out to 300 kpc.

The middle and bottom panels of Figure \ref{fig:lf} show a comparison with the SAGA results. In this comparison, we remove the massive ELVES hosts that would not satisfy the SAGA host $M_{K_s}$, isolation, and/or halo mass criteria. The SAGA host $M_{K_s}$ requirement ($M_{K_s}>-24.6$ mag) removes M104 and M31, the isolation criteria (no satellite with $M_{K_s}$ within $1.6$ mag of the host) removes NGC 1808, NGC 5194, NGC 3379, and M81, and the halo mass cut ($M_h<10^{13}$~\msun) removes CenA and NGC 3627. In applying the halo mass cut, we remove any host that has either of the two halo mass estimates provided in the group catalog of \citet{kourkchi2017} above $10^{13}$~\msun\footnote{We believe that the halo mass estimates from the group catalog of \citet{kourkchi2017} are more likely accurate for LV hosts than the group catalog \citep{lim2017} that the SAGA Survey uses.}. One halo mass estimate comes from group kinematics while the second comes from total $K$-band luminosity. If the two estimates differ by more than $0.5$ dex, we consider the luminosity-based estimate as the more robust. Note that we still include the six ELVES hosts that are actually \textit{below} SAGA's host $M_{K_s}$ range (NGC 628, NGC 3344, NGC 3556, NGC 4517, NGC 4631, and NGC 4736). There is no change to the conclusions if these are cut as well, albeit the ELVES statistics become poorer. In comparing with SAGA, we additionally only include satellites more than 15 kpc projected from their hosts as this is the estimated inner radial limit of the SAGA satellite lists (Y.-Y. Mao, private communication). The SAGA results make use of our own S\'{e}rsic photometry for the SAGA satellites, which we describe below (\S\ref{sec:sat_sf}).

In order to account for possible differences in the host $M_{K_s}$ distribution even with the SAGA selection criteria applied, we have tried scaling the host abundances to a standard value of $M_{K_s}=-23.5$ mag. We use the abundance-$M_{K_s}$ relation of \citet{font2021} to reduce the weight of more massive hosts and increase the weight of less massive hosts in the average LF stack. We find that this re-scaling has only a minor effect on the average LFs and is thus not shown.

\begin{deluxetable*}{ccccc}
\tablecaption{Bright vs. faint satellite abundances per host\label{tab:abund}}
\tablehead{
\colhead{Radial coverage} & \colhead{Survey} & \colhead{$N_\mathrm{hosts}$} & \colhead{$\langle N_\mathrm{sat} \rangle$ with $-15<M_V<-12$ mag}  & \colhead{$\langle N_\mathrm{sat}\rangle$ with $-19<M_V<-15$ mag}    }  
\startdata
$r_\mathrm{proj}<150$ kpc & SAGA &  36 & 0.73$\pm$0.12 & 0.89$\pm$0.16\\ 
 & ELVES & 22 & 2.2$\pm$0.31 & 0.82$\pm$0.2\\ 
  &     &  (16) & (1.94$\pm$0.37) & (0.56$\pm$0.22)\\ 
\hline
$r_\mathrm{proj}<300$ kpc & SAGA & 36 & 1.82$\pm$0.2 & 2.15$\pm$0.25\\ 
 & ELVES & 16 & 3.68$\pm$0.61 & 0.88$\pm$0.23\\ 
\enddata
\tablecomments{Average number of satellites per host for different radial and luminosity ranges. Note that the ELVES host sample is slightly different between the $r_\mathrm{proj}<150$ and $<300$ kpc cases. The values in parentheses give the $r_\mathrm{proj}<150$ kpc abundances for the same host sample as in the $r_\mathrm{proj}<300$ kpc case. Many of the hosts complete to only 150 kpc are among the richest ELVES hosts. }
\end{deluxetable*}

Table \ref{tab:abund} lists the average satellite abundance per host in different radial and luminosity ranges. The bright ($-19<M_V<-15$ mag) ELVES satellites appear quite centrally concentrated with a similar abundance per host regardless of radial range. As stated in the table, the host sample is slightly different between the $r_\mathrm{proj}<150$ and $<300$ kpc cases, as the former includes 6 more hosts complete to only 150 or 200 kpc. In the case where we restrict the $r_\mathrm{proj}<150$ kpc sample to be the same as in the $r_\mathrm{proj}<300$ kpc case, there are fewer bright satellites in the inner radial range (these values are given in parentheses). 

Interestingly the average SAGA LF is higher than that of ELVES at bright ($M_V<-15$ mag) luminosities for the 300 kpc case. It is unclear what is causing this higher abundance. Dwarfs at these luminosities in the Local Volume are hard to miss and most, if not all, will have been cataloged years ago. The Updated Nearby Galaxy Catalog (UNGC) of \citet{karachentsev2013} will be essentially complete for these dwarfs, and we show in Appendix \ref{app:prev_work} that ELVES is complete to UNGC entries within our coverage footprints. Future work will be needed to understand this difference, including a detailed analysis of the different host samples and expected interloper fractions (i.e. false positives amongst distance confirmed satellites). 

On the other hand, the SAGA abundance is lower at magnitudes fainter than $M_V\sim-15$ mag in both considered radial ranges. This has the result of the SAGA average LF being much shallower than the average ELVES LF. Note we are including the correction for incompleteness in the spectroscopic follow-up of SAGA \citep[see \S 5.3 of ][]{mao2020}, which helps close the gap at faint luminosities but does not close it fully. There appears to be $\sim1.5-2$ more  satellites in the $-15<M_V<-12$ mag range per ELVES host than in SAGA with most ($\sim1.5$) of the difference being in the inner 150 kpc regions.

The dearth of faint satellites in SAGA and the dropping average LF are suggestive of incompleteness beyond what is accounted for in the model for incomplete redshift follow-up. In particular, these results point to incompleteness in the DECaLS \citep{lang2016, decals} object detection and targeting catalogs used in SAGA. We have experimented with applying various surface brightness cuts to the ELVES satellite lists to see if a simple cut in surface brightness can explain the discrepancy. When applying such a cut, the ELVES LF does indeed flatten and even drop, but it does not reach the SAGA points unless a fairly bright surface brightness cut ($\langle \mu_V \rangle_e \sim24.5$ mag arcsec$^{-2}$) is applied. This is unrealistically high as there are a number of SAGA satellites fainter than this. We therefore suspect that while surface brightness incompleteness in the DECaLS catalogs is part of the problem, there are likely other issues as well. For instance, shredding can cause objects that otherwise would be brighter than SAGA's $M_r=-12.3$ mag limit to drop below. We emphasize that this is not due to an inherent limitation of Legacy Survey or Dark Energy Survey data but instead due to the specific objection detection algorithms used in making the SAGA targeting catalogs. After all, in ELVES we use this data to detect dwarfs as faint as $\mu_{0,V} \sim 27$ mag arcsec$^{-2}$.

In addition to observational incompleteness, it is conceivable that host-to-host scatter and/or physical differences in the hosts are playing a role in the discrepancy. For instance, it is possible that, due to some circumstance of the LV's cosmic environment, LV hosts are more satellite abundant than average hosts of their mass. \citet{neuzil2020} argue that the LV is an outlier in terms of overall dwarf (satellite and field) abundance compared to $\Lambda$CDM simulations. This will be something that might be explored with the final, 100-host SAGA sample. We note that there is no 16-host (to match the number of SAGA-selected ELVES hosts that are complete to 300 kpc) sub-sample of the current 36-host SAGA sample that is able to come close to the steepness and abundance of the average ELVES LF, but the complete SAGA sample might change this. In the end, multiple effects (incompleteness, host-to-host scatter, and/or differences in the host samples) are likely playing a role.

\begin{figure*}
\includegraphics[width=\textwidth]{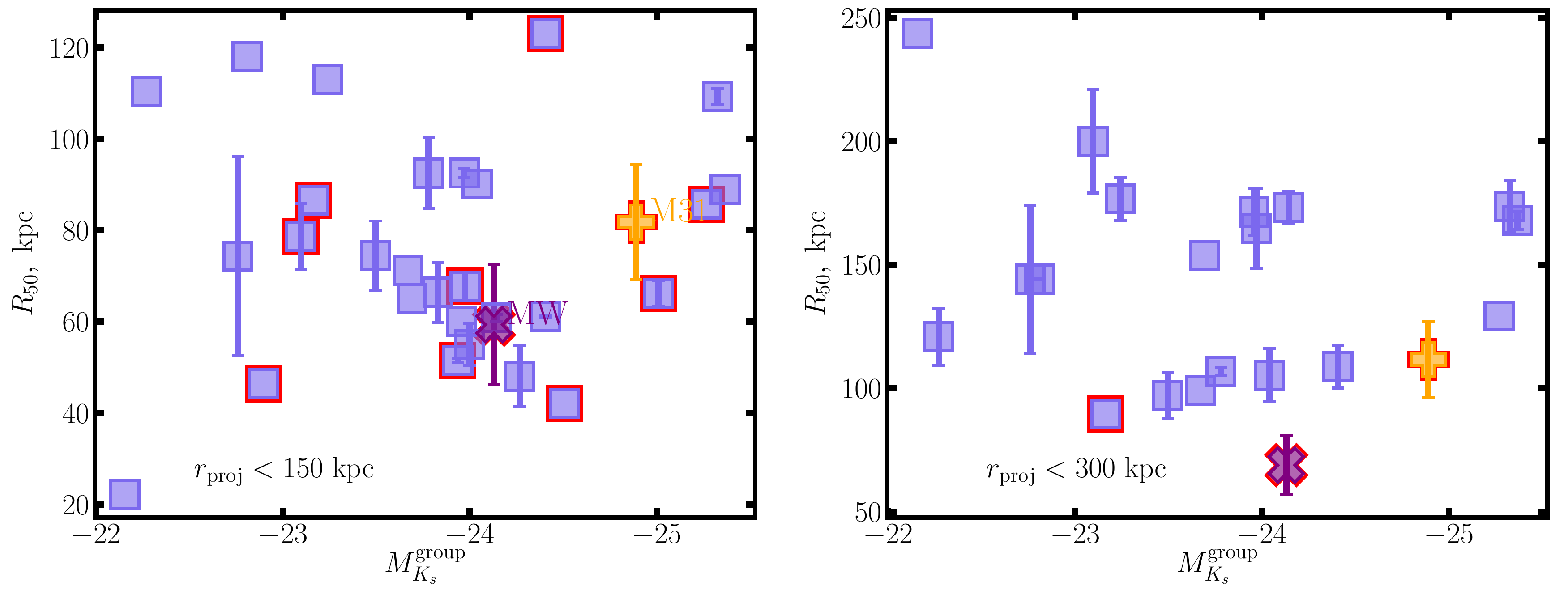}
\caption{$R_{50}$, the radius containing half of satellites for a given host, versus host $K_s$-band luminosity. The left panel shows the results restricted to $r_\mathrm{proj}<150$ kpc, and the right panel shows the results out to 300 kpc, for the hosts that are surveyed out that far. The MW and M31 points show the average projected $R_{50}$ if the satellite systems were re-observed at a distance of 7 Mpc, with errorbars indicating the spread due to different viewing angles. For the other points, the errorbars indicate the uncertainty due to unconfirmed candidate satellites without distance information. Points outlined in red were in the previous study of \citet{carlsten2020c}. }
\label{fig:r50s}
\end{figure*}

\subsection{Satellite Spatial Distribution}
\label{sec:sat_rads}

The radial distribution of satellites is an important observable to test the disruptive effect of the host's central disk and the level of physical versus numerical (due to low-resolution) disruption in simulations \citep[e.g.][]{vdb2018a, samuel2020}. In \citet{carlsten2020c}, we found that the ELVES satellite systems surveyed up to that point (that paper used 6 hosts surveyed with CFHT/MegaCam data and 6 surveyed in the literature) showed significantly more concentrated radial distributions of satellites than predicted by galaxy formation simulations. We did not have a complete explanation, but discussed both artificial disruption in the simulations or a bias in the observed sample relative to the general population of massive hosts. The 36 hosts presented in the SAGA Survey \citep{mao2020} did not show similarly concentrated profiles, suggesting the latter explanation. In this section, we explore the satellite radial profiles using the full, nearly volume-limited, ELVES sample of hosts.

Figure \ref{fig:r50s} shows $R_{50}$, the projected radius encompassing half the satellites of a given host, versus host $K_s$-band luminosity both within $r_\mathrm{proj}<150$ kpc and 300 kpc. Since this is a projected radius for the LV hosts, we `re-observe' the MW and M31 systems at a distance of 7 Mpc using the known 3D structure of these systems and considering many different sight-lines. The points show the average projected $R_{50}$ along with the 1$\sigma$ spread due to sight-line differences. The errorbars on the other hosts show the $1\sigma$ spread when stochastically including, or not, the unconfirmed candidate satellites according to their satellite probabilities from \S\ref{sec:cont_mod}. The points outlined in red were included in the \citet{carlsten2020c} sample. We note that the $R_{50}$ values for some of these hosts, particularly CenA and M31, are somewhat different from those in \citet{carlsten2020c}. This is due to slightly different satellite lists being used for these hosts. For CenA, here we use largely our own photometry to select which satellites have $M_V<-9$ (and hence be included in ELVES), while \citet{carlsten2020c} used literature photometry, leading to slightly different inclusion. For M31, here we simply use the photometry of \citet{mcconnachie2012} while \citet{carlsten2020c} used various other literature sources, again leading to a slightly different list of satellites. 

The hosts included in \citet{carlsten2020c} tend to be on the more concentrated side, with several of the most concentrated in all of ELVES falling in that group. This is especially the case for the $r_\mathrm{proj}<300$ kpc sample where \citet{carlsten2020c} only had three hosts surveyed out that far and two of them (MW and NGC 5457) are the two most concentrated hosts in all of ELVES. This would suggest that the discrepancy found in that paper was at least partly due to an unlucky and small sample of hosts. With that said, \citet{carlsten2020c} found that simulations generally had $R_{50}\sim90$ kpc and $\sim150$ kpc for the $r_\mathrm{proj}<150$ and $r_\mathrm{proj}<300$ kpc cases, respectively. Comparing with the complete ELVES sample in Figure \ref{fig:r50s}, the observed systems still appear systematically more concentrated, but an in-depth comparison with simulations is needed for a firm conclusion.

\begin{figure*}
\includegraphics[width=\textwidth]{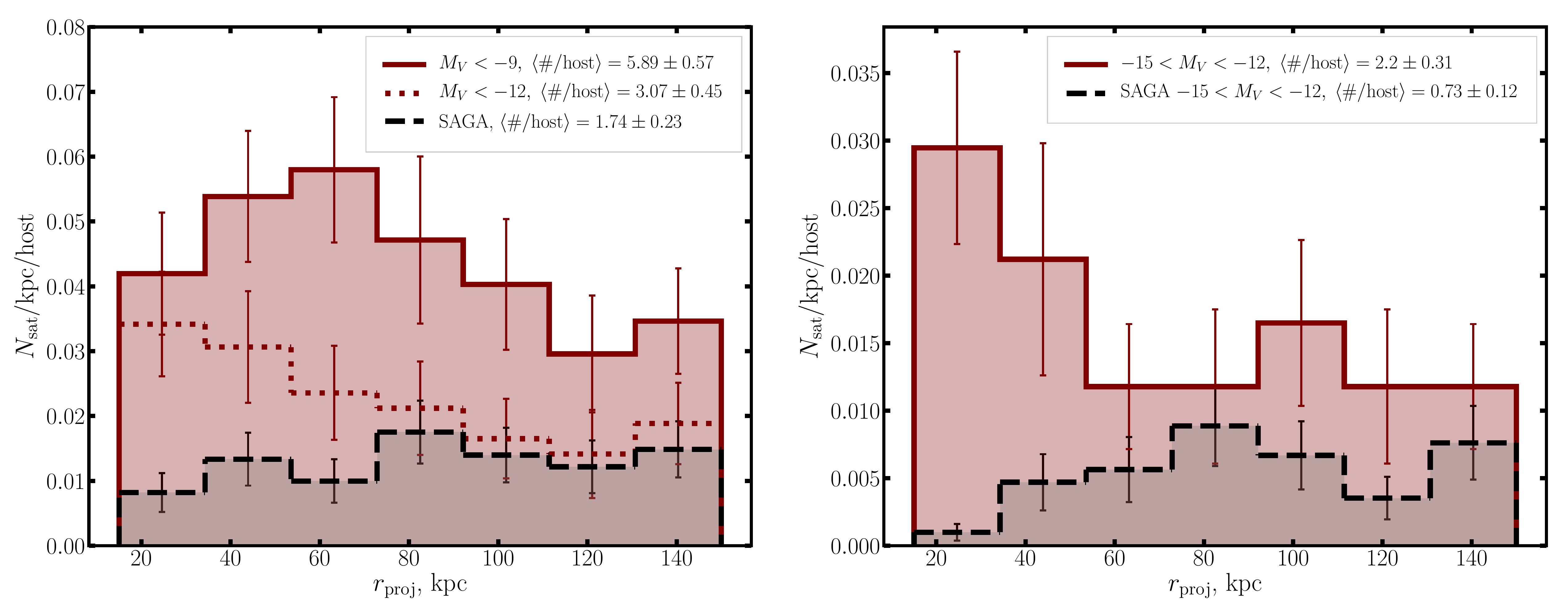}
\caption{Histogram of the projected separations of satellites from their hosts. The massive hosts that would not meet the SAGA selection criteria are not included. Left panel shows all ELVES satellites along with just the brighter satellites ($M_V<-12$ mag). Right panel shows just satellites in the intermediate luminosity range $-15<M_V<-12$ mag. The satellites from the SAGA Survey \citep{mao2020} are also shown, which exhibit fewer satellites per host and less central concentration. }
\label{fig:rproj}
\end{figure*}

Figure \ref{fig:rproj} shows a histogram of the projected satellite-host separations, comparing with the SAGA Survey results. To be more comparable to SAGA, we again remove the massive ELVES hosts that would not satisfy the SAGA selection criteria as described in \S\ref{sec:sat_abund}. In the left panel, we show both the complete ELVES satellite sample ($M_V<-9$ mag) and just the brighter, $M_V<-12$ mag, satellites to compare with SAGA. Unconfirmed candidate satellites are again included weighted by their satellite probabilities from \S\ref{sec:cont_mod}. To include all the hosts, we just plot satellites out to $150$ kpc projected. The errorbars show the standard error in the mean number of satellites per host in each radial bin. 

Overall, the ELVES systems show higher satellite abundances with $\sim1.5$ more satellites per host than in SAGA, similar to Table \ref{tab:abund}. Interestingly, most of the difference between ELVES and SAGA is at quite small projected radii, $r_\mathrm{proj}\lesssim75$ kpc.  Assessing the level of significance of the discrepancy with a 2-sample Kolmogorov–Smirnov test is not straightforward as the unconfirmed satellites are included statistically. Specific realizations of the unconfirmed satellites yield $p$-values $\lesssim 5\times 10^{-3}$ in all cases, however, indicating the discrepancy is significant. The right panel shows just the satellites in the $-15<M_V<-12$ mag range singled out in Table \ref{tab:abund}, again showing the difference in abundance is due to satellites very near the host.

The difference in radial concentration between ELVES and SAGA is again suggestive of incompleteness in SAGA. If the incompleteness in the DECaLS targeting catalogs used by SAGA is due, at least in part, to lack of low surface brightness sensitivity, this radial trend would make sense. Inner satellites in ELVES are lower in surface brightness by $\sim0.5$ mag arcsec$^{-2}$, likely due to earlier quenching or enhanced tidal effects, but we leave an exploration of this to future work.

We note that the radial disagreement mostly goes away (but the abundance disagreement remains the same as in Table \ref{tab:abund}) if we do not include the ELVES hosts only surveyed out to 150 or 200 kpc. There are six of these\footnote{For reference there are 16 hosts included in this figure complete to 300 kpc.}: NGC 891, NGC 1023, NGC 4258, NGC 4565, NGC 4631, and NGC 6744, which include the most centrally concentrated hosts from \citet{carlsten2020c}. All of these hosts are complete to at least $r_\mathrm{proj}=150$ kpc, the outer limit shown in Figure \ref{fig:rproj}, and we argue extensively in \citet{carlsten2020c} that their central concentration is not due to any observational systematic. Thus, there is no justification for their exclusion from Figure \ref{fig:rproj}, but this highlights the host-to-host scatter in satellite radial profiles and how a few extreme systems can affect the average.

\subsection{Satellite Star-forming Properties}
\label{sec:sat_sf}
Dwarf satellites experience profound effects as the result of their association with a massive host galaxy. The hot gas halo of the host galaxy can remove the gas reservoirs of the dwarfs quickly via ram pressure stripping \citep{gunn1972, grebel2003, boselli2008} or at least cut off the supply of fresh gas to the dwarfs. This is often invoked to explain why dwarfs in the LG are primarily gas-poor and quenched \citep{grcevich2009, spekkens2014, putman2021}. Furthermore, it is thought that the tidal field of the host induces a morphological transformation of an initially gas-rich dwarf irregular to a dwarf spheroidal \citep[e.g.][]{mayer2006, kazantzidis2011, kazantzidis2013}. 

However, there is growing uncertainty as to whether these results based on the LG can generalize to the overall satellite galaxy population. There is some evidence from simulations that a double-primary configuration like in the LG can, indeed, boost satellite quenching efficiency \citep{gk2019b}. Additionally, the SAGA Survey has reported much lower quenched fractions amongst their MW-analog targets \citep{geha2017, mao2020}, due to the ubiquitous presence of H$\alpha$ emission. In this section, we use the ELVES satellite systems to address the question of whether the LG is atypical or not in this regard. We also show various comparisons between the ELVES satellite sample and that of the SAGA Survey. 

Due to the fact that ELVES is a photometric survey, determining the quenched fraction of satellites is not straight-forward. Without spectroscopic follow-up data, dwarf color and morphology are the best indicators of star-forming activity available. As described in \S\ref{sec:morph}, we visually split the dwarfs into early- and late-type based on the presence of clear indicators of active star-formation, such as blue clumps, dust lanes, and irregular morphology. Based on a limited sub-sample that have H$\alpha$  and/or H\textsc{I} measurements, we find that these measurements closely corroborate the visual classification. Early-type dwarfs are almost always characterized by small or non-existent H\textsc{I} reservoirs and/or no H$\alpha$ emission, indicating quenched star formation.  However, instead of just trusting that this correlation holds true for the whole satellite sample, we also perform more `apples-to-apples' comparisons with the SAGA satellites using color.

\begin{figure*}
\includegraphics[width=\textwidth]{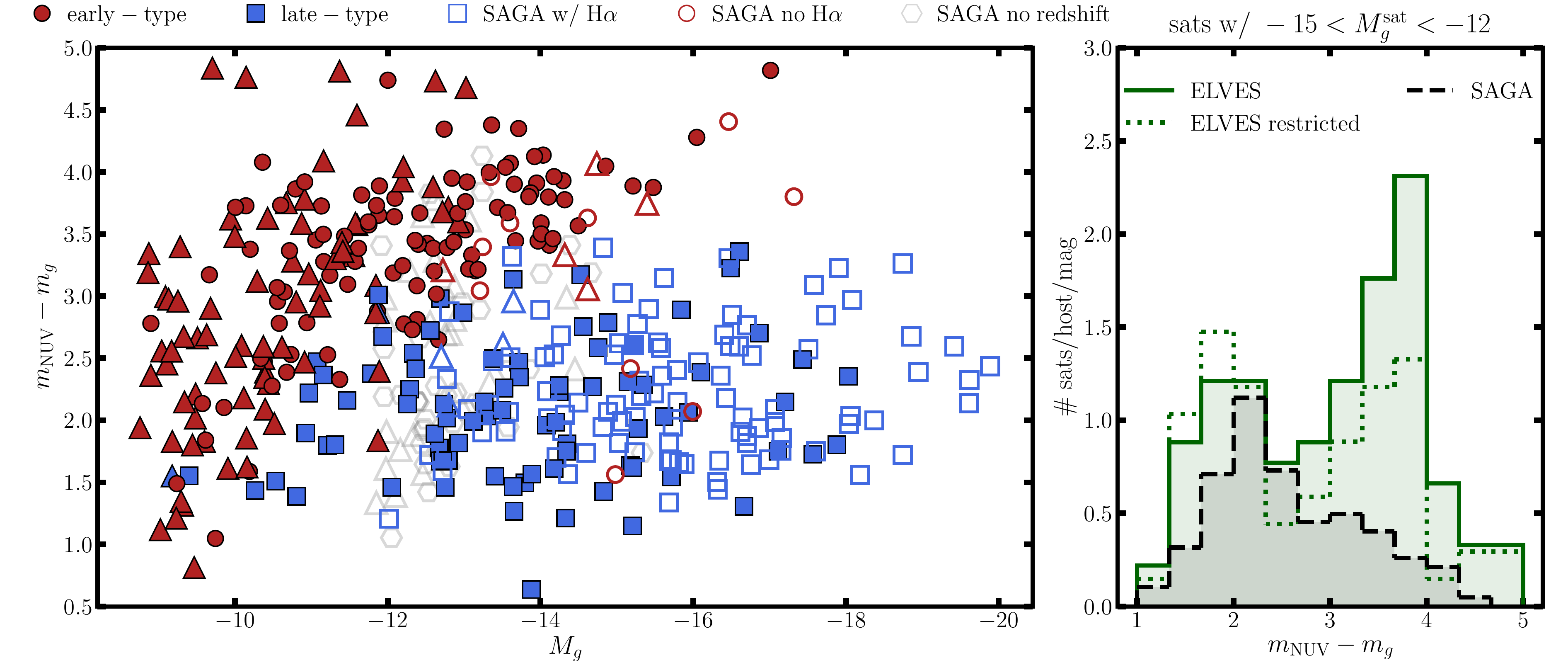}
\caption{The NUV$-g$ color versus luminosity for the ELVES satellites. The points are colored based on their morphology class (early- or late-type). The two morphological classes cleanly separate in this plane. Shown in the open symbols are the SAGA satellites from \citet{mao2020}, colored by whether or not they have H$\alpha$ emission. Dwarfs not detected in GALEX with S/N$>2$ are shown as triangles indicating $2\sigma$ upper limits to their NUV flux. In general, the quenched SAGA satellites do have similar red UV-optical color to the ELVES early-types, however, there are far more ELVES early-type satellites. This is shown in the histograms in the right panel. Both SAGA and ELVES show a bimodality in UV-optical color corresponding to quenched and star-forming satellites. However, the ELVES quenched population is significantly larger. The dotted histogram shows the ELVES sample without the hosts more massive than the SAGA selection criteria. Note that the SAGA completeness limit is $M_g\lesssim -12$ mag and that for ELVES is $M_g\lesssim -9$ mag. }
\label{fig:galex_properties}
\end{figure*}

The first comparison we show is that of UV-optical color versus dwarf luminosity in Figure \ref{fig:galex_properties}. The ELVES satellites are colored according to their morphology while the SAGA satellites are colored according to the H$\alpha$ emission criterion used in \citet{mao2020} (whether the equivalent width is more or less than $2$\AA). The photometry (optical and UV) for the SAGA satellites come from our own measurements using the same methods as for ELVES, although we find largely the same results if we use SAGA UV photometry from \citet{karunakaran2021} and optical photometry from \citet{mao2020}. We incorporate the redshift incompleteness model of SAGA by also 
including SAGA candidate satellites that \citet{mao2020} infer to have satellite likelihoods $>0.1$\footnote{As described in \S5.3 of \citet{mao2020}, the candidate satellites in SAGA without redshift split into two groups: several thousand candidates with individually very low likelihood of being satellites ($\sim0.1\%$) and $\sim70$ candidates with much higher probabilities of being a satellite ($\sim25$\%). In making this cut, we are only considering the latter group.}. These candidates have S\'{e}rsic optical and GALEX photometry measured in the same way as the confirmed SAGA satellites and are shown in the faint open symbols. Since, by definition, they have no spectral information, these points are not colored to indicate H$\alpha$ presence. 

The early- and late-type ELVES satellites clearly separate in this plane with early-type dwarfs having redder UV-optical color. Since optical color is part of the visual classification (in addition to morphology), this is largely just a reflection of UV-optical color being correlated with optical color. The SAGA satellites without H$\alpha$ generally have red UV-optical color, as well. This is unsurprising as both H$\alpha$ and UV emission are well-established tracers of recent star formation in dwarfs \citep{lee2009, karachentsev2013_sfr} and generally show good consistency. The few UV-optically blue SAGA satellites without H$\alpha$ are quite blue and irregular in optical DECaLS imaging, and it seems likely they do emit in H$\alpha$ but the emission was possibly missed by the SAGA fiber placement. Other than these exceptions, the SAGA quenched satellites appear similar to the ELVES early-type dwarfs, providing further support that the early-type morphology classification is a robust indicator of quenched star formation.

There appears to be significantly more ELVES early-type dwarfs than quenched SAGA dwarfs. This is shown in the histogram in the right panel which includes dwarfs of an intermediate luminosity ($-15<M_g<-12$ mag). Candidate SAGA satellites without distance confirmation are included weighted by their satellite probability inferred by \citet{mao2020}. No correction is made for ELVES candidate satellites without distance measurements since there are essentially none in this luminosity range. Satellites with only GALEX NUV upper limits are included in this histogram at the location of their upper limit. Since only a few of the ELVES or confirmed SAGA satellites in this luminosity range have only GALEX upper limits there is no worry that the histograms are biased to blue NUV$-g$ colors because of weak NUV flux upper limits. However, many of the SAGA candidate satellites without redshifts have only GALEX upper limits. To ensure the SAGA histogram is not unfairly biased blue, we give the candidates that have similar red optical colors to the ELVES early-types (see the end of this section for the specific cut in color-magnitude space) an NUV$-g$ color of $\sim3.7$ mag, similar to the ELVES early-types.

Both ELVES and SAGA show bimodality in UV-optical color, corresponding to quenched and star-forming populations. The dotted green histogram shows the ELVES results without the hosts that are too massive to fit the SAGA host criteria, as described in \S\ref{sec:sat_abund}.  The overall normalization of the ELVES histogram is larger (i.e. more satellites per host) than that of SAGA, similar to the findings of Figure \ref{fig:lf}, and this is especially the case with the red, quenched population. The SAGA-selected ELVES hosts have, on average, $1.38\pm0.26$ satellites per host in this luminosity range with NUV$-g>3$ while SAGA hosts only have $0.47\pm0.11$. The ELVES hosts have, on average, $1.62\pm0.28$ satellites per host in this luminosity range with NUV$-g<3$ while the SAGA hosts have $1.15\pm0.18$. Combining these, the ELVES hosts have $\sim1.5$ more satellites per host in this luminosity range with most of these satellites being UV-optically red. This difference is about the same as seen in Table \ref{tab:abund}. We do not expect the numbers to be exactly the same since no correction has been made for the fact that not all satellites have GALEX coverage. Note that also no correction has been made for the fact that some ELVES hosts are only surveyed out to a projected $150$ kpc nor for any unconfirmed ELVES satellites in this magnitude range, although there are not many (cf. Figure \ref{fig:contam}). 

The fact that the blue population for ELVES is not smaller than that of SAGA indicates that the visual classification is not simply erroneously classifying dwarfs as early-type that actually do have significant H$\alpha$ emission.

This discrepancy reinforces the possibility that SAGA is missing some number of quenched, LSB satellites, although, as before, it is conceivable that some of this can be explained by subtle differences in the host sample (e.g. large-scale environment). This might be explored with the full, 100-host sample from SAGA.

\begin{figure*}
\includegraphics[width=\textwidth]{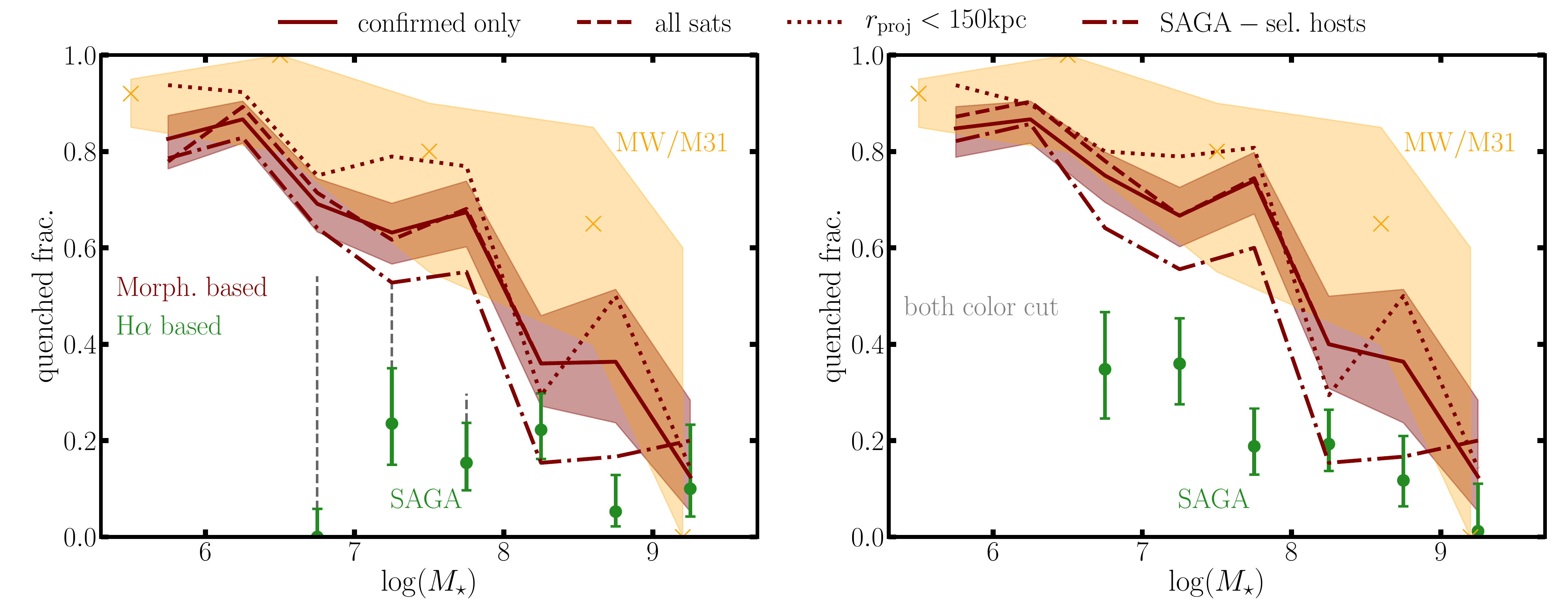}
\caption{The quenched fraction as a function of satellite stellar mass. The maroon lines show the ELVES results under various restrictions. The left panel uses the morphology to classify ELVES satellites as star-forming or quenched and the presence of H$\alpha$ emission for the SAGA satellites. The right panel uses a cut in color-magnitude space to classify satellites as star-forming or quenched for both ELVES and SAGA. For ELVES, the quenched fraction rises to roughly $80-90$\% for $M_\star\lesssim10^6$\msun, quite similar to the results for the Local Group (MW \& M31) from \citet{wetzel2015}, while the SAGA quenched fractions are much lower. The errorbars and shaded regions show Bayesian confidence regions for the binomial proportion parameter. In the left panel, the thin dotted extensions to the SAGA points show the correction for incompleteness in the spectroscopic follow-up, assuming all candidates are quenched. This incompleteness is also accounted for in the right panel, with the color being used to classify candidates as star-forming or quenched.}
\label{fig:qfrac_vs_mstar}
\end{figure*}

Figure \ref{fig:qfrac_vs_mstar} shows the quenched fraction as a function of satellite stellar mass. In the left panel, morphology is used to classify ELVES satellites as quenched or star-forming, and the presence of H$\alpha$ is used for the SAGA satellites \citep{mao2020}. In the right panel, a cut in color-magnitude space is used to classify both ELVES and SAGA satellites. The line we use to divide quenched/early-type from star-forming/late-type is $g-i=-0.067\times M_V - 0.23$. In \citet{carlsten2021a}, we found this line to cleanly separate early- from late-type dwarfs. Encouragingly, \citet{font2021b} also found that this line did a good job of separating star-forming from quenched satellites in the ARTEMIS simulations. The photometry we use for the SAGA satellites is again from our own S\'{e}rsic fits using DECaLS data, which is therefore quite comparable to ELVES photometry since $\sim3/4$ of ELVES hosts use DECaLS for photometry (see Table \ref{tab:hosts}). Additionally, we have argued above that the CFHT photometry (which the other $\sim1/4$ uses) is quite comparable to DECaLS without the need for a filter conversion.

Whether using morphology or a cut in color-magnitude space, the quenched fraction for ELVES satellites rises to $\sim85$\% for $M_\star < 10^6$\msun~ and is consistent with the results for LG satellites from \citet{wetzel2015}. This is the case regardless of whether only confirmed satellites are included, confirmed and possible satellites, only satellites within $r_\mathrm{proj}<150$ kpc, or only satellites of lower mass hosts that would satisfy the SAGA selection criteria (see \S\ref{sec:sat_abund} above). The inner satellites do show slightly higher quenched fractions, while the satellites of the less massive hosts show slightly lower fractions, but the differences are not great. In all cases, the ELVES quenched fraction is significantly higher than that of SAGA. In both panels, we include the uncertainty in the SAGA results due to incomplete spectroscopic follow-up. In the left panel, the thin dashed lines show the upper limit if \textit{all} unconfirmed satellites are quenched (but still taking into account their probability of being a satellite as calculated by SAGA). The upper limit of this uncertainty does reach the ELVES results. Thus it is possible that this difference would be resolved with complete spectroscopic follow-up in SAGA, but it would require \textit{every} new confirmed satellite to be quenched. In the right panel, the candidate satellites without spectra are classed as quenched or star-forming using their colors like all other dwarfs in that panel. The fact that the points do not reach the ELVES trend means that it is unreasonable to expect each of these candidates to be quenched as many are quite blue.

\begin{figure*}
\includegraphics[width=\textwidth]{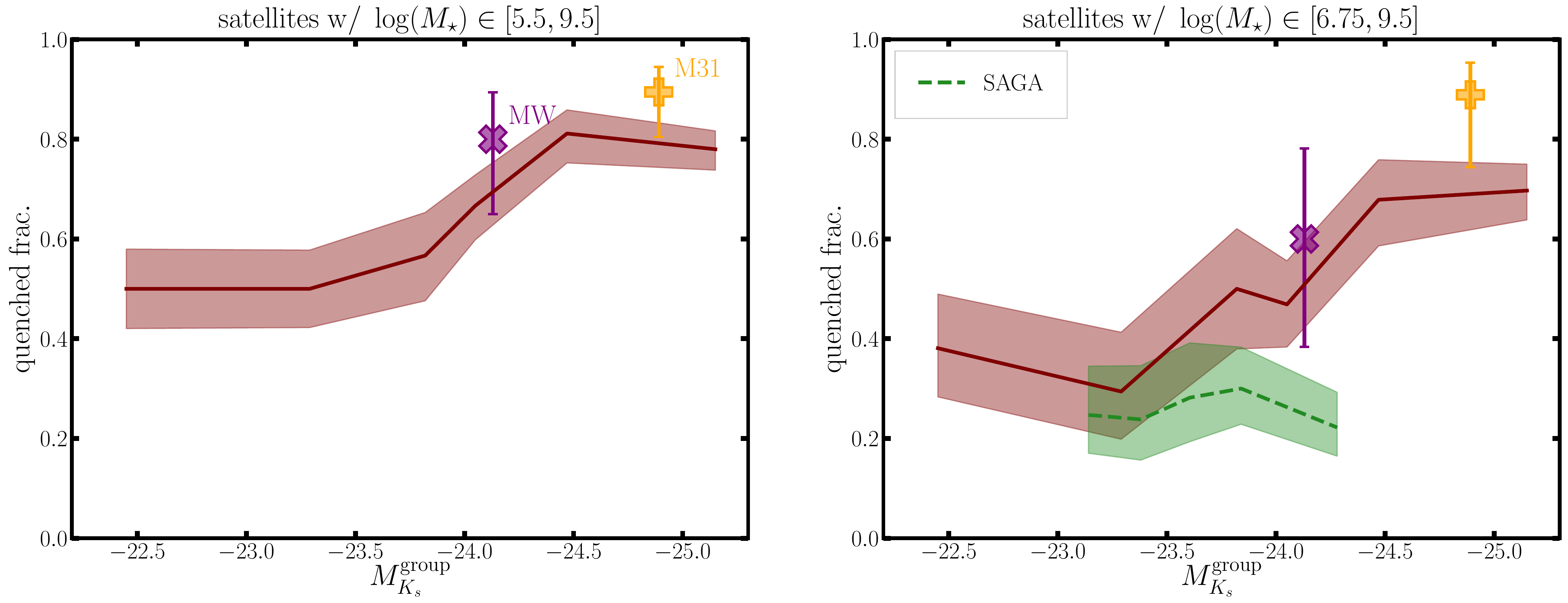}
\caption{The quenched fraction as a function of host mass. The satellite morphology is used to classify satellites as quenched or star-forming for the ELVES satellites. The two panels differ on the range of satellite stellar mass included. Individual results for the MW and M31 are shown. The right panel also shows the results for the SAGA Survey from \citep{mao2020}.}
\label{fig:qfrac_vs_host}
\end{figure*}

While we have shown that the difference between SAGA and ELVES satellites remains if we remove the ELVES hosts more massive than those selected in SAGA, it is possible that different distributions of host masses skew the comparison. Thus, in Figure \ref{fig:qfrac_vs_host}, we show the quenched fractions directly as a function of host mass (as proxied by $K$-band luminosity). The panels differ in the range of satellite stellar mass included in the quenched fraction calculation. The right panel includes satellites more massive than $\log(M_\star/M_\odot)>6.75$, corresponding to the SAGA luminosity limit of $M_r=-12.3$ mag assuming $M_\star/L_r\approx 1$ which is appropriate for a $g-r\sim0.45$ dwarf \citep{into2013} which is about the average color in ELVES. In both cases, there is a noticeable trend of higher quenched fractions for more massive hosts, a result that makes intuitive sense and has been observed for higher mass satellites in SDSS \citep{wetzel2013, wang2014}. 

The SAGA results include the incompleteness correction from \citet{mao2020}. For the satellites with redshifts, the star-forming status is determined by the presence of H$\alpha$ while the color is used for candidates without redshifts. The SAGA results do not show a strong trend of increasing quenched fraction with higher mass host\footnote{We note that a weak trend is seen if the incompleteness correction is not applied.} and are always lower than the ELVES trend, particularly for the higher mass hosts, similar to the results of Figure \ref{fig:qfrac_vs_mstar}.

Interestingly, even the lowest mass ELVES hosts ($M_{K_s}>-23$ mag, $M_\star\lesssim10^{10}$\msun) appear to be quenching their satellites quite efficiently. The current picture of ram pressure stripping is that it is due to the presence of an accretion shock-heated hot gas halo \citep[e.g.][]{birnboim2003,keres2009,gatto2013} around the host. Lower mass hosts ($M_\mathrm{halo}\lesssim10^{11}$ M$_\odot$) are not expected to form hot gas halos \citep{correa2018}, so it is not completely clear what the quenching mechanism may be or whether ram pressure stripping is enough to explain the $\sim50\%$ quenched fractions. Along this vein, recent observations have indicated that satellites of even Large Magellanic Cloud (LMC)-mass centrals can be quenched \citep{carlin2019, garling2020}.

In summary, in this section, we investigated the star formation properties of the ELVES satellite systems, particularly looking at how many of the satellites are actively forming stars versus how many are quenched. Since ELVES is only a photometric survey, we can only gauge whether a satellite is quenched from its morphology and color. However, we show that this morphology-based quenching classification is closely corroborated by UV-optical color (Figure \ref{fig:galex_properties}). Assuming this quenched classification is physical, we show that the quenched fraction of satellites in ELVES systems closely matches that of the LG with a majority of low-mass satellites being quenched (Figure \ref{fig:qfrac_vs_mstar}). In Figure \ref{fig:qfrac_vs_host}, we show that the quenched fraction increases with higher mass hosts, again with the MW and M31 being characteristic for hosts of their masses. We also compare with the results of the SAGA Survey \citep{geha2017, mao2020}. We show that, in general, the SAGA satellites without H$\alpha$ emission have similar UV-optical colors to the early-type ELVES satellites, but there are significantly more early-type satellites in ELVES. This causes the inferred quenched fractions for SAGA to be much lower than in the LG or ELVES. The dearth of red, early-type satellites in SAGA even though ELVES hosts of the same mass (Figure \ref{fig:qfrac_vs_host}) have plenty is suggestive of incompleteness in SAGA. We have shown that this discrepancy remains even when the SAGA results are corrected for incomplete spectroscopic follow-up, likely indicating incompleteness at the catalog level. This also explains the results in the previous two sections which showed lower average satellite abundance in SAGA, particularly in the inner regions of the satellite systems where satellites will likely be lower surface brightness, on average, and more easily missed.

\section{Conclusions}
\label{sec:conclusions}

This paper presents the Exploration of Local VolumE Satellites (ELVES) Survey. The main scientific goal of ELVES is to provide a robust and statistical sample of well-surveyed satellite systems that can illuminate many aspects of small-scale structure formation and galaxy evolution. In particular, ELVES surveyed the satellites of a nearly \textit{volume-limited} sample of massive, MW-like galaxies in the Local Volume. Hosts are selected via simply a cut in distance and $K_s$ luminosity (cf. \S\ref{sec:hosts}), facilitating easy, and direct, comparison with simulations. While most hosts are roughly MW-like in mass (cf. Figure \ref{fig:host_prop}), ELVES includes hosts both more and less massive than the MW, allowing insight into how host properties can affect satellite dwarf evolution.

The ELVES results include a satellite census for 30 of the 31 hosts in the volume-limited sample. Five of the satellite catalogs come from searches done by other groups in the literature while 25 systems are surveyed here in a consistent manner (cf. Table \ref{tab:hosts}). This work subsumes and extends our previous surveys presented in \citet{carlsten2020a, carlsten2020b}. For all hosts that we survey, we use the specialized low surface brightness dwarf detection algorithm presented in \citet{carlsten2020a} to detect candidate dwarf satellites via integrated light. Through extensive tests with injected artificial galaxies, we have verified that the ELVES satellite catalogs are complete to $M_V\sim-9$ mag and $\mu_{0,V}\sim26.5$ mag arcsec$^{-2}$ (cf. Figure \ref{fig:completeness_avg}). ELVES more than quadruples the number of satellite systems that have been surveyed to such low luminosity and surface brightness, opening the door to many different analyses on dwarf satellites \citep[e.g.][]{carlsten2021a, carlsten2021b}. While some hosts are only surveyed out to 150 projected kpc ($\sim R_\mathrm{vir}/2$ for MW-mass hosts), the majority are surveyed out to 300 kpc ($\sim R_\mathrm{vir}$).

Confirming that candidate satellites found in deep wide-field imaging surveys are actual satellites of a host is difficult as it requires a distance measurement to the candidate and is arguably the main complication hampering previous satellite survey work. To deal with this, ELVES makes heavy use of surface brightness fluctuation (SBF) measurements as an efficient way to determine distances to LV dwarfs with ground based data \citep[e.g.][]{sbf_calib, greco2020}. The $\sim15\%$ distance accuracy achievable with SBF can, in most cases, determine whether a satellite candidate is a real satellite or a background contaminant. However, our previous work has found that $\sim15-20\%$ of satellites confirmed via SBF are `near-field' dwarfs that are within $\sim1$ Mpc of the host but outside of the virial radius \citep{carlsten2020b}. This should be kept in mind when analyzing ELVES satellites lists. 

In total, we report 552 satellite candidates across the 25 surveyed hosts. Our surveys compare very favorably to previous surveys of the same hosts with ELVES almost always uncovering new, faint candidates. 251 of these candidates are confirmed via distance measurement with 136 of these being SBF distances. 196 are rejected via a distance measurement indicating they are in the background with almost all of these being SBF-based. This leaves 105 `unconfirmed/possible' satellites that do not yet have a conclusive distance constraint. Based on the luminosities and surface brightness of these candidates, we estimate the probability that each is a satellite so they can be easily accounted for in future analyses. Combining with the 5 previously surveyed systems from the literature, there is a total of 338 confirmed dwarf satellites with well-understood completeness. For all  ELVES-detected satellites and many of the literature satellites, we provide consistent S\'{e}rsic photometry, representing an important reference sample for studies of dwarf structure and evolution \citep[e.g.][]{carlsten2021a}.

Below, we list science highlights from the ELVES survey thus far, both in this paper and in the previous ones in this series:
\begin{itemize}
    \item \citet{carlsten2020b} performed in-depth comparisons between the observed satellite luminosity functions (LFs) of LV hosts and simulated LFs from modern cosmological simulations. That work found that popular stellar-to-halo-mass relations (SHMR) \citep[e.g.][]{gk_2017} combined with the simulations could naturally reproduce the observed satellite abundances and host-to-host scatter. That work also highlighted the potential for the full ELVES sample to actually \textit{fit} for the SHMR and thus constrain galaxy formation simulations.
    
    \item \citet{carlsten2020c} similarly compared observed satellite systems with simulated systems focusing on the radial distribution of satellites. They found the systems observed by that point (only 12 of the full ELVES sample) were significantly more concentrated than the simulations. In \S\ref{sec:sat_rads}, we briefly revisited this with the full ELVES sample and found that many of the \citet{carlsten2020c} host sample were among the most centrally concentrated of all of ELVES. This suggests the results of  \citet{carlsten2020c} could indeed be due to its limited sample size, but an in-depth comparison with simulations with the full ELVES sample is warranted.
    
    \item \citet{carlsten2021a} used most of the final ELVES sample to study the detailed structure of LV dwarf galaxies. That work found that early-type and late-type dwarfs had very similar sizes at fixed mass, indicating that the quenching and transformation process of a late-type into an early-type involves only mild, if any, structural change and likely is as simple as removing the gas from the dwarf. With that said, there were mild differences in the intrinsic (i.e. 3D) structure between early- and late-type dwarfs. That work provided a robust quantification of the mass-size relation in this low-mass regime, an important benchmark for simulations to try to recover and also useful in predicting dwarf search yields from future observational projects.
    
    \item \citet{carlsten2021b} investigated the nuclear star clusters and globular clusters (GCs) of ELVES dwarfs and compared with those of dwarfs in the Virgo Cluster. They found that dwarfs in the less dense LV environments were less likely to be nucleated, had lower mass nuclei, and had fewer GCs at fixed dwarf mass than dwarfs in Virgo. This correspondence in environmental dependence of both nuclei and GCs strengthens the case that nuclei form from coalescing GCs in this mass range \citep[e.g.][]{neumayer2020} and shows that GC formation and/or survival is strongly dependent on the large-scale environment that the host galaxy forms in \citep{peng2008}.
    
    \item In \S\ref{sec:sat_sf} of this work, we explore the star formation properties of ELVES satellites. Since ELVES is only a photometric survey, we use galaxy morphology (early- vs. late-type) as the primary indicator of whether a satellite is star-forming or quenched. We show that this distinction is strongly corroborated by UV-optical color (Figure \ref{fig:galex_properties}) with early-type satellites having quite red UV-optical colors, similar to those of satellites in the SAGA Survey \citep{mao2020} that show no H$\alpha$ emission. We show that ELVES satellites are strongly quenched ($\sim90$\%) at low masses ($M_\star\lesssim10^6$\msun), similar to the known situation in the Local Group (Figure \ref{fig:qfrac_vs_mstar}). In this way, the required rapid quenching timescales needed to reproduce the Local Group results hold up for other hosts in the LV. In Figure \ref{fig:qfrac_vs_host}, we show that the quenched fractions are higher for higher mass ELVES hosts but do not appear to get lower than $\sim50\%$ for even the lowest mass ($M_\star\lesssim10^{10}$\msun) hosts.

    \item Throughout this paper, we perform many comparisons with the SAGA Survey \citep{geha2017, mao2020}, a complementary spectroscopic survey of many nearby satellite systems. SAGA is sensitive to only the bright ($M_V\lesssim-12$ mag) satellites but will eventually survey more hosts (100) than is possible in the LV. Overall, we show that the ELVES results indicate higher satellite abundance than SAGA for hosts of the same mass and a much steeper average LF. This is particularly the case for lower luminosity ($M_V\gtrsim-15$ mag) satellites (Figure \ref{fig:lf}) and the inner regions of satellite systems (Figure \ref{fig:rproj}). ELVES hosts have on average $\sim1.5$ satellites per host more than SAGA hosts. Similarly, ELVES hosts exhibit a significant population of red, early-type satellites that are much rarer in the SAGA results (Figure \ref{fig:galex_properties}), causing SAGA to infer lower quenched fractions than are found in ELVES (Figure \ref{fig:qfrac_vs_mstar}). At least some part of this difference is likely caused by incompleteness in the SAGA satellite lists, particularly in the DECaLS targeting catalogs used. Incompleteness in only the spectroscopic follow-up, which is modelled within SAGA, is not enough to explain this discrepancy. However, it is also possible that host-to-host scatter is contributing some of the discrepancy. The full 100-host SAGA sample will be important in quantifying this and exploring whether this  might be caused by some physical differences in the hosts \citep[e.g. different large scale environments, ][]{neuzil2020}. 
    
\end{itemize}

The field of dwarf satellite research has a bright future with the next-generation observational facilities coming online in the upcoming decade. Vera Rubin Observatory's Legacy Survey of Space and Time and the Roman Space Telescope will allow for searches and, even more importantly, distance measurements of classical-mass (e.g. $M_\star>10^5$ \msun) satellites out to 20-30 Mpc \citep{greco2020}, pushing beyond the statistics of ELVES and what is possible in the LV. Additionally both facilities will allow for dwarf searches in the LV to extend into the ultra-faint regime \citep{mutlu2021b}, opening up a window into another, very interesting regime of dwarf galaxy evolution. These facilities will also allow for a complete census of low-mass dwarfs in the field (i.e. non-satellite dwarfs) in the LV which, even with SBF, is infeasible to obtain with current facilities. This will, in turn, allow for much more in-depth studies on dwarf galaxy evolution and the effects that massive hosts have on their dwarf satellite retinue.

\section*{Acknowledgements}
We thank Yao-Yuan Mao and Marla Geha for helpful discussions regarding the SAGA Survey results. Support for this work was provided by NASA through Hubble Fellowship grant \#51386.01 awarded to R.L.B. by the Space Telescope Science Institute, which is operated by the Association of  Universities for Research in Astronomy, Inc., for NASA, under contract NAS 5-26555. J.P.G. is supported by an NSF Astronomy and Astrophysics Postdoctoral Fellowship under award AST-1801921. J.E.G. is partially supported by the National Science Foundation grant AST-1713828. S.G.C acknowledges support by the National Science Foundation Graduate Research Fellowship Program under Grant No. \#DGE-1656466. S.D. is supported by NASA through Hubble Fellowship grant HST-HF2-51454.001-A awarded by the Space Telescope Science Institute, which is operated by the Association of Universities for Research in Astronomy, Incorporated, under NASA contract NAS5-26555.

Based on observations obtained with MegaPrime/MegaCam, a joint project of CFHT and CEA/IRFU, at the Canada-France-Hawaii Telescope (CFHT) which is operated by the National Research Council (NRC) of Canada, the Institut National des Science de l'Univers of the Centre National de la Recherche Scientifique (CNRS) of France, and the University of Hawaii. This research was made possible through the use of the AAVSO Photometric All-Sky Survey (APASS) \citep{apass}, funded by the Robert Martin Ayers Sciences Fund and NSF AST-1412587

This research has made use of the NASA/IPAC Extragalactic Database (NED),
which is operated by the Jet Propulsion Laboratory, California Institute of Technology, under contract with the National Aeronautics and Space Administration. \\
This research has made use of the SIMBAD database, operated at CDS, Strasbourg, France \citep{Wenger_2000}.

The Legacy Surveys consist of three individual and complementary projects: the Dark Energy Camera Legacy Survey (DECaLS; Proposal ID \#2014B-0404; PIs: David Schlegel and Arjun Dey), the Beijing-Arizona Sky Survey (BASS; NOAO Prop. ID \#2015A-0801; PIs: Zhou Xu and Xiaohui Fan), and the Mayall z-band Legacy Survey (MzLS; Prop. ID \#2016A-0453; PI: Arjun Dey). DECaLS, BASS and MzLS together include data obtained, respectively, at the Blanco telescope, Cerro Tololo Inter-American Observatory, NSF’s NOIRLab; the Bok telescope, Steward Observatory, University of Arizona; and the Mayall telescope, Kitt Peak National Observatory, NOIRLab. The Legacy Surveys project is honored to be permitted to conduct astronomical research on Iolkam Du’ag (Kitt Peak), a mountain with particular significance to the Tohono O’odham Nation.

NOIRLab is operated by the Association of Universities for Research in Astronomy (AURA) under a cooperative agreement with the National Science Foundation.

This project used data obtained with the Dark Energy Camera (DECam), which was constructed by the Dark Energy Survey (DES) collaboration. Funding for the DES Projects has been provided by the U.S. Department of Energy, the U.S. National Science Foundation, the Ministry of Science and Education of Spain, the Science and Technology Facilities Council of the United Kingdom, the Higher Education Funding Council for England, the National Center for Supercomputing Applications at the University of Illinois at Urbana-Champaign, the Kavli Institute of Cosmological Physics at the University of Chicago, Center for Cosmology and Astro-Particle Physics at the Ohio State University, the Mitchell Institute for Fundamental Physics and Astronomy at Texas A\&M University, Financiadora de Estudos e Projetos, Fundacao Carlos Chagas Filho de Amparo, Financiadora de Estudos e Projetos, Fundacao Carlos Chagas Filho de Amparo a Pesquisa do Estado do Rio de Janeiro, Conselho Nacional de Desenvolvimento Cientifico e Tecnologico and the Ministerio da Ciencia, Tecnologia e Inovacao, the Deutsche Forschungsgemeinschaft and the Collaborating Institutions in the Dark Energy Survey. The Collaborating Institutions are Argonne National Laboratory, the University of California at Santa Cruz, the University of Cambridge, Centro de Investigaciones Energeticas, Medioambientales y Tecnologicas-Madrid, the University of Chicago, University College London, the DES-Brazil Consortium, the University of Edinburgh, the Eidgenossische Technische Hochschule (ETH) Zurich, Fermi National Accelerator Laboratory, the University of Illinois at Urbana-Champaign, the Institut de Ciencies de l’Espai (IEEC/CSIC), the Institut de Fisica d’Altes Energies, Lawrence Berkeley National Laboratory, the Ludwig Maximilians Universitat Munchen and the associated Excellence Cluster Universe, the University of Michigan, NSF’s NOIRLab, the University of Nottingham, the Ohio State University, the University of Pennsylvania, the University of Portsmouth, SLAC National Accelerator Laboratory, Stanford University, the University of Sussex, and Texas A\&M University.

BASS is a key project of the Telescope Access Program (TAP), which has been funded by the National Astronomical Observatories of China, the Chinese Academy of Sciences (the Strategic Priority Research Program “The Emergence of Cosmological Structures” Grant \# XDB09000000), and the Special Fund for Astronomy from the Ministry of Finance. The BASS is also supported by the External Cooperation Program of Chinese Academy of Sciences (Grant \# 114A11KYSB20160057), and Chinese National Natural Science Foundation (Grant \# 11433005).

The Legacy Survey team makes use of data products from the Near-Earth Object Wide-field Infrared Survey Explorer (NEOWISE), which is a project of the Jet Propulsion Laboratory/California Institute of Technology. NEOWISE is funded by the National Aeronautics and Space Administration.

The Legacy Surveys imaging of the DESI footprint is supported by the Director, Office of Science, Office of High Energy Physics of the U.S. Department of Energy under Contract No. DE-AC02-05CH1123, by the National Energy Research Scientific Computing Center, a DOE Office of Science User Facility under the same contract; and by the U.S. National Science Foundation, Division of Astronomical Sciences under Contract No. AST-0950945 to NOAO.

The Hyper Suprime-Cam (HSC) collaboration includes the astronomical communities of Japan and Taiwan, and Princeton University. The HSC instrumentation and software were developed by the National Astronomical Observatory of Japan (NAOJ), the Kavli Institute for the Physics and Mathematics of the Universe (Kavli IPMU), the University of Tokyo, the High Energy Accelerator Research Organization (KEK), the Academia Sinica Institute for Astronomy and Astrophysics in Taiwan (ASIAA), and Princeton University. Funding was contributed by the FIRST program from the Japanese Cabinet Office, the Ministry of Education, Culture, Sports, Science and Technology (MEXT), the Japan Society for the Promotion of Science (JSPS), Japan Science and Technology Agency (JST), the Toray Science Foundation, NAOJ, Kavli IPMU, KEK, ASIAA, and Princeton University. 

This paper makes use of software developed for Vera C. Rubin Observatory. We thank the Rubin Observatory for making their code available as free software at http://pipelines.lsst.io/.

This paper is based on data collected at the Subaru Telescope and retrieved from the HSC data archive system, which is operated by the Subaru Telescope and Astronomy Data Center (ADC) at NAOJ. Data analysis was in part carried out with the cooperation of Center for Computational Astrophysics (CfCA), NAOJ. We are honored and grateful for the opportunity of observing the Universe from Maunakea, which has the cultural, historical and natural significance in Hawaii.

\software{ \texttt{astropy} \citep{astropy, bradley2020}, \texttt{sep} \citep{sep}, \texttt{imfit} \citep{imfit}, \texttt{SWarp} \citep{swarp}, \texttt{Scamp} \citep{scamp}, \texttt{SExtractor} \citep{sextractor}, \texttt{astrometry.net} \citep{astrometry_net}   } 

\bibliographystyle{aasjournal}
\bibliography{calib}

\begin{thebibliography}{}
\expandafter\ifx\csname natexlab\endcsname\relax\def\natexlab#1{#1}\fi
\providecommand{\url}[1]{\href{#1}{#1}}

\bibitem[{{Abolfathi} {et~al.}(2018){Abolfathi}, {Aguado}, {Aguilar}, {Allende
  Prieto}, {Almeida}, {Ananna}, {Anders}, {Anderson}, {Andrews}, {Anguiano}, \&
  et~al.}]{sdss_df14}
{Abolfathi}, B., {Aguado}, D.~S., {Aguilar}, G., {et~al.} 2018, \apjs, 235, 42

\bibitem[{{Aihara} {et~al.}(2018){Aihara}, {Arimoto}, {Armstrong}, {Arnouts},
  {Bahcall}, {Bickerton}, {Bosch}, {Bundy}, {Capak}, {Chan}, {Chiba}, {Coupon},
  {Egami}, {Enoki}, {Finet}, {Fujimori}, {Fujimoto}, {Furusawa}, {Furusawa},
  {Goto}, {Goulding}, {Greco}, {Greene}, {Gunn}, {Hamana}, {Harikane},
  {Hashimoto}, {Hattori}, {Hayashi}, {Hayashi}, {He{\l}miniak}, {Higuchi},
  {Hikage}, {Ho}, {Hsieh}, {Huang}, {Huang}, {Ikeda}, {Imanishi}, {Inoue},
  {Iwasawa}, {Iwata}, {Jaelani}, {Jian}, {Kamata}, {Karoji}, {Kashikawa},
  {Katayama}, {Kawanomoto}, {Kayo}, {Koda}, {Koike}, {Kojima}, {Komiyama},
  {Konno}, {Koshida}, {Koyama}, {Kusakabe}, {Leauthaud}, {Lee}, {Lin}, {Lin},
  {Lupton}, {Mandelbaum}, {Matsuoka}, {Medezinski}, {Mineo}, {Miyama},
  {Miyatake}, {Miyazaki}, {Momose}, {More}, {More}, {Moritani}, {Moriya},
  {Morokuma}, {Mukae}, {Murata}, {Murayama}, {Nagao}, {Nakata}, {Niida},
  {Niikura}, {Nishizawa}, {Obuchi}, {Oguri}, {Oishi}, {Okabe}, {Okamoto},
  {Okura}, {Ono}, {Onodera}, {Onoue}, {Osato}, {Ouchi}, {Price}, {Pyo}, {Sako},
  {Sawicki}, {Shibuya}, {Shimasaku}, {Shimono}, {Shirasaki}, {Silverman},
  {Simet}, {Speagle}, {Spergel}, {Strauss}, {Sugahara}, {Sugiyama}, {Suto},
  {Suyu}, {Suzuki}, {Tait}, {Takada}, {Takata}, {Tamura}, {Tanaka}, {Tanaka},
  {Tanaka}, {Tanaka}, {Terai}, {Terashima}, {Toba}, {Tominaga}, {Toshikawa},
  {Turner}, {Uchida}, {Uchiyama}, {Umetsu}, {Uraguchi}, {Urata}, {Usuda},
  {Utsumi}, {Wang}, {Wang}, {Wong}, {Yabe}, {Yamada}, {Yamanoi}, {Yasuda},
  {Yeh}, {Yonehara}, \& {Yuma}}]{aihara2018}
{Aihara}, H., {Arimoto}, N., {Armstrong}, R., {et~al.} 2018, \pasj, 70, S4

\bibitem[{{Aihara} {et~al.}(2019){Aihara}, {AlSayyad}, {Ando}, {Armstrong},
  {Bosch}, {Egami}, {Furusawa}, {Furusawa}, {Goulding}, {Harikane}, {Hikage},
  {Ho}, {Hsieh}, {Huang}, {Ikeda}, {Imanishi}, {Ito}, {Iwata}, {Jaelani},
  {Kakuma}, {Kawana}, {Kikuta}, {Kobayashi}, {Koike}, {Komiyama}, {Li},
  {Liang}, {Lin}, {Luo}, {Lupton}, {Lust}, {MacArthur}, {Matsuoka}, {Mineo},
  {Miyatake}, {Miyazaki}, {More}, {Murata}, {Namiki}, {Nishizawa}, {Oguri},
  {Okabe}, {Okamoto}, {Okura}, {Ono}, {Onodera}, {Onoue}, {Osato}, {Ouchi},
  {Shibuya}, {Strauss}, {Sugiyama}, {Suto}, {Takada}, {Takagi}, {Takata},
  {Takita}, {Tanaka}, {Terai}, {Toba}, {Uchiyama}, {Utsumi}, {Wang}, {Wang}, \&
  {Yamada}}]{aihara2019}
{Aihara}, H., {AlSayyad}, Y., {Ando}, M., {et~al.} 2019, \pasj, 71, 114

\bibitem[{{Akhlaghi} \& {Ichikawa}(2015)}]{noise_chisel}
{Akhlaghi}, M., \& {Ichikawa}, T. 2015, \apjs, 220, 1

\bibitem[{{Akins} {et~al.}(2021){Akins}, {Christensen}, {Brooks}, {Munshi},
  {Applebaum}, {Engelhardt}, \& {Chamberland}}]{akins2021}
{Akins}, H.~B., {Christensen}, C.~R., {Brooks}, A.~M., {et~al.} 2021, \apj,
  909, 139

\bibitem[{{Anand} {et~al.}(2021){Anand}, {Rizzi}, {Tully}, {Shaya},
  {Karachentsev}, {Makarov}, {Makarova}, {Wu}, {Dolphin}, \&
  {Kourkchi}}]{anand2021}
{Anand}, G.~S., {Rizzi}, L., {Tully}, R.~B., {et~al.} 2021, arXiv e-prints,
  arXiv:2104.02649

\bibitem[{{Applebaum} {et~al.}(2020){Applebaum}, {Brooks}, {Christensen},
  {Munshi}, {Quinn}, {Shen}, \& {Tremmel}}]{applebaum2020}
{Applebaum}, E., {Brooks}, A.~M., {Christensen}, C.~R., {et~al.} 2020, arXiv
  e-prints, arXiv:2008.11207

\bibitem[{{Astropy Collaboration} {et~al.}(2018){Astropy Collaboration},
  {Price-Whelan}, {Sip{\H o}cz}, {G{\"u}nther}, {Lim}, {Crawford}, {Conseil},
  {Shupe}, {Craig}, {Dencheva}, {Ginsburg}, {VanderPlas}, {Bradley},
  {P{\'e}rez-Su{\'a}rez}, {de Val-Borro}, {Aldcroft}, {Cruz}, {Robitaille},
  {Tollerud}, {Ardelean}, {Babej}, {Bach}, {Bachetti}, {Bakanov}, {Bamford},
  {Barentsen}, {Barmby}, {Baumbach}, {Berry}, {Biscani}, {Boquien}, {Bostroem},
  {Bouma}, {Brammer}, {Bray}, {Breytenbach}, {Buddelmeijer}, {Burke},
  {Calderone}, {Cano Rodr{\'{\i}}guez}, {Cara}, {Cardoso}, {Cheedella},
  {Copin}, {Corrales}, {Crichton}, {D'Avella}, {Deil}, {Depagne}, {Dietrich},
  {Donath}, {Droettboom}, {Earl}, {Erben}, {Fabbro}, {Ferreira}, {Finethy},
  {Fox}, {Garrison}, {Gibbons}, {Goldstein}, {Gommers}, {Greco}, {Greenfield},
  {Groener}, {Grollier}, {Hagen}, {Hirst}, {Homeier}, {Horton}, {Hosseinzadeh},
  {Hu}, {Hunkeler}, {Ivezi{\'c}}, {Jain}, {Jenness}, {Kanarek}, {Kendrew},
  {Kern}, {Kerzendorf}, {Khvalko}, {King}, {Kirkby}, {Kulkarni}, {Kumar},
  {Lee}, {Lenz}, {Littlefair}, {Ma}, {Macleod}, {Mastropietro}, {McCully},
  {Montagnac}, {Morris}, {Mueller}, {Mumford}, {Muna}, {Murphy}, {Nelson},
  {Nguyen}, {Ninan}, {N{\"o}the}, {Ogaz}, {Oh}, {Parejko}, {Parley}, {Pascual},
  {Patil}, {Patil}, {Plunkett}, {Prochaska}, {Rastogi}, {Reddy Janga},
  {Sabater}, {Sakurikar}, {Seifert}, {Sherbert}, {Sherwood-Taylor}, {Shih},
  {Sick}, {Silbiger}, {Singanamalla}, {Singer}, {Sladen}, {Sooley},
  {Sornarajah}, {Streicher}, {Teuben}, {Thomas}, {Tremblay}, {Turner},
  {Terr{\'o}n}, {van Kerkwijk}, {de la Vega}, {Watkins}, {Weaver}, {Whitmore},
  {Woillez}, {Zabalza}, \& {Astropy Contributors}}]{astropy}
{Astropy Collaboration}, {Price-Whelan}, A.~M., {Sip{\H o}cz}, B.~M., {et~al.}
  2018, \aj, 156, 123

\bibitem[{{Barbary}(2016)}]{sep}
{Barbary}, K. 2016, JOSS, 1, 58

\bibitem[{{Beaton} {et~al.}(2018){Beaton}, {Bono}, {Braga}, {Dall'Ora},
  {Fiorentino}, {Jang}, {Mart{\'{\i}}nez-V{\'a}zquez}, {Matsunaga}, {Monelli},
  {Neeley}, \& {Salaris}}]{beaton2018}
{Beaton}, R.~L., {Bono}, G., {Braga}, V.~F., {et~al.} 2018, \ssr, 214, 113

\bibitem[{{Beaton} {et~al.}(2019){Beaton}, {Seibert}, {Hatt}, {Freedman},
  {Hoyt}, {Jang}, {Lee}, {Madore}, {Monson}, {Neeley}, {Rich}, \&
  {Scowcroft}}]{beaton2019}
{Beaton}, R.~L., {Seibert}, M., {Hatt}, D., {et~al.} 2019, \apj, 885, 141

\bibitem[{{Bennet} {et~al.}(2019){Bennet}, {Sand}, {Crnojevi{\'c}}, {Spekkens},
  {Karunakaran}, {Zaritsky}, \& {Mutlu-Pakdil}}]{bennet2019}
{Bennet}, P., {Sand}, D.~J., {Crnojevi{\'c}}, D., {et~al.} 2019, arXiv
  e-prints, arXiv:1906.03230

\bibitem[{{Bennet} {et~al.}(2020){Bennet}, {Sand}, {Crnojevi{\'c}}, {Spekkens},
  {Karunakaran}, {Zaritsky}, \& {Mutlu-Pakdil}}]{bennet2020}
---. 2020, \apjl, 893, L9

\bibitem[{{Bennet} {et~al.}(2017){Bennet}, {Sand}, {Crnojevi{\'c}}, {Spekkens},
  {Zaritsky}, \& {Karunakaran}}]{bennet2017}
---. 2017, \apj, 850, 109

\bibitem[{{Bertin}(2006)}]{scamp}
{Bertin}, E. 2006, in Astronomical Society of the Pacific Conference Series,
  Vol. 351, Astronomical Data Analysis Software and Systems XV, ed.
  C.~{Gabriel}, C.~{Arviset}, D.~{Ponz}, \& S.~{Enrique}, 112

\bibitem[{{Bertin}(2010)}]{swarp}
{Bertin}, E. 2010, {SWarp: Resampling and Co-adding FITS Images Together},
  Astrophysics Source Code Library, , , ascl:1010.068

\bibitem[{{Bertin} \& {Arnouts}(1996)}]{sextractor}
{Bertin}, E., \& {Arnouts}, S. 1996, \aaps, 117, 393

\bibitem[{{Birnboim} \& {Dekel}(2003)}]{birnboim2003}
{Birnboim}, Y., \& {Dekel}, A. 2003, \mnras, 345, 349

\bibitem[{{Bland-Hawthorn} \& {Gerhard}(2016)}]{bland2016}
{Bland-Hawthorn}, J., \& {Gerhard}, O. 2016, \araa, 54, 529

\bibitem[{{Blanton} {et~al.}(2011){Blanton}, {Kazin}, {Muna}, {Weaver}, \&
  {Price-Whelan}}]{blanton2011}
{Blanton}, M.~R., {Kazin}, E., {Muna}, D., {Weaver}, B.~A., \& {Price-Whelan},
  A. 2011, \aj, 142, 31

\bibitem[{{Bosch} {et~al.}(2018){Bosch}, {Armstrong}, {Bickerton}, {Furusawa},
  {Ikeda}, {Koike}, {Lupton}, {Mineo}, {Price}, {Takata}, {Tanaka}, {Yasuda},
  {AlSayyad}, {Becker}, {Coulton}, {Coupon}, {Garmilla}, {Huang}, {Krughoff},
  {Lang}, {Leauthaud}, {Lim}, {Lust}, {MacArthur}, {Mand elbaum}, {Miyatake},
  {Miyazaki}, {Murata}, {More}, {Okura}, {Owen}, {Swinbank}, {Strauss},
  {Yamada}, \& {Yamanoi}}]{bosch2018}
{Bosch}, J., {Armstrong}, R., {Bickerton}, S., {et~al.} 2018, \pasj, 70, S5

\bibitem[{{Boselli} {et~al.}(2008){Boselli}, {Boissier}, {Cortese}, \&
  {Gavazzi}}]{boselli2008}
{Boselli}, A., {Boissier}, S., {Cortese}, L., \& {Gavazzi}, G. 2008, \apj, 674,
  742

\bibitem[{{Boylan-Kolchin} {et~al.}(2011){Boylan-Kolchin}, {Bullock}, \&
  {Kaplinghat}}]{bk2011}
{Boylan-Kolchin}, M., {Bullock}, J.~S., \& {Kaplinghat}, M. 2011, \mnras, 415,
  L40

\bibitem[{{Boylan-Kolchin} {et~al.}(2012){Boylan-Kolchin}, {Bullock}, \&
  {Kaplinghat}}]{bk2012}
---. 2012, \mnras, 422, 1203

\bibitem[{{Bradley} {et~al.}(2020){Bradley}, {Sip{\H{o}}cz}, {Robitaille},
  {Tollerud}, {Vin{\'\i}cius}, {Deil}, {Barbary}, {Wilson}, {Busko},
  {G{\"u}nther}, {Cara}, {Conseil}, {Bostroem}, {Droettboom}, {Bray}, {Andersen
  Bratholm}, {Lim}, {Barentsen}, {Craig}, {Pascual}, {Perren}, {Greco},
  {Donath}, {De Val-Borro}, {Kerzendorf}, {Bach}, {Weaver}, {D'Eugenio},
  {Souchereau}, \& {Ferreira}}]{bradley2020}
{Bradley}, L., {Sip{\H{o}}cz}, B., {Robitaille}, T., {et~al.} 2020,
  {astropy/photutils: 1.0.0}, v1.0.0,  Zenodo, doi:10.5281/zenodo.4044744

\bibitem[{{Brooks} {et~al.}(2013){Brooks}, {Kuhlen}, {Zolotov}, \&
  {Hooper}}]{brooks2013}
{Brooks}, A.~M., {Kuhlen}, M., {Zolotov}, A., \& {Hooper}, D. 2013, \apj, 765,
  22

\bibitem[{{Brooks} \& {Zolotov}(2014)}]{brooks2014}
{Brooks}, A.~M., \& {Zolotov}, A. 2014, \apj, 786, 87

\bibitem[{{Bullock} \& {Boylan-Kolchin}(2017)}]{bullock2017}
{Bullock}, J.~S., \& {Boylan-Kolchin}, M. 2017, \araa, 55, 343

\bibitem[{{Byun} {et~al.}(2020){Byun}, {Sheen}, {Park}, {Ho}, {Lee}, {Kim},
  {Jeong}, {Park}, {Seon}, {Lee}, {Lee}, {Cha}, {Ko}, \& {Kim}}]{byun2020}
{Byun}, W., {Sheen}, Y.-K., {Park}, H.~S., {et~al.} 2020, \apj, 891, 18

\bibitem[{{Cantiello} {et~al.}(2018){Cantiello}, {Blakeslee}, {Ferrarese},
  {C{\^o}t{\'e}}, {Roediger}, {Raimondo}, {Peng}, {Gwyn}, {Durrell}, \&
  {Cuillandre}}]{cantiello2018}
{Cantiello}, M., {Blakeslee}, J.~P., {Ferrarese}, L., {et~al.} 2018, \apj, 856,
  126

\bibitem[{{Carlin} {et~al.}(2016){Carlin}, {Sand}, {Price}, {Willman},
  {Karunakaran}, {Spekkens}, {Bell}, {Brodie}, {Crnojevi{\'c}}, {Forbes},
  {Hargis}, {Kirby}, {Lupton}, {Peter}, {Romanowsky}, \& {Strader}}]{madcash}
{Carlin}, J.~L., {Sand}, D.~J., {Price}, P., {et~al.} 2016, \apjl, 828, L5

\bibitem[{{Carlin} {et~al.}(2019){Carlin}, {Garling}, {Peter}, {Crnojevi{\'c}},
  {Forbes}, {Hargis}, {Mutlu-Pakdil}, {Pucha}, {Romanowsky}, \&
  {Sand}}]{carlin2019}
{Carlin}, J.~L., {Garling}, C.~T., {Peter}, A. H.~G., {et~al.} 2019, arXiv
  e-prints, arXiv:1906.08260

\bibitem[{{Carlin} {et~al.}(2021){Carlin}, {Mutlu-Pakdil}, {Crnojevi{\'c}},
  {Garling}, {Karunakaran}, {Peter}, {Tollerud}, {Forbes}, {Hargis}, {Lim},
  {Romanowsky}, {Sand}, {Spekkens}, \& {Strader}}]{carlin2021}
{Carlin}, J.~L., {Mutlu-Pakdil}, B., {Crnojevi{\'c}}, D., {et~al.} 2021, \apj,
  909, 211

\bibitem[{{Carlsten} {et~al.}(2019{\natexlab{a}}){Carlsten}, {Beaton}, {Greco},
  \& {Greene}}]{sbf_m101}
{Carlsten}, S.~G., {Beaton}, R.~L., {Greco}, J.~P., \& {Greene}, J.~E.
  2019{\natexlab{a}}, \apjl, 878, L16

\bibitem[{{Carlsten} {et~al.}(2019{\natexlab{b}}){Carlsten}, {Beaton}, {Greco},
  \& {Greene}}]{sbf_calib}
---. 2019{\natexlab{b}}, \apj, 879, 13

\bibitem[{{Carlsten} {et~al.}(2020{\natexlab{a}}){Carlsten}, {Greco}, {Beaton},
  \& {Greene}}]{carlsten2020a}
{Carlsten}, S.~G., {Greco}, J.~P., {Beaton}, R.~L., \& {Greene}, J.~E.
  2020{\natexlab{a}}, \apj, 891, 144

\bibitem[{{Carlsten} {et~al.}(2021{\natexlab{a}}){Carlsten}, {Greene},
  {Beaton}, \& {Greco}}]{carlsten2021b}
{Carlsten}, S.~G., {Greene}, J.~E., {Beaton}, R.~L., \& {Greco}, J.~P.
  2021{\natexlab{a}}, arXiv e-prints, arXiv:2105.03440

\bibitem[{{Carlsten} {et~al.}(2021{\natexlab{b}}){Carlsten}, {Greene}, {Greco},
  {Beaton}, \& {Kado-Fong}}]{carlsten2021a}
{Carlsten}, S.~G., {Greene}, J.~E., {Greco}, J.~P., {Beaton}, R.~L., \&
  {Kado-Fong}, E. 2021{\natexlab{b}}, arXiv e-prints, arXiv:2105.03435

\bibitem[{{Carlsten} {et~al.}(2021{\natexlab{c}}){Carlsten}, {Greene}, {Peter},
  {Beaton}, \& {Greco}}]{carlsten2020b}
{Carlsten}, S.~G., {Greene}, J.~E., {Peter}, A. H.~G., {Beaton}, R.~L., \&
  {Greco}, J.~P. 2021{\natexlab{c}}, \apj, 908, 109

\bibitem[{{Carlsten} {et~al.}(2020{\natexlab{b}}){Carlsten}, {Greene}, {Peter},
  {Greco}, \& {Beaton}}]{carlsten2020c}
{Carlsten}, S.~G., {Greene}, J.~E., {Peter}, A. H.~G., {Greco}, J.~P., \&
  {Beaton}, R.~L. 2020{\natexlab{b}}, \apj, 902, 124

\bibitem[{{Carlsten} {et~al.}(2018){Carlsten}, {Strauss}, {Lupton}, {Meyers},
  \& {Miyazaki}}]{scott_psf}
{Carlsten}, S.~G., {Strauss}, M.~A., {Lupton}, R.~H., {Meyers}, J.~E., \&
  {Miyazaki}, S. 2018, \mnras, 479, 1491

\bibitem[{{Chambers} {et~al.}(2016){Chambers}, {Magnier}, {Metcalfe},
  {Flewelling}, {Huber}, {Waters}, {Denneau}, {Draper}, {Farrow}, {Finkbeiner},
  {Holmberg}, {Koppenhoefer}, {Price}, {Saglia}, {Schlafly}, {Smartt},
  {Sweeney}, {Wainscoat}, {Burgett}, {Grav}, {Heasley}, {Hodapp}, {Jedicke},
  {Kaiser}, {Kudritzki}, {Luppino}, {Lupton}, {Monet}, {Morgan}, {Onaka},
  {Stubbs}, {Tonry}, {Banados}, {Bell}, {Bender}, {Bernard}, {Botticella},
  {Casertano}, {Chastel}, {Chen}, {Chen}, {Cole}, {Deacon}, {Frenk},
  {Fitzsimmons}, {Gezari}, {Goessl}, {Goggia}, {Goldman}, {Grebel}, {Hambly},
  {Hasinger}, {Heavens}, {Heckman}, {Henderson}, {Henning}, {Holman}, {Hopp},
  {Ip}, {Isani}, {Keyes}, {Koekemoer}, {Kotak}, {Long}, {Lucey}, {Liu},
  {Martin}, {McLean}, {Morganson}, {Murphy}, {Nieto-Santisteban}, {Norberg},
  {Peacock}, {Pier}, {Postman}, {Primak}, {Rae}, {Rest}, {Riess}, {Riffeser},
  {Rix}, {Roser}, {Schilbach}, {Schultz}, {Scolnic}, {Szalay}, {Seitz},
  {Shiao}, {Small}, {Smith}, {Soderblom}, {Taylor}, {Thakar}, {Thiel},
  {Thilker}, {Urata}, {Valenti}, {Walter}, {Watters}, {Werner}, {White},
  {Wood-Vasey}, \& {Wyse}}]{panstarrs}
{Chambers}, K.~C., {Magnier}, E.~A., {Metcalfe}, N., {et~al.} 2016, ArXiv
  e-prints, arXiv:1612.05560

\bibitem[{{Chiboucas} {et~al.}(2013){Chiboucas}, {Jacobs}, {Tully}, \&
  {Karachentsev}}]{chiboucas2013}
{Chiboucas}, K., {Jacobs}, B.~A., {Tully}, R.~B., \& {Karachentsev}, I.~D.
  2013, \aj, 146, 126

\bibitem[{{Chiboucas} {et~al.}(2009){Chiboucas}, {Karachentsev}, \&
  {Tully}}]{chiboucas2009}
{Chiboucas}, K., {Karachentsev}, I.~D., \& {Tully}, R.~B. 2009, \aj, 137, 3009

\bibitem[{{Choi} {et~al.}(2016){Choi}, {Dotter}, {Conroy}, {Cantiello},
  {Paxton}, \& {Johnson}}]{mist_models}
{Choi}, J., {Dotter}, A., {Conroy}, C., {et~al.} 2016, \apj, 823, 102

\bibitem[{{Cohen} {et~al.}(2018){Cohen}, {van Dokkum}, {Danieli}, {Romanowsky},
  {Abraham}, {Merritt}, {Zhang}, {Mowla}, {Kruijssen}, {Conroy}, \&
  {Wasserman}}]{cohen2018}
{Cohen}, Y., {van Dokkum}, P., {Danieli}, S., {et~al.} 2018, \apj, 868, 96

\bibitem[{{Cook} {et~al.}(2014{\natexlab{a}}){Cook}, {Dale}, {Johnson}, {Van
  Zee}, {Lee}, {Kennicutt}, {Calzetti}, {Staudaher}, \&
  {Engelbracht}}]{cook2014a}
{Cook}, D.~O., {Dale}, D.~A., {Johnson}, B.~D., {et~al.} 2014{\natexlab{a}},
  \mnras, 445, 881

\bibitem[{{Cook} {et~al.}(2014{\natexlab{b}}){Cook}, {Dale}, {Johnson}, {Van
  Zee}, {Lee}, {Kennicutt}, {Calzetti}, {Staudaher}, \&
  {Engelbracht}}]{cook2014b}
---. 2014{\natexlab{b}}, \mnras, 445, 899

\bibitem[{{Correa} {et~al.}(2018){Correa}, {Schaye}, {Wyithe}, {Duffy},
  {Theuns}, {Crain}, \& {Bower}}]{correa2018}
{Correa}, C.~A., {Schaye}, J., {Wyithe}, J. S.~B., {et~al.} 2018, \mnras, 473,
  538

\bibitem[{{Crnojevi{\'c}} {et~al.}(2014){Crnojevi{\'c}}, {Sand}, {Caldwell},
  {Guhathakurta}, {McLeod}, {Seth}, {Simon}, {Strader}, \&
  {Toloba}}]{crnojevic2014}
{Crnojevi{\'c}}, D., {Sand}, D.~J., {Caldwell}, N., {et~al.} 2014, \apj, 795,
  L35

\bibitem[{{Crnojevi{\'c}} {et~al.}(2016){Crnojevi{\'c}}, {Sand}, {Spekkens},
  {Caldwell}, {Guhathakurta}, {McLeod}, {Seth}, {Simon}, {Strader}, \&
  {Toloba}}]{crnojevic2016}
{Crnojevi{\'c}}, D., {Sand}, D.~J., {Spekkens}, K., {et~al.} 2016, \apj, 823,
  19

\bibitem[{{Crnojevi{\'c}} {et~al.}(2019){Crnojevi{\'c}}, {Sand}, {Bennet},
  {Pasetto}, {Spekkens}, {Caldwell}, {Guhathakurta}, {McLeod}, {Seth}, \&
  {Simon}}]{crnojevic2019}
{Crnojevi{\'c}}, D., {Sand}, D.~J., {Bennet}, P., {et~al.} 2019, \apj, 872, 80

\bibitem[{{Dalcanton} {et~al.}(1997){Dalcanton}, {Spergel}, {Gunn}, {Schmidt},
  \& {Schneider}}]{dalcanton1997}
{Dalcanton}, J.~J., {Spergel}, D.~N., {Gunn}, J.~E., {Schmidt}, M., \&
  {Schneider}, D.~P. 1997, \aj, 114, 635

\bibitem[{{Dalcanton} {et~al.}(2009){Dalcanton}, {Williams}, {Seth}, {Dolphin},
  {Holtzman}, {Rosema}, {Skillman}, {Cole}, {Girardi}, {Gogarten},
  {Karachentsev}, {Olsen}, {Weisz}, {Christensen}, {Freeman}, {Gilbert},
  {Gallart}, {Harris}, {Hodge}, {de Jong}, {Karachentseva}, {Mateo}, {Stetson},
  {Tavarez}, {Zaritsky}, {Governato}, \& {Quinn}}]{dalcanton2009}
{Dalcanton}, J.~J., {Williams}, B.~F., {Seth}, A.~C., {et~al.} 2009, \apjs,
  183, 67

\bibitem[{{Danieli} \& {van Dokkum}(2019)}]{danieli2019}
{Danieli}, S., \& {van Dokkum}, P. 2019, \apj, 875, 155

\bibitem[{{Danieli} {et~al.}(2018){Danieli}, {van Dokkum}, \&
  {Conroy}}]{danieli_field}
{Danieli}, S., {van Dokkum}, P., \& {Conroy}, C. 2018, \apj, 856, 69

\bibitem[{{Danieli} {et~al.}(2017){Danieli}, {van Dokkum}, {Merritt},
  {Abraham}, {Zhang}, {Karachentsev}, \& {Makarova}}]{danieli2017}
{Danieli}, S., {van Dokkum}, P., {Merritt}, A., {et~al.} 2017, \apj, 837, 136

\bibitem[{{Danieli} {et~al.}(2020){Danieli}, {Lokhorst}, {Zhang}, {Merritt},
  {van Dokkum}, {Abraham}, {Conroy}, {Gilhuly}, {Greco}, {Janssens}, {Li},
  {Liu}, {Miller}, \& {Mowla}}]{danieli2020}
{Danieli}, S., {Lokhorst}, D., {Zhang}, J., {et~al.} 2020, \apj, 894, 119

\bibitem[{{Davis} {et~al.}(2021){Davis}, {Nierenberg}, {Peter}, {Garling},
  {Greco}, {Kochanek}, {Utomo}, {Casey}, {Pogge}, {Roberts}, {Sand}, \&
  {Sardone}}]{davis2021}
{Davis}, A.~B., {Nierenberg}, A.~M., {Peter}, A. H.~G., {et~al.} 2021, \mnras,
  500, 3854

\bibitem[{{de Vaucouleurs} {et~al.}(1991){de Vaucouleurs}, {de Vaucouleurs},
  {Corwin}, {Buta}, {Paturel}, \& {Fouque}}]{rc3}
{de Vaucouleurs}, G., {de Vaucouleurs}, A., {Corwin}, Herold~G., J., {et~al.}
  1991, {Third Reference Catalogue of Bright Galaxies}

\bibitem[{{Dey} {et~al.}(2019){Dey}, {Schlegel}, {Lang}, {Blum}, {Burleigh},
  {Fan}, {Findlay}, {Finkbeiner}, {Herrera}, {Juneau}, {Landriau}, {Levi},
  {McGreer}, {Meisner}, {Myers}, {Moustakas}, {Nugent}, {Patej}, {Schlafly},
  {Walker}, {Valdes}, {Weaver}, {Y{\`e}che}, {Zou}, {Zhou}, {Abareshi},
  {Abbott}, {Abolfathi}, {Aguilera}, {Alam}, {Allen}, {Alvarez}, {Annis},
  {Ansarinejad}, {Aubert}, {Beechert}, {Bell}, {BenZvi}, {Beutler}, {Bielby},
  {Bolton}, {Brice{\~n}o}, {Buckley-Geer}, {Butler}, {Calamida}, {Carlberg},
  {Carter}, {Casas}, {Castander}, {Choi}, {Comparat}, {Cukanovaite}, {Delubac},
  {DeVries}, {Dey}, {Dhungana}, {Dickinson}, {Ding}, {Donaldson}, {Duan},
  {Duckworth}, {Eftekharzadeh}, {Eisenstein}, {Etourneau}, {Fagrelius},
  {Farihi}, {Fitzpatrick}, {Font-Ribera}, {Fulmer}, {G{\"a}nsicke},
  {Gaztanaga}, {George}, {Gerdes}, {Gontcho}, {Gorgoni}, {Green}, {Guy},
  {Harmer}, {Hernand ez}, {Honscheid}, {Huang}, {James}, {Jannuzi}, {Jiang},
  {Joyce}, {Karcher}, {Karkar}, {Kehoe}, {Kneib}, {Kueter-Young}, {Lan},
  {Lauer}, {Le Guillou}, {Le Van Suu}, {Lee}, {Lesser}, {Perreault Levasseur},
  {Li}, {Mann}, {Marshall}, {Mart{\'\i}nez-V{\'a}zquez}, {Martini}, {du Mas des
  Bourboux}, {McManus}, {Meier}, {M{\'e}nard}, {Metcalfe},
  {Mu{\~n}oz-Guti{\'e}rrez}, {Najita}, {Napier}, {Narayan}, {Newman}, {Nie},
  {Nord}, {Norman}, {Olsen}, {Paat}, {Palanque-Delabrouille}, {Peng},
  {Poppett}, {Poremba}, {Prakash}, {Rabinowitz}, {Raichoor}, {Rezaie},
  {Robertson}, {Roe}, {Ross}, {Ross}, {Rudnick}, {Safonova}, {Saha},
  {S{\'a}nchez}, {Savary}, {Schweiker}, {Scott}, {Seo}, {Shan}, {Silva},
  {Slepian}, {Soto}, {Sprayberry}, {Staten}, {Stillman}, {Stupak}, {Summers},
  {Sien Tie}, {Tirado}, {Vargas-Maga{\~n}a}, {Vivas}, {Wechsler}, {Williams},
  {Yang}, {Yang}, {Yapici}, {Zaritsky}, {Zenteno}, {Zhang}, {Zhang}, {Zhou}, \&
  {Zhou}}]{decals}
{Dey}, A., {Schlegel}, D.~J., {Lang}, D., {et~al.} 2019, \aj, 157, 168

\bibitem[{{Drlica-Wagner} {et~al.}(2020){Drlica-Wagner}, {Bechtol}, {Mau},
  {McNanna}, {Nadler}, {Pace}, {Li}, {Pieres}, {Rozo}, {Simon}, {Walker},
  {Wechsler}, {Abbott}, {Allam}, {Annis}, {Bertin}, {Brooks}, {Burke},
  {Rosell}, {Carrasco Kind}, {Carretero}, {Costanzi}, {da Costa}, {De Vicente},
  {Desai}, {Diehl}, {Doel}, {Eifler}, {Everett}, {Flaugher}, {Frieman},
  {Garc{\'\i}a-Bellido}, {Gaztanaga}, {Gruen}, {Gruendl}, {Gschwend},
  {Gutierrez}, {Honscheid}, {James}, {Krause}, {Kuehn}, {Kuropatkin}, {Lahav},
  {Maia}, {Marshall}, {Melchior}, {Menanteau}, {Miquel}, {Palmese}, {Plazas},
  {Sanchez}, {Scarpine}, {Schubnell}, {Serrano}, {Sevilla-Noarbe}, {Smith},
  {Suchyta}, {Tarle}, \& {DES Collaboration}}]{drlica2020}
{Drlica-Wagner}, A., {Bechtol}, K., {Mau}, S., {et~al.} 2020, \apj, 893, 47

\bibitem[{{Drlica-Wagner} {et~al.}(2021){Drlica-Wagner}, {Carlin}, {Nidever},
  {Ferguson}, {Kuropatkin}, {Adam{\'o}w}, {Cerny}, {Choi}, {Esteves},
  {Mart{\'\i}nez-V{\'a}zquez}, {Mau}, {Miller}, {Mutlu-Pakdil}, {Neilsen},
  {Olsen}, {Pace}, {Riley}, {Sakowska}, {Sand}, {Santana-Silva}, {Tollerud},
  {Tucker}, {Vivas}, {Zaborowski}, {Zenteno}, {Abbott}, {Allam}, {Bechtol},
  {Bell}, {Bell}, {Bilaji}, {Bom}, {Carballo-Bello}, {Crnojevi{\'c}}, {Cioni},
  {Diaz-Ocampo}, {de Boer}, {Erkal}, {Gruendl}, {Hernandez-Lang}, {Hughes},
  {James}, {Johnson}, {Li}, {Mao}, {Mart{\'\i}nez-Delgado}, {Massana},
  {McNanna}, {Morgan}, {Nadler}, {No{\"e}l}, {Palmese}, {Peter}, {Rykoff},
  {S{\'a}nchez}, {Shipp}, {Simon}, {Smercina}, {Soares-Santos}, {Stringfellow},
  {Tavangar}, {van der Marel}, {Walker}, {Wechsler}, {Wu}, {Yanny},
  {Fitzpatrick}, {Huang}, {Jacques}, {Nikutta}, {Scott}, \& {Astro Data
  Lab}}]{drlica2021}
{Drlica-Wagner}, A., {Carlin}, J.~L., {Nidever}, D.~L., {et~al.} 2021, \apjs,
  256, 2

\bibitem[{{Eigenthaler} {et~al.}(2018){Eigenthaler}, {Puzia}, {Taylor},
  {Ordenes-Brice{\~n}o}, {Mu{\~n}oz}, {Ribbeck}, {Alamo-Mart{\'\i}nez},
  {Zhang}, {{\'A}ngel}, {Capaccioli}, {C{\^o}t{\'e}}, {Ferrarese}, {Galaz},
  {Grebel}, {Hempel}, {Hilker}, {Lan{\c{c}}on}, {Mieske}, {Miller}, {Paolillo},
  {Powalka}, {Richtler}, {Roediger}, {Rong}, {S{\'a}nchez-Janssen}, \&
  {Spengler}}]{eigenthaler2018}
{Eigenthaler}, P., {Puzia}, T.~H., {Taylor}, M.~A., {et~al.} 2018, \apj, 855,
  142

\bibitem[{{Engler} {et~al.}(2021){Engler}, {Pillepich}, {Pasquali}, {Nelson},
  {Rodriguez-Gomez}, {Chua}, {Grebel}, {Springel}, {Marinacci}, {Weinberger},
  {Vogelsberger}, \& {Hernquist}}]{engler2021}
{Engler}, C., {Pillepich}, A., {Pasquali}, A., {et~al.} 2021, arXiv e-prints,
  arXiv:2101.12215

\bibitem[{{Erwin}(2015)}]{imfit}
{Erwin}, P. 2015, \apj, 799, 226

\bibitem[{{Ferrarese} {et~al.}(2012){Ferrarese}, {C{\^o}t{\'e}}, {Cuilland re},
  {Gwyn}, {Peng}, {MacArthur}, {Duc}, {Boselli}, {Mei}, {Erben}, {McConnachie},
  {Durrell}, {Mihos}, {Jord{\'a}n}, {Lan{\c{c}}on}, {Puzia}, {Emsellem},
  {Balogh}, {Blakeslee}, {van Waerbeke}, {Gavazzi}, {Vollmer}, {Kavelaars},
  {Woods}, {Ball}, {Boissier}, {Courteau}, {Ferriere}, {Gavazzi},
  {Hildebrandt}, {Hudelot}, {Huertas-Company}, {Liu}, {McLaughlin}, {Mellier},
  {Milkeraitis}, {Schade}, {Balkowski}, {Bournaud}, {Carlberg}, {Chapman},
  {Hoekstra}, {Peng}, {Sawicki}, {Simard}, {Taylor}, {Tully}, {van Driel},
  {Wilson}, {Burdullis}, {Mahoney}, \& {Manset}}]{ferrarese2012}
{Ferrarese}, L., {C{\^o}t{\'e}}, P., {Cuilland re}, J.-C., {et~al.} 2012,
  \apjs, 200, 4

\bibitem[{{Ferrarese} {et~al.}(2016){Ferrarese}, {C{\^o}t{\'e}},
  {S{\'a}nchez-Janssen}, {Roediger}, {McConnachie}, {Durrell}, {MacArthur},
  {Blakeslee}, {Duc}, {Boissier}, {Boselli}, {Courteau}, {Cuillandre},
  {Emsellem}, {Gwyn}, {Guhathakurta}, {Jord{\'a}n}, {Lan{\c{c}}on}, {Liu},
  {Mei}, {Mihos}, {Navarro}, {Peng}, {Puzia}, {Taylor}, {Toloba}, \&
  {Zhang}}]{ferrarese2016}
{Ferrarese}, L., {C{\^o}t{\'e}}, P., {S{\'a}nchez-Janssen}, R., {et~al.} 2016,
  \apj, 824, 10

\bibitem[{{Ferrarese} {et~al.}(2020){Ferrarese}, {C{\^o}t{\'e}}, {MacArthur},
  {Durrell}, {Gwyn}, {Duc}, {S{\'a}nchez-Janssen}, {Santos}, {Blakeslee},
  {Boselli}, {Boyer}, {Cantiello}, {Courteau}, {Cuillandre}, {Emsellem},
  {Erben}, {Gavazzi}, {Guhathakurta}, {Huertas-Company}, {Jord{\'a}n},
  {Lan{\c{c}}on}, {Liu}, {Mei}, {Mihos}, {Peng}, {Puzia}, {Roediger}, {Schade},
  {Taylor}, {Toloba}, \& {Zhang}}]{ferrarese2020}
{Ferrarese}, L., {C{\^o}t{\'e}}, P., {MacArthur}, L.~A., {et~al.} 2020, \apj,
  890, 128

\bibitem[{{Fillingham} {et~al.}(2015){Fillingham}, {Cooper}, {Wheeler},
  {Garrison-Kimmel}, {Boylan-Kolchin}, \& {Bullock}}]{fillingham2015}
{Fillingham}, S.~P., {Cooper}, M.~C., {Wheeler}, C., {et~al.} 2015, \mnras,
  454, 2039

\bibitem[{{Flewelling} {et~al.}(2016){Flewelling}, {Magnier}, {Chambers},
  {Heasley}, {Holmberg}, {Huber}, {Sweeney}, {Waters}, {Calamida}, {Casertano},
  {Chen}, {Farrow}, {Hasinger}, {Henderson}, {Long}, {Metcalfe}, {Narayan},
  {Nieto-Santisteban}, {Norberg}, {Rest}, {Saglia}, {Szalay}, {Thakar},
  {Tonry}, {Valenti}, {Werner}, {White}, {Denneau}, {Draper}, {Hodapp},
  {Jedicke}, {Kaiser}, {Kudritzki}, {Price}, {Wainscoat}, {Builders},
  {Chastel}, {McLean}, {Postman}, \& {Shiao}}]{panstarrs2}
{Flewelling}, H.~A., {Magnier}, E.~A., {Chambers}, K.~C., {et~al.} 2016, arXiv
  e-prints, arXiv:1612.05243

\bibitem[{{Font} {et~al.}(2021{\natexlab{a}}){Font}, {McCarthy}, \&
  {Belokurov}}]{font2021}
{Font}, A.~S., {McCarthy}, I.~G., \& {Belokurov}, V. 2021{\natexlab{a}},
  \mnras, 505, 783

\bibitem[{{Font} {et~al.}(2021{\natexlab{b}}){Font}, {McCarthy}, {Belokurov},
  {Brown}, \& {Stafford}}]{font2021b}
{Font}, A.~S., {McCarthy}, I.~G., {Belokurov}, V., {Brown}, S.~T., \&
  {Stafford}, S.~G. 2021{\natexlab{b}}, arXiv e-prints, arXiv:2109.06215

\bibitem[{{Font} {et~al.}(2020){Font}, {McCarthy}, {Poole-Mckenzie},
  {Stafford}, {Brown}, {Schaye}, {Crain}, {Theuns}, \& {Schaller}}]{font2020}
{Font}, A.~S., {McCarthy}, I.~G., {Poole-Mckenzie}, R., {et~al.} 2020, \mnras,
  498, 1765

\bibitem[{{Gaia Collaboration} {et~al.}(2018){Gaia Collaboration}, {Brown},
  {Vallenari}, {Prusti}, {de Bruijne}, {Babusiaux}, {Bailer-Jones}, {Biermann},
  {Evans}, {Eyer}, \& et~al.}]{gaia_dr2}
{Gaia Collaboration}, {Brown}, A.~G.~A., {Vallenari}, A., {et~al.} 2018, \aap,
  616, A1

\bibitem[{{Garling} {et~al.}(2020){Garling}, {Peter}, {Kochanek}, {Sand}, \&
  {Crnojevi{\'c}}}]{garling2020}
{Garling}, C.~T., {Peter}, A. H.~G., {Kochanek}, C.~S., {Sand}, D.~J., \&
  {Crnojevi{\'c}}, D. 2020, \mnras, 492, 1713

\bibitem[{{Garling} {et~al.}(2021){Garling}, {Peter}, {Kochanek}, {Sand}, \&
  {Crnojevi{\'c}}}]{garling2021}
---. 2021, arXiv e-prints, arXiv:2105.01082

\bibitem[{{Garrison-Kimmel} {et~al.}(2014){Garrison-Kimmel}, {Boylan-Kolchin},
  {Bullock}, \& {Lee}}]{elvis}
{Garrison-Kimmel}, S., {Boylan-Kolchin}, M., {Bullock}, J.~S., \& {Lee}, K.
  2014, \mnras, 438, 2578

\bibitem[{{Garrison-Kimmel} {et~al.}(2017){Garrison-Kimmel}, {Bullock},
  {Boylan-Kolchin}, \& {Bardwell}}]{gk_2017}
{Garrison-Kimmel}, S., {Bullock}, J.~S., {Boylan-Kolchin}, M., \& {Bardwell},
  E. 2017, \mnras, 464, 3108

\bibitem[{{Garrison-Kimmel} {et~al.}(2019{\natexlab{a}}){Garrison-Kimmel},
  {Hopkins}, {Wetzel}, {Bullock}, {Boylan-Kolchin}, {Kere{\v{s}}},
  {Faucher-Gigu{\`e}re}, {El-Badry}, {Lamberts}, {Quataert}, \& {Sand
  erson}}]{gk2019}
{Garrison-Kimmel}, S., {Hopkins}, P.~F., {Wetzel}, A., {et~al.}
  2019{\natexlab{a}}, \mnras, 487, 1380

\bibitem[{{Garrison-Kimmel} {et~al.}(2019{\natexlab{b}}){Garrison-Kimmel},
  {Wetzel}, {Hopkins}, {Sanderson}, {El-Badry}, {Graus}, {Chan}, {Feldmann},
  {Boylan-Kolchin}, {Hayward}, {Bullock}, {Fitts}, {Samuel}, {Wheeler},
  {Kere{\v{s}}}, \& {Faucher-Gigu{\`e}re}}]{gk2019b}
{Garrison-Kimmel}, S., {Wetzel}, A., {Hopkins}, P.~F., {et~al.}
  2019{\natexlab{b}}, \mnras, 489, 4574

\bibitem[{{Gatto} {et~al.}(2013){Gatto}, {Fraternali}, {Read}, {Marinacci},
  {Lux}, \& {Walch}}]{gatto2013}
{Gatto}, A., {Fraternali}, F., {Read}, J.~I., {et~al.} 2013, \mnras, 433, 2749

\bibitem[{{Geha} {et~al.}(2017){Geha}, {Wechsler}, {Mao}, {Tollerud}, {Weiner},
  {Bernstein}, {Hoyle}, {Marchi}, {Marshall}, {Mu{\~n}oz}, \& {Lu}}]{geha2017}
{Geha}, M., {Wechsler}, R.~H., {Mao}, Y.-Y., {et~al.} 2017, \apj, 847, 4

\bibitem[{{Gil de Paz} {et~al.}(2007){Gil de Paz}, {Boissier}, {Madore},
  {Seibert}, {Joe}, {Boselli}, {Wyder}, {Thilker}, {Bianchi}, {Rey}, {Rich},
  {Barlow}, {Conrow}, {Forster}, {Friedman}, {Martin}, {Morrissey}, {Neff},
  {Schiminovich}, {Small}, {Donas}, {Heckman}, {Lee}, {Milliard}, {Szalay}, \&
  {Yi}}]{galex2007}
{Gil de Paz}, A., {Boissier}, S., {Madore}, B.~F., {et~al.} 2007, \apjs, 173,
  185

\bibitem[{{Grcevich} \& {Putman}(2009)}]{grcevich2009}
{Grcevich}, J., \& {Putman}, M.~E. 2009, \apj, 696, 385

\bibitem[{{Grebel} {et~al.}(2003){Grebel}, {Gallagher}, \&
  {Harbeck}}]{grebel2003}
{Grebel}, E.~K., {Gallagher}, John~S., I., \& {Harbeck}, D. 2003, \aj, 125,
  1926

\bibitem[{{Greco} {et~al.}(2018{\natexlab{a}}){Greco}, {Goulding}, {Greene},
  {Strauss}, {Huang}, {Kim}, \& {Komiyama}}]{greco2018_two}
{Greco}, J.~P., {Goulding}, A.~D., {Greene}, J.~E., {et~al.}
  2018{\natexlab{a}}, \apj, 866, 112

\bibitem[{{Greco} {et~al.}(2020){Greco}, {van Dokkum}, {Danieli}, {Carlsten},
  \& {Conroy}}]{greco2020}
{Greco}, J.~P., {van Dokkum}, P., {Danieli}, S., {Carlsten}, S.~G., \&
  {Conroy}, C. 2020, arXiv e-prints, arXiv:2004.07273

\bibitem[{{Greco} {et~al.}(2018{\natexlab{b}}){Greco}, {Greene}, {Strauss},
  {Macarthur}, {Flowers}, {Goulding}, {Huang}, {Kim}, {Komiyama}, {Leauthaud},
  {Leisman}, {Lupton}, {Sif{\'o}n}, \& {Wang}}]{greco2018}
{Greco}, J.~P., {Greene}, J.~E., {Strauss}, M.~A., {et~al.} 2018{\natexlab{b}},
  \apj, 857, 104

\bibitem[{{Gunn} \& {Gott}(1972)}]{gunn1972}
{Gunn}, J.~E., \& {Gott}, J.~Richard, I. 1972, \apj, 176, 1

\bibitem[{{Habas} {et~al.}(2020){Habas}, {Marleau}, {Duc}, {Durrell}, {Paudel},
  {Poulain}, {S{\'a}nchez-Janssen}, {Sreejith}, {Ramasawmy}, {Stemock},
  {Leach}, {Cuillandre}, {Gwyn}, {Agnello}, {B{\'\i}lek}, {Fensch},
  {M{\"u}ller}, {Peng}, \& {van der Burg}}]{habas2020}
{Habas}, R., {Marleau}, F.~R., {Duc}, P.-A., {et~al.} 2020, \mnras, 491, 1901

\bibitem[{{Haigh} {et~al.}(2021){Haigh}, {Chamba}, {Venhola}, {Peletier},
  {Doorenbos}, {Watkins}, \& {Wilkinson}}]{haigh2021}
{Haigh}, C., {Chamba}, N., {Venhola}, A., {et~al.} 2021, \aap, 645, A107

\bibitem[{{Hausammann} {et~al.}(2019){Hausammann}, {Revaz}, \&
  {Jablonka}}]{hausammann2019}
{Hausammann}, L., {Revaz}, Y., \& {Jablonka}, P. 2019, \aap, 624, A11

\bibitem[{{Haynes} {et~al.}(2018){Haynes}, {Giovanelli}, {Kent}, {Adams},
  {Balonek}, {Craig}, {Fertig}, {Finn}, {Giovanardi}, {Hallenbeck}, {Hess},
  {Hoffman}, {Huang}, {Jones}, {Koopmann}, {Kornreich}, {Leisman}, {Miller},
  {Moorman}, {O'Connor}, {O'Donoghue}, {Papastergis}, {Troischt}, {Stark}, \&
  {Xiao}}]{haynes2018}
{Haynes}, M.~P., {Giovanelli}, R., {Kent}, B.~R., {et~al.} 2018, \apj, 861, 49

\bibitem[{{Henden} \& {Munari}(2014)}]{apass}
{Henden}, A., \& {Munari}, U. 2014, Contributions of the Astronomical
  Observatory Skalnate Pleso, 43, 518

\bibitem[{{Hern{\'a}ndez-Toledo} {et~al.}(2011){Hern{\'a}ndez-Toledo},
  {M{\'e}ndez-Hern{\'a}ndez}, {Aceves}, \& {Olgu{\'\i}n}}]{ht11}
{Hern{\'a}ndez-Toledo}, H.~M., {M{\'e}ndez-Hern{\'a}ndez}, H., {Aceves}, H., \&
  {Olgu{\'\i}n}, L. 2011, \aj, 141, 74

\bibitem[{{Huchra} {et~al.}(2012){Huchra}, {Macri}, {Masters}, {Jarrett},
  {Berlind}, {Calkins}, {Crook}, {Cutri}, {Erdo{\v{g}}du}, {Falco}, {George},
  {Hutcheson}, {Lahav}, {Mader}, {Mink}, {Martimbeau}, {Schneider},
  {Skrutskie}, {Tokarz}, \& {Westover}}]{huchra2012}
{Huchra}, J.~P., {Macri}, L.~M., {Masters}, K.~L., {et~al.} 2012, \apjs, 199,
  26

\bibitem[{{Humphreys} {et~al.}(2013){Humphreys}, {Reid}, {Moran}, {Greenhill},
  \& {Argon}}]{humphreys2013}
{Humphreys}, E.~M.~L., {Reid}, M.~J., {Moran}, J.~M., {Greenhill}, L.~J., \&
  {Argon}, A.~L. 2013, \apj, 775, 13

\bibitem[{{Ibata} {et~al.}(2013){Ibata}, {Lewis}, {Conn}, {Irwin},
  {McConnachie}, {Chapman}, {Collins}, {Fardal}, {Ferguson}, {Ibata}, {Mackey},
  {Martin}, {Navarro}, {Rich}, {Valls-Gabaud}, \& {Widrow}}]{ibataGPOA}
{Ibata}, R.~A., {Lewis}, G.~F., {Conn}, A.~R., {et~al.} 2013, \nat, 493, 62

\bibitem[{{Into} \& {Portinari}(2013)}]{into2013}
{Into}, T., \& {Portinari}, L. 2013, \mnras, 430, 2715

\bibitem[{{Irwin} {et~al.}(2009){Irwin}, {Hoffman}, {Spekkens}, {Haynes},
  {Giovanelli}, {Linder}, {Catinella}, {Momjian}, {Koribalski}, \&
  {Davies}}]{irwin2009}
{Irwin}, J.~A., {Hoffman}, G.~L., {Spekkens}, K., {et~al.} 2009, \apj, 692,
  1447

\bibitem[{{Jacobs} {et~al.}(2009){Jacobs}, {Rizzi}, {Tully}, {Shaya},
  {Makarov}, \& {Makarova}}]{jacobs2009}
{Jacobs}, B.~A., {Rizzi}, L., {Tully}, R.~B., {et~al.} 2009, \aj, 138, 332

\bibitem[{{Javanmardi} {et~al.}(2016){Javanmardi}, {Martinez-Delgado},
  {Kroupa}, {Henkel}, {Crawford}, {Teuwen}, {Gabany}, {Hanson}, {Chonis}, \&
  {Neyer}}]{javanmardi_m101}
{Javanmardi}, B., {Martinez-Delgado}, D., {Kroupa}, P., {et~al.} 2016, \aap,
  588, A89

\bibitem[{{Jennings} {et~al.}(2015){Jennings}, {Romanowsky}, {Brodie}, {Janz},
  {Norris}, {Forbes}, {Martinez-Delgado}, {Fagioli}, \& {Penny}}]{jennings2015}
{Jennings}, Z.~G., {Romanowsky}, A.~J., {Brodie}, J.~P., {et~al.} 2015, \apjl,
  812, L10

\bibitem[{{Jensen} {et~al.}(2003){Jensen}, {Tonry}, {Barris}, {Thompson},
  {Liu}, {Rieke}, {Ajhar}, \& {Blakeslee}}]{jensen2003}
{Jensen}, J.~B., {Tonry}, J.~L., {Barris}, B.~J., {et~al.} 2003, \apj, 583, 712

\bibitem[{{Jerjen} {et~al.}(2001){Jerjen}, {Rekola}, {Takalo}, {Coleman}, \&
  {Valtonen}}]{jerjen_field}
{Jerjen}, H., {Rekola}, R., {Takalo}, L., {Coleman}, M., \& {Valtonen}, M.
  2001, \aap, 380, 90

\bibitem[{{Jiang} {et~al.}(2020){Jiang}, {Dekel}, {Freundlich}, {van den
  Bosch}, {Green}, {Hopkins}, {Benson}, \& {Du}}]{jiang2020}
{Jiang}, F., {Dekel}, A., {Freundlich}, J., {et~al.} 2020, arXiv e-prints,
  arXiv:2005.05974

\bibitem[{{Kado-Fong} {et~al.}(2020){Kado-Fong}, {Greene}, {Huang}, {Beaton},
  {Goulding}, \& {Komiyama}}]{kadofong2020}
{Kado-Fong}, E., {Greene}, J.~E., {Huang}, S., {et~al.} 2020, \apj, 900, 163

\bibitem[{{Karachentsev} \& {Kaisina}(2013)}]{karachentsev2013_sfr}
{Karachentsev}, I.~D., \& {Kaisina}, E.~I. 2013, \aj, 146, 46

\bibitem[{{Karachentsev} {et~al.}(2004){Karachentsev}, {Karachentseva},
  {Huchtmeier}, \& {Makarov}}]{karachentsev2004}
{Karachentsev}, I.~D., {Karachentseva}, V.~E., {Huchtmeier}, W.~K., \&
  {Makarov}, D.~I. 2004, \aj, 127, 2031

\bibitem[{{Karachentsev} \& {Kudrya}(2014)}]{karachentsev2014_masses}
{Karachentsev}, I.~D., \& {Kudrya}, Y.~N. 2014, \aj, 148, 50

\bibitem[{{Karachentsev} {et~al.}(2013){Karachentsev}, {Makarov}, \&
  {Kaisina}}]{karachentsev2013}
{Karachentsev}, I.~D., {Makarov}, D.~I., \& {Kaisina}, E.~I. 2013, \aj, 145,
  101

\bibitem[{{Karachentsev} {et~al.}(2020){Karachentsev}, {Makarova}, {Brent
  Tully}, {Anand}, {Rizzi}, \& {Shaya}}]{karachentsev2020}
{Karachentsev}, I.~D., {Makarova}, L.~N., {Brent Tully}, R., {et~al.} 2020,
  \aap, 643, A124

\bibitem[{{Karachentsev} {et~al.}(2015{\natexlab{a}}){Karachentsev}, {Tully},
  {Makarova}, {Makarov}, \& {Rizzi}}]{karachentsev2015b}
{Karachentsev}, I.~D., {Tully}, R.~B., {Makarova}, L.~N., {Makarov}, D.~I., \&
  {Rizzi}, L. 2015{\natexlab{a}}, \apj, 805, 144

\bibitem[{{Karachentsev} {et~al.}(2014){Karachentsev}, {Tully}, {Wu}, {Shaya},
  \& {Dolphin}}]{karachentsev2014}
{Karachentsev}, I.~D., {Tully}, R.~B., {Wu}, P.-F., {Shaya}, E.~J., \&
  {Dolphin}, A.~E. 2014, \apj, 782, 4

\bibitem[{{Karachentsev} {et~al.}(2006){Karachentsev}, {Dolphin}, {Tully},
  {Sharina}, {Makarova}, {Makarov}, {Karachentseva}, {Sakai}, \&
  {Shaya}}]{karachentsev2006}
{Karachentsev}, I.~D., {Dolphin}, A., {Tully}, R.~B., {et~al.} 2006, \aj, 131,
  1361

\bibitem[{{Karachentsev} {et~al.}(2007){Karachentsev}, {Tully}, {Dolphin},
  {Sharina}, {Makarova}, {Makarov}, {Sakai}, {Shaya}, {Kashibadze},
  {Karachentseva}, \& {Rizzi}}]{karachentsev2007}
{Karachentsev}, I.~D., {Tully}, R.~B., {Dolphin}, A., {et~al.} 2007, \aj, 133,
  504

\bibitem[{{Karachentsev} {et~al.}(2015{\natexlab{b}}){Karachentsev}, {Riepe},
  {Zilch}, {Blauensteiner}, {Elvov}, {Hochleitner}, {Hubl}, {Kerschhuber},
  {K{\"u}ppers}, \& {Neyer}}]{karachentsev2015}
{Karachentsev}, I.~D., {Riepe}, P., {Zilch}, T., {et~al.} 2015{\natexlab{b}},
  Astrophysical Bulletin, 70, 379

\bibitem[{{Karachentseva}(1968)}]{karachentseva1968}
{Karachentseva}, V.~E. 1968, Communications of the Byurakan Astrophysical
  Observatory, 39, 61

\bibitem[{{Karunakaran} {et~al.}(2020{\natexlab{a}}){Karunakaran}, {Spekkens},
  {Bennet}, {Sand}, {Crnojevi{\'c}}, \& {Zaritsky}}]{karunakaran2020}
{Karunakaran}, A., {Spekkens}, K., {Bennet}, P., {et~al.} 2020{\natexlab{a}},
  \aj, 159, 37

\bibitem[{{Karunakaran} {et~al.}(2020{\natexlab{b}}){Karunakaran}, {Spekkens},
  {Zaritsky}, {Donnerstein}, {Kadowaki}, \& {Dey}}]{karunakaran_smudges}
{Karunakaran}, A., {Spekkens}, K., {Zaritsky}, D., {et~al.} 2020{\natexlab{b}},
  arXiv e-prints, arXiv:2005.14202

\bibitem[{{Karunakaran} {et~al.}(2021){Karunakaran}, {Spekkens}, {Oman},
  {Simpson}, {Fattahi}, {Sand}, {Bennet}, {Crnojevi{\'c}}, {Frenk},
  {G{\'o}mez}, {Grand}, {Jones}, {Marinacci}, {Mutlu-Pakdil}, {Navarro}, \&
  {Zaritsky}}]{karunakaran2021}
{Karunakaran}, A., {Spekkens}, K., {Oman}, K.~A., {et~al.} 2021, arXiv
  e-prints, arXiv:2105.09321

\bibitem[{{Kazantzidis} {et~al.}(2011){Kazantzidis}, {{\L}okas}, {Callegari},
  {Mayer}, \& {Moustakas}}]{kazantzidis2011}
{Kazantzidis}, S., {{\L}okas}, E.~L., {Callegari}, S., {Mayer}, L., \&
  {Moustakas}, L.~A. 2011, \apj, 726, 98

\bibitem[{{Kazantzidis} {et~al.}(2013){Kazantzidis}, {{\L}okas}, \&
  {Mayer}}]{kazantzidis2013}
{Kazantzidis}, S., {{\L}okas}, E.~L., \& {Mayer}, L. 2013, \apjl, 764, L29

\bibitem[{{Kennicutt} {et~al.}(2008){Kennicutt}, {Lee}, {Funes}, {J.}, {Sakai},
  \& {Akiyama}}]{kennicutt2008}
{Kennicutt}, Robert~C., J., {Lee}, J.~C., {Funes}, J.~G., {et~al.} 2008, \apjs,
  178, 247

\bibitem[{{Kere{\v{s}}} {et~al.}(2009){Kere{\v{s}}}, {Katz}, {Fardal},
  {Dav{\'e}}, \& {Weinberg}}]{keres2009}
{Kere{\v{s}}}, D., {Katz}, N., {Fardal}, M., {Dav{\'e}}, R., \& {Weinberg},
  D.~H. 2009, \mnras, 395, 160

\bibitem[{{Kim} {et~al.}(2011){Kim}, {Kim}, {Hwang}, {Lee}, {Chun}, \&
  {Ann}}]{kim2011}
{Kim}, E., {Kim}, M., {Hwang}, N., {et~al.} 2011, \mnras, 412, 1881

\bibitem[{{Kim} {et~al.}(2018){Kim}, {Peter}, \& {Hargis}}]{kim2018}
{Kim}, S.~Y., {Peter}, A. H.~G., \& {Hargis}, J.~R. 2018, \prl, 121, 211302

\bibitem[{{Kim} \& {Lee}(2021)}]{kim2021}
{Kim}, Y.~J., \& {Lee}, M.~G. 2021, arXiv e-prints, arXiv:2110.02522

\bibitem[{{Klypin} {et~al.}(1999){Klypin}, {Kravtsov}, {Valenzuela}, \&
  {Prada}}]{klypin1999}
{Klypin}, A., {Kravtsov}, A.~V., {Valenzuela}, O., \& {Prada}, F. 1999, \apj,
  522, 82

\bibitem[{{Kondapally} {et~al.}(2018){Kondapally}, {Russell}, {Conselice}, \&
  {Penny}}]{kondapally2018}
{Kondapally}, R., {Russell}, G.~A., {Conselice}, C.~J., \& {Penny}, S.~J. 2018,
  \mnras, 481, 1759

\bibitem[{{Koposov} {et~al.}(2008){Koposov}, {Belokurov}, {Evans}, {Hewett},
  {Irwin}, {Gilmore}, {Zucker}, {Rix}, {Fellhauer}, {Bell}, \&
  {Glushkova}}]{koposov2008}
{Koposov}, S., {Belokurov}, V., {Evans}, N.~W., {et~al.} 2008, \apj, 686, 279

\bibitem[{{Kormendy} {et~al.}(2009){Kormendy}, {Fisher}, {Cornell}, \&
  {Bender}}]{kormendy2009}
{Kormendy}, J., {Fisher}, D.~B., {Cornell}, M.~E., \& {Bender}, R. 2009, \apjs,
  182, 216

\bibitem[{{Kourkchi} \& {Tully}(2017)}]{kourkchi2017}
{Kourkchi}, E., \& {Tully}, R.~B. 2017, \apj, 843, 16

\bibitem[{{La Marca} {et~al.}(2021){La Marca}, {Peletier}, {Iodice},
  {Paolillo}, {Choque Challapa}, {Venhola}, {Forbes}, {Cantiello}, {Hilker},
  {Rejkuba}, {Arnaboldi}, {Spavone}, {D'Ago}, {Raj}, {Ragusa}, {Mirabile},
  {Rampazzo}, {Spiniello}, {Mieske}, \& {Schipani}}]{lamarca2021}
{La Marca}, A., {Peletier}, R., {Iodice}, E., {et~al.} 2021, arXiv e-prints,
  arXiv:2112.00711

\bibitem[{{Lang} {et~al.}(2010){Lang}, {Hogg}, {Mierle}, {Blanton}, \&
  {Roweis}}]{astrometry_net}
{Lang}, D., {Hogg}, D.~W., {Mierle}, K., {Blanton}, M., \& {Roweis}, S. 2010,
  \aj, 139, 1782

\bibitem[{{Lang} {et~al.}(2016){Lang}, {Hogg}, \& {Mykytyn}}]{lang2016}
{Lang}, D., {Hogg}, D.~W., \& {Mykytyn}, D. 2016, {The Tractor: Probabilistic
  astronomical source detection and measurement}, , , ascl:1604.008

\bibitem[{{Laureijs} {et~al.}(2011){Laureijs}, {Amiaux}, {Arduini},
  {Augu{\`e}res}, {Brinchmann}, {Cole}, {Cropper}, {Dabin}, {Duvet}, {Ealet},
  {Garilli}, {Gondoin}, {Guzzo}, {Hoar}, {Hoekstra}, {Holmes}, {Kitching},
  {Maciaszek}, {Mellier}, {Pasian}, {Percival}, {Rhodes}, {Saavedra Criado},
  {Sauvage}, {Scaramella}, {Valenziano}, {Warren}, {Bender}, {Castander},
  {Cimatti}, {Le F{\`e}vre}, {Kurki-Suonio}, {Levi}, {Lilje}, {Meylan},
  {Nichol}, {Pedersen}, {Popa}, {Rebolo Lopez}, {Rix}, {Rottgering},
  {Zeilinger}, {Grupp}, {Hudelot}, {Massey}, {Meneghetti}, {Miller}, {Paltani},
  {Paulin-Henriksson}, {Pires}, {Saxton}, {Schrabback}, {Seidel}, {Walsh},
  {Aghanim}, {Amendola}, {Bartlett}, {Baccigalupi}, {Beaulieu}, {Benabed},
  {Cuby}, {Elbaz}, {Fosalba}, {Gavazzi}, {Helmi}, {Hook}, {Irwin}, {Kneib},
  {Kunz}, {Mannucci}, {Moscardini}, {Tao}, {Teyssier}, {Weller}, {Zamorani},
  {Zapatero Osorio}, {Boulade}, {Foumond}, {Di Giorgio}, {Guttridge}, {James},
  {Kemp}, {Martignac}, {Spencer}, {Walton}, {Bl{\"u}mchen}, {Bonoli},
  {Bortoletto}, {Cerna}, {Corcione}, {Fabron}, {Jahnke}, {Ligori}, {Madrid},
  {Martin}, {Morgante}, {Pamplona}, {Prieto}, {Riva}, {Toledo}, {Trifoglio},
  {Zerbi}, {Abdalla}, {Douspis}, {Grenet}, {Borgani}, {Bouwens}, {Courbin},
  {Delouis}, {Dubath}, {Fontana}, {Frailis}, {Grazian}, {Koppenh{\"o}fer},
  {Mansutti}, {Melchior}, {Mignoli}, {Mohr}, {Neissner}, {Noddle}, {Poncet},
  {Scodeggio}, {Serrano}, {Shane}, {Starck}, {Surace}, {Taylor},
  {Verdoes-Kleijn}, {Vuerli}, {Williams}, {Zacchei}, {Altieri}, {Escudero
  Sanz}, {Kohley}, {Oosterbroek}, {Astier}, {Bacon}, {Bardelli}, {Baugh},
  {Bellagamba}, {Benoist}, {Bianchi}, {Biviano}, {Branchini}, {Carbone},
  {Cardone}, {Clements}, {Colombi}, {Conselice}, {Cresci}, {Deacon}, {Dunlop},
  {Fedeli}, {Fontanot}, {Franzetti}, {Giocoli}, {Garcia-Bellido}, {Gow},
  {Heavens}, {Hewett}, {Heymans}, {Holland}, {Huang}, {Ilbert}, {Joachimi},
  {Jennins}, {Kerins}, {Kiessling}, {Kirk}, {Kotak}, {Krause}, {Lahav}, {van
  Leeuwen}, {Lesgourgues}, {Lombardi}, {Magliocchetti}, {Maguire}, {Majerotto},
  {Maoli}, {Marulli}, {Maurogordato}, {McCracken}, {McLure}, {Melchiorri},
  {Merson}, {Moresco}, {Nonino}, {Norberg}, {Peacock}, {Pello}, {Penny},
  {Pettorino}, {Di Porto}, {Pozzetti}, {Quercellini}, {Radovich}, {Rassat},
  {Roche}, {Ronayette}, {Rossetti}, {Sartoris}, {Schneider}, {Semboloni},
  {Serjeant}, {Simpson}, {Skordis}, {Smadja}, {Smartt}, {Spano}, {Spiro},
  {Sullivan}, {Tilquin}, {Trotta}, {Verde}, {Wang}, {Williger}, {Zhao},
  {Zoubian}, \& {Zucca}}]{euclid}
{Laureijs}, R., {Amiaux}, J., {Arduini}, S., {et~al.} 2011, arXiv e-prints,
  arXiv:1110.3193

\bibitem[{{Lee} {et~al.}(2009){Lee}, {Gil de Paz}, {Tremonti}, {Kennicutt},
  {Salim}, {Bothwell}, {Calzetti}, {Dalcanton}, {Dale}, {Engelbracht}, {Funes},
  {Johnson}, {Sakai}, {Skillman}, {van Zee}, {Walter}, \& {Weisz}}]{lee2009}
{Lee}, J.~C., {Gil de Paz}, A., {Tremonti}, C., {et~al.} 2009, \apj, 706, 599

\bibitem[{{Lee} {et~al.}(2011){Lee}, {Gil de Paz}, {Kennicutt}, {Bothwell},
  {Dalcanton}, {Jos{\'e} G. Funes S.}, {Johnson}, {Sakai}, {Skillman},
  {Tremonti}, \& {van Zee}}]{lee2011}
{Lee}, J.~C., {Gil de Paz}, A., {Kennicutt}, Robert~C., J., {et~al.} 2011,
  \apjs, 192, 6

\bibitem[{{Lee} \& {Jang}(2013)}]{lee2013}
{Lee}, M.~G., \& {Jang}, I.~S. 2013, \apj, 773, 13

\bibitem[{{Leroy} {et~al.}(2019){Leroy}, {Sandstrom}, {Lang}, {Lewis}, {Salim},
  {Behrens}, {Chastenet}, {Chiang}, {Gallagher}, {Kessler}, \&
  {Utomo}}]{leroy2019}
{Leroy}, A.~K., {Sandstrom}, K.~M., {Lang}, D., {et~al.} 2019, \apjs, 244, 24

\bibitem[{{Libeskind} {et~al.}(2020){Libeskind}, {Carlesi}, {Grand},
  {Khalatyan}, {Knebe}, {Pakmor}, {Pilipenko}, {Pawlowski}, {Sparre}, {Tempel},
  {Wang}, {Courtois}, {Gottl{\"o}ber}, {Hoffman}, {Minchev}, {Pfrommer},
  {Sorce}, {Springel}, {Steinmetz}, {Tully}, {Vogelsberger}, \&
  {Yepes}}]{libeskind2020}
{Libeskind}, N.~I., {Carlesi}, E., {Grand}, R. J.~J., {et~al.} 2020, \mnras,
  498, 2968

\bibitem[{{Licquia} {et~al.}(2015){Licquia}, {Newman}, \&
  {Brinchmann}}]{licquia2015}
{Licquia}, T.~C., {Newman}, J.~A., \& {Brinchmann}, J. 2015, \apj, 809, 96

\bibitem[{{Lim} {et~al.}(2017){Lim}, {Mo}, {Lu}, {Wang}, \& {Yang}}]{lim2017}
{Lim}, S.~H., {Mo}, H.~J., {Lu}, Y., {Wang}, H., \& {Yang}, X. 2017, \mnras,
  470, 2982

\bibitem[{{Mao} {et~al.}(2021){Mao}, {Geha}, {Wechsler}, {Weiner}, {Tollerud},
  {Nadler}, \& {Kallivayalil}}]{mao2020}
{Mao}, Y.-Y., {Geha}, M., {Wechsler}, R.~H., {et~al.} 2021, \apj, 907, 85

\bibitem[{{Martin} {et~al.}(2005){Martin}, {Fanson}, {Schiminovich},
  {Morrissey}, {Friedman}, {Barlow}, {Conrow}, {Grange}, {Jelinsky},
  {Milliard}, {Siegmund}, {Bianchi}, {Byun}, {Donas}, {Forster}, {Heckman},
  {Lee}, {Madore}, {Malina}, {Neff}, {Rich}, {Small}, {Surber}, {Szalay},
  {Welsh}, \& {Wyder}}]{galex}
{Martin}, D.~C., {Fanson}, J., {Schiminovich}, D., {et~al.} 2005, \apjl, 619,
  L1

\bibitem[{{Mateo}(1998)}]{mateo1998}
{Mateo}, M.~L. 1998, \araa, 36, 435

\bibitem[{{Mayer} {et~al.}(2006){Mayer}, {Mastropietro}, {Wadsley}, {Stadel},
  \& {Moore}}]{mayer2006}
{Mayer}, L., {Mastropietro}, C., {Wadsley}, J., {Stadel}, J., \& {Moore}, B.
  2006, \mnras, 369, 1021

\bibitem[{{McConnachie}(2012)}]{mcconnachie2012}
{McConnachie}, A.~W. 2012, \aj, 144, 4

\bibitem[{{McGaugh} \& {Schombert}(2014)}]{mcgaugh2014}
{McGaugh}, S.~S., \& {Schombert}, J.~M. 2014, \aj, 148, 77

\bibitem[{{McQuinn} {et~al.}(2016{\natexlab{a}}){McQuinn}, {Skillman},
  {Dolphin}, {Berg}, \& {Kennicutt}}]{mcquinn2016a}
{McQuinn}, K. B.~W., {Skillman}, E.~D., {Dolphin}, A.~E., {Berg}, D., \&
  {Kennicutt}, R. 2016{\natexlab{a}}, \apj, 826, 21

\bibitem[{{McQuinn} {et~al.}(2016{\natexlab{b}}){McQuinn}, {Skillman},
  {Dolphin}, {Berg}, \& {Kennicutt}}]{mcquinn2016b}
---. 2016{\natexlab{b}}, \aj, 152, 144

\bibitem[{{McQuinn} {et~al.}(2017){McQuinn}, {Skillman}, {Dolphin}, {Berg}, \&
  {Kennicutt}}]{mcquinn2017}
---. 2017, \aj, 154, 51

\bibitem[{{Melchior} {et~al.}(2018){Melchior}, {Moolekamp}, {Jerdee},
  {Armstrong}, {Sun}, {Bosch}, \& {Lupton}}]{melchior2018}
{Melchior}, P., {Moolekamp}, F., {Jerdee}, M., {et~al.} 2018, Astronomy and
  Computing, 24, 129

\bibitem[{{Merritt} {et~al.}(2014){Merritt}, {van Dokkum}, \&
  {Abraham}}]{merritt2014}
{Merritt}, A., {van Dokkum}, P., \& {Abraham}, R. 2014, \apjl, 787, L37

\bibitem[{{Meyer} {et~al.}(2004){Meyer}, {Zwaan}, {Webster}, {Staveley-Smith},
  {Ryan-Weber}, {Drinkwater}, {Barnes}, {Howlett}, {Kilborn}, {Stevens},
  {Waugh}, {Pierce}, {Bhathal}, {de Blok}, {Disney}, {Ekers}, {Freeman},
  {Garcia}, {Gibson}, {Harnett}, {Henning}, {Jerjen}, {Kesteven}, {Knezek},
  {Koribalski}, {Mader}, {Marquarding}, {Minchin}, {O'Brien}, {Oosterloo},
  {Price}, {Putman}, {Ryder}, {Sadler}, {Stewart}, {Stootman}, \&
  {Wright}}]{hipass2004}
{Meyer}, M.~J., {Zwaan}, M.~A., {Webster}, R.~L., {et~al.} 2004, \mnras, 350,
  1195

\bibitem[{{Moore} {et~al.}(1999){Moore}, {Ghigna}, {Governato}, {Lake},
  {Quinn}, {Stadel}, \& {Tozzi}}]{moore1999}
{Moore}, B., {Ghigna}, S., {Governato}, F., {et~al.} 1999, \apj, 524, L19

\bibitem[{{Morrissey} {et~al.}(2007){Morrissey}, {Conrow}, {Barlow}, {Small},
  {Seibert}, {Wyder}, {Budav{\'a}ri}, {Arnouts}, {Friedman}, {Forster},
  {Martin}, {Neff}, {Schiminovich}, {Bianchi}, {Donas}, {Heckman}, {Lee},
  {Madore}, {Milliard}, {Rich}, {Szalay}, {Welsh}, \& {Yi}}]{morrissey2007}
{Morrissey}, P., {Conrow}, T., {Barlow}, T.~A., {et~al.} 2007, \apjs, 173, 682

\bibitem[{{Mu{\~n}oz} {et~al.}(2015){Mu{\~n}oz}, {Eigenthaler}, {Puzia},
  {Taylor}, {Ordenes-Brice{\~n}o}, {Alamo-Mart{\'{\i}}nez}, {Ribbeck},
  {{\'A}ngel}, {Capaccioli}, {C{\^o}t{\'e}}, {Ferrarese}, {Galaz}, {Hempel},
  {Hilker}, {Jord{\'a}n}, {Lan{\c c}on}, {Mieske}, {Paolillo}, {Richtler},
  {S{\'a}nchez-Janssen}, \& {Zhang}}]{munoz2015}
{Mu{\~n}oz}, R.~P., {Eigenthaler}, P., {Puzia}, T.~H., {et~al.} 2015, \apjl,
  813, L15

\bibitem[{{M{\"u}ller} {et~al.}(2019{\natexlab{a}}){M{\"u}ller}, {Ibata},
  {Rejkuba}, \& {Posti}}]{muller2019_n891}
{M{\"u}ller}, O., {Ibata}, R., {Rejkuba}, M., \& {Posti}, L.
  2019{\natexlab{a}}, \aap, 629, L2

\bibitem[{{M{\"u}ller} \& {Jerjen}(2020)}]{muller2020_low_mass}
{M{\"u}ller}, O., \& {Jerjen}, H. 2020, \aap, 644, A91

\bibitem[{{M{\"u}ller} {et~al.}(2015){M{\"u}ller}, {Jerjen}, \&
  {Binggeli}}]{muller2015}
{M{\"u}ller}, O., {Jerjen}, H., \& {Binggeli}, B. 2015, \aap, 583, A79

\bibitem[{{M{\"u}ller} {et~al.}(2017{\natexlab{a}}){M{\"u}ller}, {Jerjen}, \&
  {Binggeli}}]{muller2017}
---. 2017{\natexlab{a}}, \aap, 597, A7

\bibitem[{{M{\"u}ller} {et~al.}(2018{\natexlab{a}}){M{\"u}ller}, {Jerjen}, \&
  {Binggeli}}]{muller2018}
---. 2018{\natexlab{a}}, \aap, 615, A105

\bibitem[{{M{\"u}ller} {et~al.}(2018{\natexlab{b}}){M{\"u}ller}, {Pawlowski},
  {Jerjen}, \& {Lelli}}]{muller_plane}
{M{\"u}ller}, O., {Pawlowski}, M.~S., {Jerjen}, H., \& {Lelli}, F.
  2018{\natexlab{b}}, Science, 359, 534

\bibitem[{{M{\"u}ller} {et~al.}(2018{\natexlab{c}}){M{\"u}ller}, {Rejkuba}, \&
  {Jerjen}}]{muller2018_trgb}
{M{\"u}ller}, O., {Rejkuba}, M., \& {Jerjen}, H. 2018{\natexlab{c}}, \aap, 615,
  A96

\bibitem[{{M{\"u}ller} {et~al.}(2019{\natexlab{b}}){M{\"u}ller}, {Rejkuba},
  {Pawlowski}, {Ibata}, {Lelli}, {Hilker}, \& {Jerjen}}]{muller2019}
{M{\"u}ller}, O., {Rejkuba}, M., {Pawlowski}, M.~S., {et~al.}
  2019{\natexlab{b}}, \aap, 629, A18

\bibitem[{{M{\"u}ller} {et~al.}(2017{\natexlab{b}}){M{\"u}ller}, {Scalera},
  {Binggeli}, \& {Jerjen}}]{muller101}
{M{\"u}ller}, O., {Scalera}, R., {Binggeli}, B., \& {Jerjen}, H.
  2017{\natexlab{b}}, \aap, 602, A119

\bibitem[{{M{\"u}ller} {et~al.}(2021){M{\"u}ller}, {Fahrion}, {Rejkuba},
  {Hilker}, {Lelli}, {Lutz}, {Pawlowski}, {Coccato}, {Anand}, \&
  {Jerjen}}]{muller2021}
{M{\"u}ller}, O., {Fahrion}, K., {Rejkuba}, M., {et~al.} 2021, \aap, 645, A92

\bibitem[{{Munshi} {et~al.}(2021){Munshi}, {Brooks}, {Applebaum},
  {Christensen}, {Sligh}, \& {Quinn}}]{munshi2021}
{Munshi}, F., {Brooks}, A., {Applebaum}, E., {et~al.} 2021, arXiv e-prints,
  arXiv:2101.05822

\bibitem[{{Mutlu-Pakdil} {et~al.}(2021{\natexlab{a}}){Mutlu-Pakdil}, {Sand},
  {Crnojevi{\'c}}, {Jones}, {Caldwell}, {Guhathakurta}, {Seth}, {Simon},
  {Spekkens}, {Strader}, \& {Toloba}}]{mutlu2021}
{Mutlu-Pakdil}, B., {Sand}, D.~J., {Crnojevi{\'c}}, D., {et~al.}
  2021{\natexlab{a}}, arXiv e-prints, arXiv:2108.09312

\bibitem[{{Mutlu-Pakdil} {et~al.}(2021{\natexlab{b}}){Mutlu-Pakdil}, {Sand},
  {Crnojevi{\'c}}, {Drlica-Wagner}, {Caldwell}, {Guhathakurta}, {Seth},
  {Simon}, {Strader}, \& {Toloba}}]{mutlu2021b}
---. 2021{\natexlab{b}}, \apj, 918, 88

\bibitem[{{Nadler} {et~al.}(2019){Nadler}, {Mao}, {Green}, \&
  {Wechsler}}]{nadler2019}
{Nadler}, E.~O., {Mao}, Y.-Y., {Green}, G.~M., \& {Wechsler}, R.~H. 2019, \apj,
  873, 34

\bibitem[{{Nadler} {et~al.}(2020){Nadler}, {Wechsler}, {Bechtol}, {Mao},
  {Green}, {Drlica-Wagner}, {McNanna}, {Mau}, {Pace}, {Simon}, {Kravtsov},
  {Dodelson}, {Li}, {Riley}, {Wang}, {Abbott}, {Aguena}, {Allam}, {Annis},
  {Avila}, {Bernstein}, {Bertin}, {Brooks}, {Burke}, {Rosell}, {Kind},
  {Carretero}, {Costanzi}, {da Costa}, {De Vicente}, {Desai}, {Evrard},
  {Flaugher}, {Fosalba}, {Frieman}, {Garc{\'\i}a-Bellido}, {Gaztanaga},
  {Gerdes}, {Gruen}, {Gschwend}, {Gutierrez}, {Hartley}, {Hinton}, {Honscheid},
  {Krause}, {Kuehn}, {Kuropatkin}, {Lahav}, {Maia}, {Marshall}, {Menanteau},
  {Miquel}, {Palmese}, {Paz-Chinch{\'o}n}, {Plazas}, {Romer}, {Sanchez},
  {Santiago}, {Scarpine}, {Serrano}, {Smith}, {Soares-Santos}, {Suchyta},
  {Tarle}, {Thomas}, {Varga}, {Walker}, \& {DES Collaboration}}]{nadler2020}
{Nadler}, E.~O., {Wechsler}, R.~H., {Bechtol}, K., {et~al.} 2020, \apj, 893, 48

\bibitem[{{Nelson} {et~al.}(2019){Nelson}, {Springel}, {Pillepich},
  {Rodriguez-Gomez}, {Torrey}, {Genel}, {Vogelsberger}, {Pakmor}, {Marinacci},
  {Weinberger}, {Kelley}, {Lovell}, {Diemer}, \& {Hernquist}}]{tng1}
{Nelson}, D., {Springel}, V., {Pillepich}, A., {et~al.} 2019, Computational
  Astrophysics and Cosmology, 6, 2

\bibitem[{{Neumayer} {et~al.}(2020){Neumayer}, {Seth}, \&
  {B{\"o}ker}}]{neumayer2020}
{Neumayer}, N., {Seth}, A., \& {B{\"o}ker}, T. 2020, \aapr, 28, 4

\bibitem[{{Neuzil} {et~al.}(2020){Neuzil}, {Mansfield}, \&
  {Kravtsov}}]{neuzil2020}
{Neuzil}, M.~K., {Mansfield}, P., \& {Kravtsov}, A.~V. 2020, \mnras, 494, 2600

\bibitem[{{Nierenberg} {et~al.}(2016){Nierenberg}, {Treu}, {Menci}, {Lu},
  {Torrey}, \& {Vogelsberger}}]{nierenberg2016}
{Nierenberg}, A.~M., {Treu}, T., {Menci}, N., {et~al.} 2016, \mnras, 462, 4473

\bibitem[{{Park} {et~al.}(2019){Park}, {Moon}, {Zaritsky}, {Kim}, {Lee}, {Cha},
  \& {Lee}}]{park2019}
{Park}, H.~S., {Moon}, D.-S., {Zaritsky}, D., {et~al.} 2019, \apj, 885, 88

\bibitem[{{Park} {et~al.}(2017){Park}, {Moon}, {Zaritsky}, {Pak}, {Lee}, {Kim},
  {Kim}, \& {Cha}}]{park2017}
---. 2017, \apj, 848, 19

\bibitem[{{Pawlowski}(2018)}]{pawlowski2018}
{Pawlowski}, M.~S. 2018, Modern Physics Letters A, 33, 1830004

\bibitem[{{Pawlowski} \& {Kroupa}(2020)}]{pawlowski2020}
{Pawlowski}, M.~S., \& {Kroupa}, P. 2020, \mnras, 491, 3042

\bibitem[{{Pawlowski} {et~al.}(2012){Pawlowski}, {Pflamm-Altenburg}, \&
  {Kroupa}}]{pawlowskiVPOS}
{Pawlowski}, M.~S., {Pflamm-Altenburg}, J., \& {Kroupa}, P. 2012, \mnras, 423,
  1109

\bibitem[{{Peacock} {et~al.}(2015){Peacock}, {Strader}, {Romanowsky}, \&
  {Brodie}}]{peacock2015}
{Peacock}, M.~B., {Strader}, J., {Romanowsky}, A.~J., \& {Brodie}, J.~P. 2015,
  \apj, 800, 13

\bibitem[{{Peng} {et~al.}(2008){Peng}, {Jord{\'a}n}, {C{\^o}t{\'e}},
  {Takamiya}, {West}, {Blakeslee}, {Chen}, {Ferrarese}, {Mei}, {Tonry}, \&
  {West}}]{peng2008}
{Peng}, E.~W., {Jord{\'a}n}, A., {C{\^o}t{\'e}}, P., {et~al.} 2008, \apj, 681,
  197

\bibitem[{{Pillepich} {et~al.}(2018){Pillepich}, {Nelson}, {Hernquist},
  {Springel}, {Pakmor}, {Torrey}, {Weinberger}, {Genel}, {Naiman}, {Marinacci},
  \& {Vogelsberger}}]{tng2}
{Pillepich}, A., {Nelson}, D., {Hernquist}, L., {et~al.} 2018, \mnras, 475, 648

\bibitem[{{Prole} {et~al.}(2021){Prole}, {van der Burg}, {Hilker}, \&
  {Spitler}}]{prole2021}
{Prole}, D.~J., {van der Burg}, R.~F.~J., {Hilker}, M., \& {Spitler}, L.~R.
  2021, \mnras, 500, 2049

\bibitem[{{Putman} {et~al.}(2021){Putman}, {Zheng}, {Price-Whelan}, {Grcevich},
  {Johnson}, {Tollerud}, \& {Peek}}]{putman2021}
{Putman}, M.~E., {Zheng}, Y., {Price-Whelan}, A.~M., {et~al.} 2021, \apj, 913,
  53

\bibitem[{{Radburn-Smith} {et~al.}(2011){Radburn-Smith}, {de Jong}, {Seth},
  {Bailin}, {Bell}, {Brown}, {Bullock}, {Courteau}, {Dalcanton}, {Ferguson},
  {Goudfrooij}, {Holfeltz}, {Holwerda}, {Purcell}, {Sick}, {Streich}, {Vlajic},
  \& {Zucker}}]{rs11}
{Radburn-Smith}, D.~J., {de Jong}, R.~S., {Seth}, A.~C., {et~al.} 2011, \apjs,
  195, 18

\bibitem[{{Radburn-Smith} {et~al.}(2014){Radburn-Smith}, {de Jong}, {Streich},
  {Bell}, {Dalcanton}, {Dolphin}, {Stilp}, {Monachesi}, {Holwerda}, \&
  {Bailin}}]{radburn2014}
{Radburn-Smith}, D.~J., {de Jong}, R.~S., {Streich}, D., {et~al.} 2014, \apj,
  780, 105

\bibitem[{{Rekola} {et~al.}(2005){Rekola}, {Jerjen}, \&
  {Flynn}}]{jerjen_field2}
{Rekola}, R., {Jerjen}, H., \& {Flynn}, C. 2005, \aap, 437, 823

\bibitem[{{Rines} {et~al.}(2003){Rines}, {Geller}, {Kurtz}, \&
  {Diaferio}}]{cairns}
{Rines}, K., {Geller}, M.~J., {Kurtz}, M.~J., \& {Diaferio}, A. 2003, \aj, 126,
  2152

\bibitem[{{Roberts} {et~al.}(2021){Roberts}, {Nierenberg}, \&
  {Peter}}]{roberts2021}
{Roberts}, D.~M., {Nierenberg}, A.~M., \& {Peter}, A. H.~G. 2021, \mnras, 502,
  1205

\bibitem[{{Robotham} {et~al.}(2018){Robotham}, {Davies}, {Driver}, {Koushan},
  {Taranu}, {Casura}, \& {Liske}}]{profound}
{Robotham}, A.~S.~G., {Davies}, L.~J.~M., {Driver}, S.~P., {et~al.} 2018,
  \mnras, 476, 3137

\bibitem[{{Sabbi} {et~al.}(2018){Sabbi}, {Calzetti}, {Ubeda}, {Adamo},
  {Cignoni}, {Thilker}, {Aloisi}, {Elmegreen}, {Elmegreen}, {Gouliermis},
  {Grebel}, {Messa}, {Smith}, {Tosi}, {Dolphin}, {Andrews}, {Ashworth},
  {Bright}, {Brown}, {Chandar}, {Christian}, {Clayton}, {Cook}, {Dale}, {de
  Mink}, {Dobbs}, {Evans}, {Fumagalli}, {Gallagher}, {Grasha}, {Herrero},
  {Hunter}, {Johnson}, {Kahre}, {Kennicutt}, {Kim}, {Krumholz}, {Lee},
  {Lennon}, {Martin}, {Nair}, {Nota}, {{\"O}stlin}, {Pellerin}, {Prieto},
  {Regan}, {Ryon}, {Sacchi}, {Schaerer}, {Schiminovich}, {Shabani}, {Van Dyk},
  {Walterbos}, {Whitmore}, \& {Wofford}}]{sabbi2018}
{Sabbi}, E., {Calzetti}, D., {Ubeda}, L., {et~al.} 2018, \apjs, 235, 23

\bibitem[{{Sales} {et~al.}(2013){Sales}, {Wang}, {White}, \&
  {Navarro}}]{sales2013}
{Sales}, L.~V., {Wang}, W., {White}, S. D.~M., \& {Navarro}, J.~F. 2013,
  \mnras, 428, 573

\bibitem[{{Samuel} {et~al.}(2020){Samuel}, {Wetzel}, {Tollerud},
  {Garrison-Kimmel}, {Loebman}, {El-Badry}, {Hopkins}, {Boylan-Kolchin},
  {Faucher-Gigu{\`e}re}, {Bullock}, {Benincasa}, \& {Bailin}}]{samuel2020}
{Samuel}, J., {Wetzel}, A., {Tollerud}, E., {et~al.} 2020, \mnras, 491, 1471

\bibitem[{{Sand} {et~al.}(2014){Sand}, {Crnojevi{\'c}}, {Strader}, {Toloba},
  {Simon}, {Caldwell}, {Guhathakurta}, {McLeod}, \& {Seth}}]{sand2014}
{Sand}, D.~J., {Crnojevi{\'c}}, D., {Strader}, J., {et~al.} 2014, \apj, 793, L7

\bibitem[{{Sawala} {et~al.}(2016){Sawala}, {Frenk}, {Fattahi}, {Navarro},
  {Bower}, {Crain}, {Dalla Vecchia}, {Furlong}, {Helly}, {Jenkins}, {Oman},
  {Schaller}, {Schaye}, {Theuns}, {Trayford}, \& {White}}]{sawala2016}
{Sawala}, T., {Frenk}, C.~S., {Fattahi}, A., {et~al.} 2016, \mnras, 457, 1931

\bibitem[{{Schlafly} \& {Finkbeiner}(2011)}]{sfd2}
{Schlafly}, E.~F., \& {Finkbeiner}, D.~P. 2011, \apj, 737, 103

\bibitem[{{Schlegel} {et~al.}(1998){Schlegel}, {Finkbeiner}, \& {Davis}}]{sfd}
{Schlegel}, D.~J., {Finkbeiner}, D.~P., \& {Davis}, M. 1998, \apj, 500, 525

\bibitem[{{Sick} {et~al.}(2015){Sick}, {Courteau}, {Cuilland re}, {Dalcanton},
  {de Jong}, {McDonald}, {Simard}, \& {Tully}}]{sick2015}
{Sick}, J., {Courteau}, S., {Cuilland re}, J.-C., {et~al.} 2015, in IAU
  Symposium, Vol. 311, Galaxy Masses as Constraints of Formation Models, ed.
  M.~{Cappellari} \& S.~{Courteau}, 82--85

\bibitem[{{Simon}(2019)}]{simon2019}
{Simon}, J.~D. 2019, \araa, 57, 375

\bibitem[{{Simpson} {et~al.}(2018){Simpson}, {Grand}, {G{\'o}mez}, {Marinacci},
  {Pakmor}, {Springel}, {Campbell}, \& {Frenk}}]{simpson2018}
{Simpson}, C.~M., {Grand}, R. J.~J., {G{\'o}mez}, F.~A., {et~al.} 2018, \mnras,
  478, 548

\bibitem[{{Skrutskie} {et~al.}(2006){Skrutskie}, {Cutri}, {Stiening},
  {Weinberg}, {Schneider}, {Carpenter}, {Beichman}, {Capps}, {Chester},
  {Elias}, {Huchra}, {Liebert}, {Lonsdale}, {Monet}, {Price}, {Seitzer},
  {Jarrett}, {Kirkpatrick}, {Gizis}, {Howard}, {Evans}, {Fowler}, {Fullmer},
  {Hurt}, {Light}, {Kopan}, {Marsh}, {McCallon}, {Tam}, {Van Dyk}, \&
  {Wheelock}}]{2mass}
{Skrutskie}, M.~F., {Cutri}, R.~M., {Stiening}, R., {et~al.} 2006, \aj, 131,
  1163

\bibitem[{{Smercina} {et~al.}(2018){Smercina}, {Bell}, {Price}, {D Souza},
  {Slater}, {Bailin}, {Monachesi}, \& {Nidever}}]{smercina2018}
{Smercina}, A., {Bell}, E.~F., {Price}, P.~A., {et~al.} 2018, \apj, 863, 152

\bibitem[{{Smercina} {et~al.}(2021){Smercina}, {Bell}, {Samuel}, \&
  {D'Souza}}]{smercina2021}
{Smercina}, A., {Bell}, E.~F., {Samuel}, J., \& {D'Souza}, R. 2021, arXiv
  e-prints, arXiv:2107.04591

\bibitem[{{Smercina} {et~al.}(2017){Smercina}, {Bell}, {Slater}, {Price},
  {Bailin}, \& {Monachesi}}]{smercina2017}
{Smercina}, A., {Bell}, E.~F., {Slater}, C.~T., {et~al.} 2017, \apjl, 843, L6

\bibitem[{{Spekkens} {et~al.}(2014){Spekkens}, {Urbancic}, {Mason}, {Willman},
  \& {Aguirre}}]{spekkens2014}
{Spekkens}, K., {Urbancic}, N., {Mason}, B.~S., {Willman}, B., \& {Aguirre},
  J.~E. 2014, \apjl, 795, L5

\bibitem[{{Spencer} {et~al.}(2014){Spencer}, {Loebman}, \&
  {Yoachim}}]{spencer2014}
{Spencer}, M., {Loebman}, S., \& {Yoachim}, P. 2014, \apj, 788, 146

\bibitem[{{Spergel} {et~al.}(2015){Spergel}, {Gehrels}, {Baltay}, {Bennett},
  {Breckinridge}, {Donahue}, {Dressler}, {Gaudi}, {Greene}, {Guyon}, {Hirata},
  {Kalirai}, {Kasdin}, {Macintosh}, {Moos}, {Perlmutter}, {Postman},
  {Rauscher}, {Rhodes}, {Wang}, {Weinberg}, {Benford}, {Hudson}, {Jeong},
  {Mellier}, {Traub}, {Yamada}, {Capak}, {Colbert}, {Masters}, {Penny},
  {Savransky}, {Stern}, {Zimmerman}, {Barry}, {Bartusek}, {Carpenter}, {Cheng},
  {Content}, {Dekens}, {Demers}, {Grady}, {Jackson}, {Kuan}, {Kruk}, {Melton},
  {Nemati}, {Parvin}, {Poberezhskiy}, {Peddie}, {Ruffa}, {Wallace}, {Whipple},
  {Wollack}, \& {Zhao}}]{spergel2015}
{Spergel}, D., {Gehrels}, N., {Baltay}, C., {et~al.} 2015, arXiv e-prints,
  arXiv:1503.03757

\bibitem[{{Stierwalt} {et~al.}(2009){Stierwalt}, {Haynes}, {Giovanelli},
  {Kent}, {Martin}, {Saintonge}, {Karachentsev}, \&
  {Karachentseva}}]{stierwalt2009}
{Stierwalt}, S., {Haynes}, M.~P., {Giovanelli}, R., {et~al.} 2009, \aj, 138,
  338

\bibitem[{{Su} {et~al.}(2021){Su}, {Salo}, {Janz}, {Laurikainen}, {Venhola},
  {Peletier}, {Iodice}, {Hilker}, {Cantiello}, {Napolitano}, {Spavone}, {Raj},
  {van de Ven}, {Mieske}, {Paolillo}, {Capaccioli}, {Valentijn}, \&
  {Watkins}}]{su2021}
{Su}, A.~H., {Salo}, H., {Janz}, J., {et~al.} 2021, \aap, 647, A100

\bibitem[{{Tanaka} {et~al.}(2018){Tanaka}, {Chiba}, {Hayashi}, {Komiyama},
  {Okamoto}, {Cooper}, {Okamoto}, \& {Spitler}}]{tanaka2018}
{Tanaka}, M., {Chiba}, M., {Hayashi}, K., {et~al.} 2018, \apj, 865, 125

\bibitem[{{Tanaka} {et~al.}(2017){Tanaka}, {Chiba}, \& {Komiyama}}]{tanaka2017}
{Tanaka}, M., {Chiba}, M., \& {Komiyama}, Y. 2017, \apj, 842, 127

\bibitem[{{Tanoglidis} {et~al.}(2021){Tanoglidis}, {Drlica-Wagner}, {Wei},
  {Li}, {S{\'a}nchez}, {Zhang}, {Peter}, {Feldmeier-Krause}, {Prat}, {Casey},
  {Palmese}, {S{\'a}nchez}, {DeRose}, {Conselice}, {Gagnon}, {Abbott},
  {Aguena}, {Allam}, {Avila}, {Bechtol}, {Bertin}, {Bhargava}, {Brooks},
  {Burke}, {Rosell}, {Kind}, {Carretero}, {Chang}, {Costanzi}, {da Costa}, {De
  Vicente}, {Desai}, {Diehl}, {Doel}, {Eifler}, {Everett}, {Evrard},
  {Flaugher}, {Frieman}, {Garc{\'\i}a-Bellido}, {Gerdes}, {Gruendl},
  {Gschwend}, {Gutierrez}, {Hartley}, {Hollowood}, {Huterer}, {James},
  {Krause}, {Kuehn}, {Kuropatkin}, {Maia}, {March}, {Marshall}, {Menanteau},
  {Miquel}, {Ogando}, {Paz-Chinch{\'o}n}, {Romer}, {Roodman}, {Sanchez},
  {Scarpine}, {Serrano}, {Sevilla-Noarbe}, {Smith}, {Suchyta}, {Tarle},
  {Thomas}, {Tucker}, {Walker}, \& {DES Collaboration}}]{tanoglidis2021}
{Tanoglidis}, D., {Drlica-Wagner}, A., {Wei}, K., {et~al.} 2021, \apjs, 252, 18

\bibitem[{{Tikhonov} {et~al.}(2015){Tikhonov}, {Lebedev}, \&
  {Galazutdinova}}]{t15}
{Tikhonov}, N.~A., {Lebedev}, V.~S., \& {Galazutdinova}, O.~A. 2015, Astronomy
  Letters, 41, 239

\bibitem[{{Toloba} {et~al.}(2016){Toloba}, {Sand}, {Spekkens}, {Crnojevi{\'c}},
  {Simon}, {Guhathakurta}, {Strader}, {Caldwell}, {McLeod}, \&
  {Seth}}]{toloba2016}
{Toloba}, E., {Sand}, D.~J., {Spekkens}, K., {et~al.} 2016, \apjl, 816, L5

\bibitem[{{Tonry} \& {Schneider}(1988)}]{tonry1988}
{Tonry}, J., \& {Schneider}, D.~P. 1988, \aj, 96, 807

\bibitem[{{Tonry} {et~al.}(2001){Tonry}, {Dressler}, {Blakeslee}, {Ajhar},
  {Fletcher}, {Luppino}, {Metzger}, \& {Moore}}]{tonry2001}
{Tonry}, J.~L., {Dressler}, A., {Blakeslee}, J.~P., {et~al.} 2001, \apj, 546,
  681

\bibitem[{{Trentham} \& {Tully}(2009)}]{trentham2009}
{Trentham}, N., \& {Tully}, R.~B. 2009, \mnras, 398, 722

\bibitem[{{Tully} {et~al.}(2016){Tully}, {Courtois}, \& {Sorce}}]{tully2016}
{Tully}, R.~B., {Courtois}, H.~M., \& {Sorce}, J.~G. 2016, \aj, 152, 50

\bibitem[{{Tully} {et~al.}(2019){Tully}, {Karachentsev}, {Rizzi}, \&
  {Shaya}}]{tully2019}
{Tully}, R.~B., {Karachentsev}, I.~D., {Rizzi}, L., \& {Shaya}, E.~J. 2019,
  {Every Known Nearby Galaxy}, HST Proposal, ,

\bibitem[{{van den Bosch} {et~al.}(2018){van den Bosch}, {Ogiya}, {Hahn}, \&
  {Burkert}}]{vdb2018a}
{van den Bosch}, F.~C., {Ogiya}, G., {Hahn}, O., \& {Burkert}, A. 2018, \mnras,
  474, 3043

\bibitem[{{Venhola} {et~al.}(2018){Venhola}, {Peletier}, {Laurikainen}, {Salo},
  {Iodice}, {Mieske}, {Hilker}, {Wittmann}, {Lisker}, {Paolillo}, {Cantiello},
  {Janz}, {Spavone}, {D'Abrusco}, {Ven}, {Napolitano}, {Kleijn}, {Maddox},
  {Capaccioli}, {Grado}, {Valentijn}, {Falc{\'o}n-Barroso}, \&
  {Limatola}}]{venhola2018}
{Venhola}, A., {Peletier}, R., {Laurikainen}, E., {et~al.} 2018, \aap, 620,
  A165

\bibitem[{{Venhola} {et~al.}(2021){Venhola}, {Peletier}, {Salo}, {Laurikainen},
  {Janz}, {Haigh}, {Wilkinson}, {Iodice}, {Hilker}, {Mieske}, {Cantiello}, \&
  {Spavone}}]{venhola2021}
{Venhola}, A., {Peletier}, R.~F., {Salo}, H., {et~al.} 2021, arXiv e-prints,
  arXiv:2111.01855

\bibitem[{{Wang} {et~al.}(2014){Wang}, {Sales}, {Henriques}, \&
  {White}}]{wang2014}
{Wang}, W., {Sales}, L.~V., {Henriques}, B. M.~B., \& {White}, S. D.~M. 2014,
  \mnras, 442, 1363

\bibitem[{{Wang} \& {White}(2012)}]{wang2012}
{Wang}, W., \& {White}, S. D.~M. 2012, \mnras, 424, 2574

\bibitem[{{Wang} {et~al.}(2021){Wang}, {Takada}, {Li}, {Carlsten}, {Lan},
  {Shi}, {Miyatake}, {More}, {Beaton}, {Lupton}, {Lin}, {Qiu}, \&
  {Luo}}]{wang2021}
{Wang}, W., {Takada}, M., {Li}, X., {et~al.} 2021, \mnras, 500, 3776

\bibitem[{{Wenger} {et~al.}(2000){Wenger}, {Ochsenbein}, {Egret}, {Dubois},
  {Bonnarel}, {Borde}, {Genova}, {Jasniewicz}, {Lalo{\"e}}, {Lesteven}, \&
  {Monier}}]{Wenger_2000}
{Wenger}, M., {Ochsenbein}, F., {Egret}, D., {et~al.} 2000, \aaps, 143, 9

\bibitem[{{Wetzel} {et~al.}(2016){Wetzel}, {Hopkins}, {Kim},
  {Faucher-Gigu{\`e}re}, {Kere{\v s}}, \& {Quataert}}]{wetzel2016}
{Wetzel}, A.~R., {Hopkins}, P.~F., {Kim}, J.-h., {et~al.} 2016, \apjl, 827, L23

\bibitem[{{Wetzel} {et~al.}(2013){Wetzel}, {Tinker}, {Conroy}, \& {van den
  Bosch}}]{wetzel2013}
{Wetzel}, A.~R., {Tinker}, J.~L., {Conroy}, C., \& {van den Bosch}, F.~C. 2013,
  \mnras, 432, 336

\bibitem[{{Wetzel} {et~al.}(2015){Wetzel}, {Tollerud}, \& {Weisz}}]{wetzel2015}
{Wetzel}, A.~R., {Tollerud}, E.~J., \& {Weisz}, D.~R. 2015, \apjl, 808, L27

\bibitem[{{Willmer}(2018)}]{willmer2018}
{Willmer}, C. N.~A. 2018, \apjs, 236, 47

\bibitem[{{Wu} {et~al.}(2021){Wu}, {Peek}, {Tollerud}, {Mao}, {Nadler}, {Geha},
  {Wechsler}, {Kallivayalil}, \& {Weiner}}]{wu2021}
{Wu}, J.~F., {Peek}, J.~E.~G., {Tollerud}, E.~J., {et~al.} 2021, arXiv
  e-prints, arXiv:2112.01542

\bibitem[{{Wyder} {et~al.}(2007){Wyder}, {Martin}, {Schiminovich}, {Seibert},
  {Budav{\'a}ri}, {Treyer}, {Barlow}, {Forster}, {Friedman}, {Morrissey},
  {Neff}, {Small}, {Bianchi}, {Donas}, {Heckman}, {Lee}, {Madore}, {Milliard},
  {Rich}, {Szalay}, {Welsh}, \& {Yi}}]{wyder2007}
{Wyder}, T.~K., {Martin}, D.~C., {Schiminovich}, D., {et~al.} 2007, \apjs, 173,
  293

\bibitem[{{Xi} {et~al.}(2018){Xi}, {Taylor}, {Massey}, {Rhodes}, {Koekemoer},
  \& {Salvato}}]{xi2018}
{Xi}, C., {Taylor}, J.~E., {Massey}, R.~J., {et~al.} 2018, \mnras, 478, 5336

\bibitem[{{Zaritsky} {et~al.}(2019){Zaritsky}, {Donnerstein}, {Dey},
  {Kadowaki}, {Zhang}, {Karunakaran}, {Mart{\'\i}nez-Delgado}, {Rahman}, \&
  {Spekkens}}]{zaritsky2019}
{Zaritsky}, D., {Donnerstein}, R., {Dey}, A., {et~al.} 2019, \apjs, 240, 1

\bibitem[{{Zou} {et~al.}(2017){Zou}, {Zhou}, {Fan}, {Zhang}, {Zhou}, {Nie},
  {Peng}, {McGreer}, {Jiang}, {Dey}, {Fan}, {He}, {Jiang}, {Lang}, {Lesser},
  {Ma}, {Mao}, {Schlegel}, \& {Wang}}]{bass1}
{Zou}, H., {Zhou}, X., {Fan}, X., {et~al.} 2017, \pasp, 129, 064101

\bibitem[{{Zou} {et~al.}(2018){Zou}, {Zhang}, {Zhou}, {Peng}, {Nie}, {Zhou},
  {Fan}, {Jiang}, {McGreer}, {Dey}, {Fan}, {Findlay}, {Gao}, {Gu}, {Guo}, {He},
  {Jin}, {Kong}, {Lang}, {Lei}, {Lesser}, {Li}, {Ma}, {Meng}, {Maxwell},
  {Myers}, {Rui}, {Schlegel}, {Sun}, {Wu}, {Wang}, \& {Yuan}}]{bass2}
{Zou}, H., {Zhang}, T., {Zhou}, Z., {et~al.} 2018, \apjs, 237, 37

\end{thebibliography}

\appendix

\begin{deluxetable}{cc}
\tablecaption{LV Galaxies not Included in Host List.\label{tab:host_rejects}}
\tablewidth{\textwidth}
\tablehead{
\colhead{Name} & \colhead{Reason} }
\startdata
M82  &  member in M81 group \\ 
NGC3184  &  Mean NED SN1a distance $>$ 12 Mpc \\ 
NGC3351  &  member in Leo I group \\ 
NGC3368  &  member in Leo I group \\ 
NGC3377  &  member in Leo I group \\ 
NGC3384  &  member in Leo I group \\ 
NGC3412  &  member in Leo I group \\ 
NGC3489  &  member in Leo I group \\ 
NGC3593  &  member in Leo Triplet \\ 
NGC3628  &  member in Leo Triplet \\ 
NGC4490  &  TRGB distance of \citet{sabbi2018} yields $M_K>-22.1$ \\ 
NGC4559  &  KT17 lists $M_K>-22.1$ \\ 
NGC2835  &  KT17 lists $M_K>-22.1$ \\ 
NGC5195  &  member of NGC5194 group \\ 
NGC1792  &  member of NGC1808 group \\ 
NGC4818  &  Mean NED distance $>$ 12 Mpc \\ 
NGC6684  &  SBF distance of \citet{tonry2001} has distance $>$ 12 Mpc \\ 
\enddata
\end{deluxetable}

\section{Galaxies Not Included in Host List}
\label{app:host_rejects}

Table \ref{tab:host_rejects} lists the massive LV galaxies that passed the initial cuts but were removed from the master host list either due to being secondaries in a group or a more precise distance estimate placing them outside the LV or below the $M_{K_s}$ cut.

\begin{figure*}
\includegraphics[width=\textwidth]{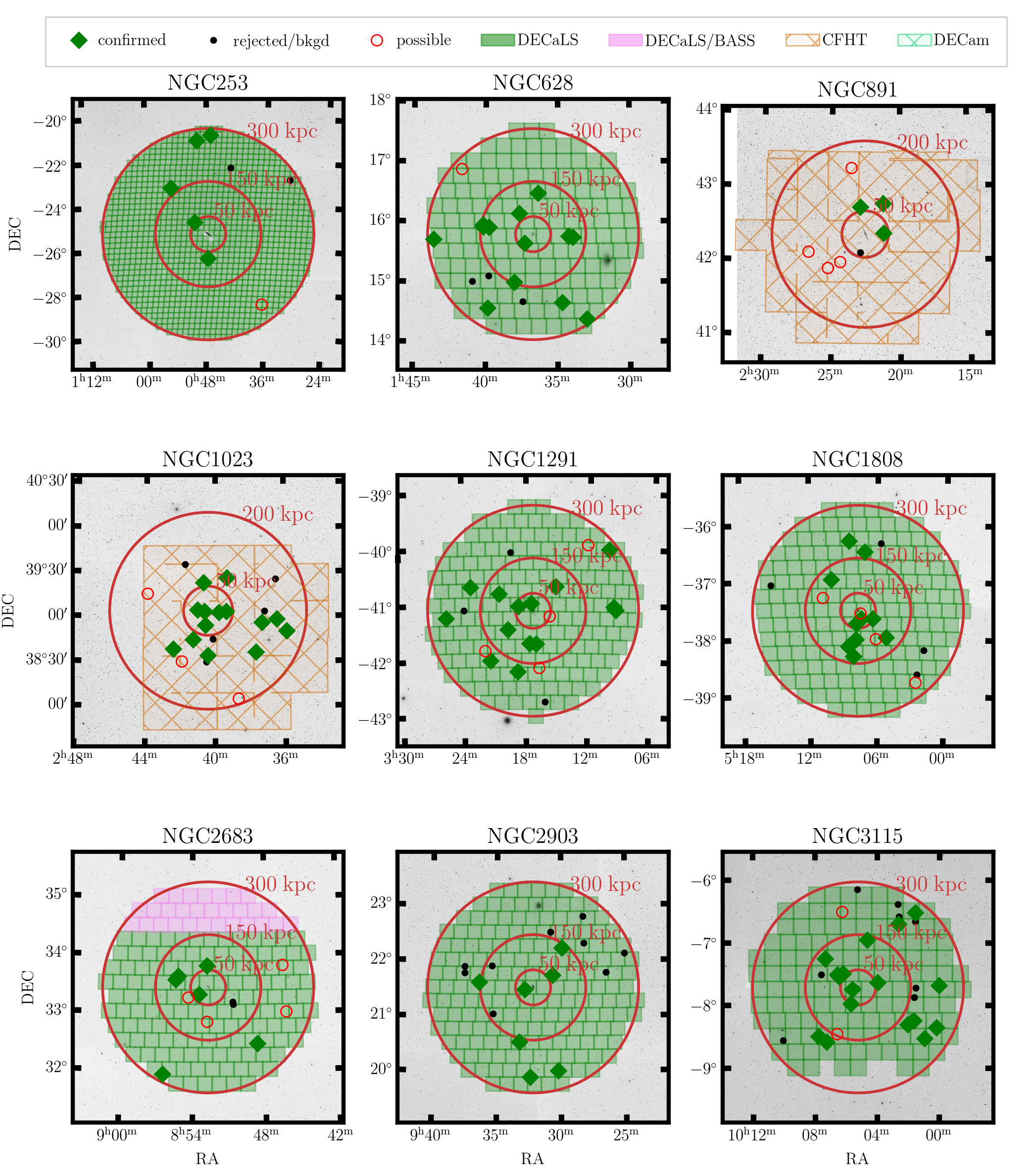}
\caption{The individual survey footprints for the hosts surveyed in this work. Candidate satellites with $M_V<-9$ mag are plotted in different ways to indicate whether they are confirmed as satellites via distance measurement, rejected as background contaminants, or whether no distance constraint is possible.}
\label{fig:area}
\end{figure*}

\begin{figure*}
\ContinuedFloat
\includegraphics[width=\textwidth]{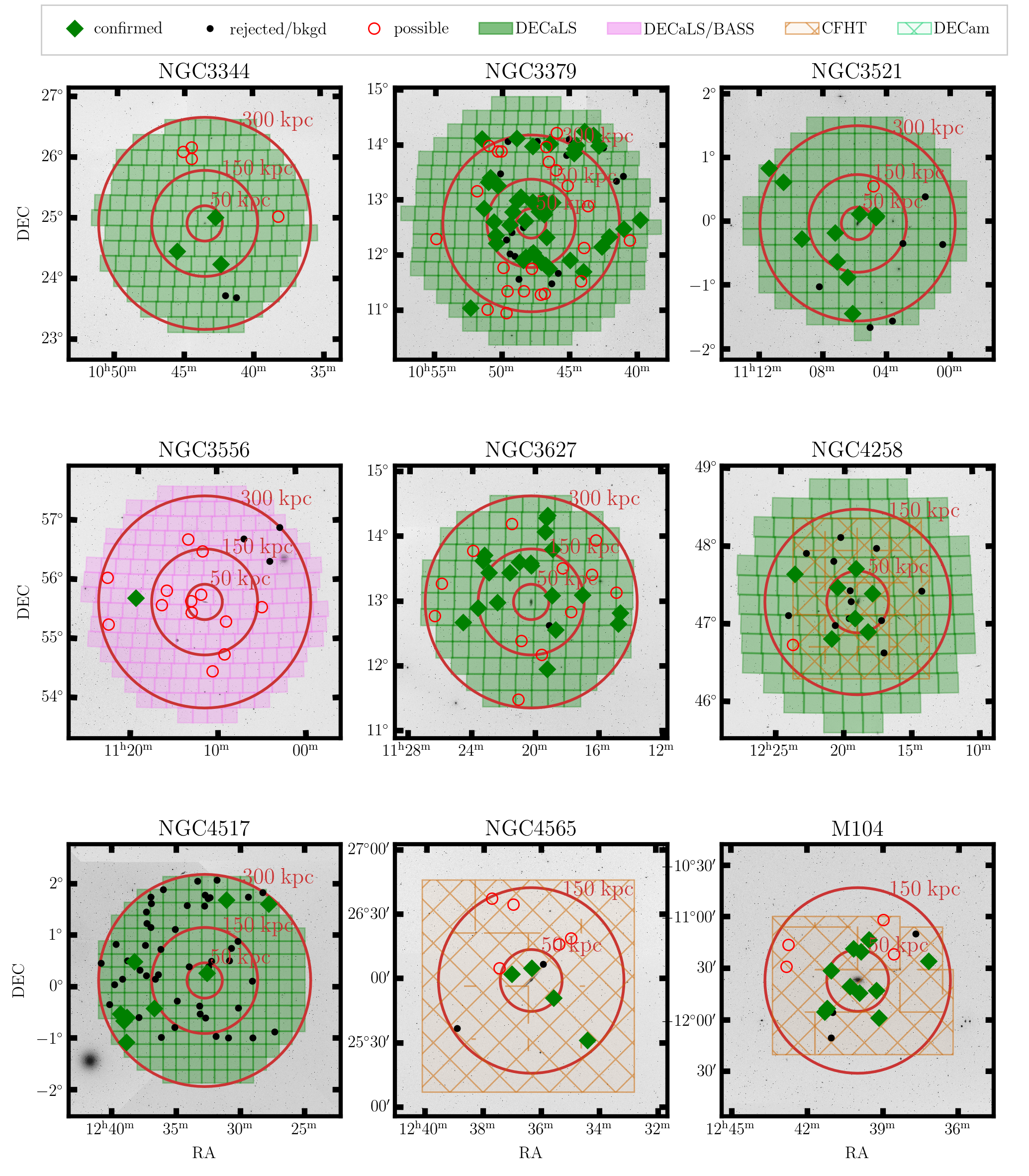}
\caption{}
\label{fig:area}
\end{figure*}

\begin{figure*}
\ContinuedFloat
\includegraphics[width=\textwidth]{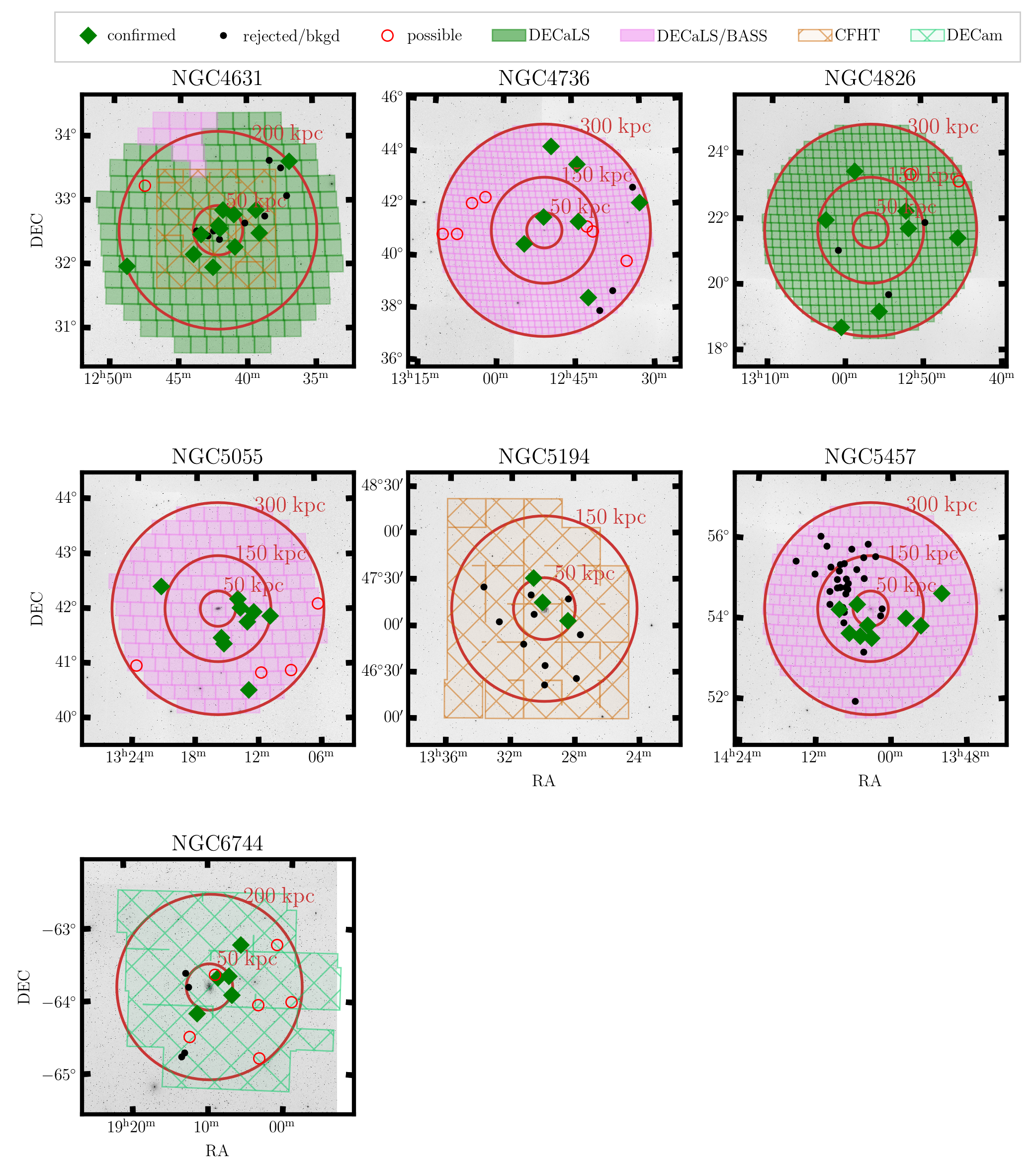}
\caption{}
\label{fig:area}
\end{figure*}

\section{Individual Host Area Coverage}
\label{app:host_areas}
Figure \ref{fig:area} shows the survey footprints and locations of candidate satellites for each individual host. The different sources of data used for candidate detection are shown. The candidate satellites are plotted in different styles depending on whether they are confirmed as satellites via distance measurement, rejected as background contaminants, or whether no distance constraint was possible. Only candidates with $M_V<-9$ mag are shown and that fall within the coverage radii given in Table \ref{tab:hosts}. The figure shows which hosts use the BASS part of DECaLS, as well as the two hosts (NGC 4258 and NGC 4631) that use DECaLS to extend the radial coverage of the CFHT/MegaCam survey.

\section{Individual Host Completeness Results}
\label{app:host_completeness}

Figure \ref{fig:completeness_idv} shows the individual recovery fractions for all 25 hosts for which we do our own candidate satellite detection. The source of the imaging data is denoted in the upper left corner. The recovery fractions do not vary much between the hosts, with two main exceptions. First is that the DECaLS hosts that use the BASS portion of DECaLS have generally $\sim0.5$ mag worse surface brightness completeness. Second is NGC 891 which uses quite shallow CFHT/MegaCam imaging.

\begin{figure*}
\includegraphics[width=\textwidth]{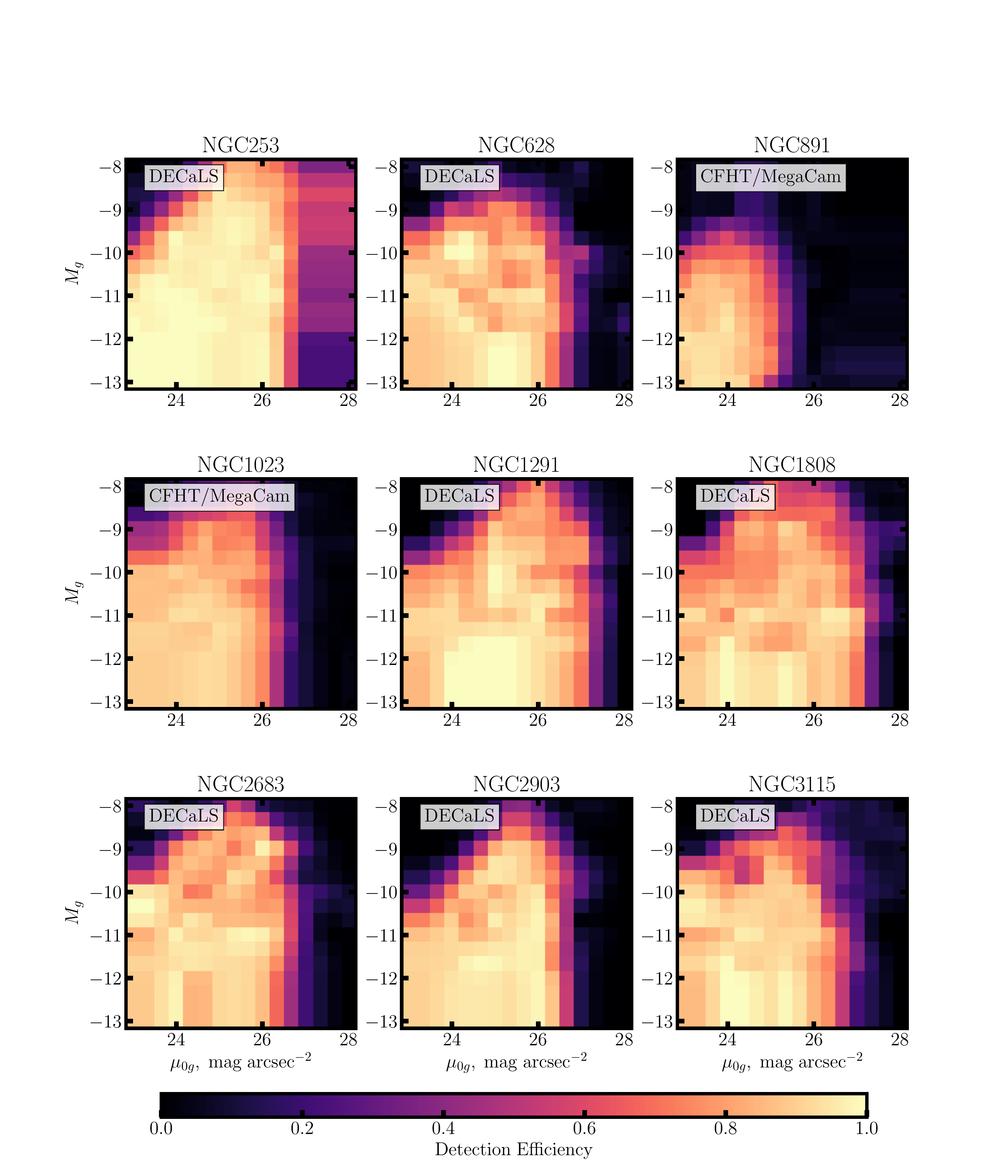}
\caption{Results of completeness tests for individual hosts. Artificial galaxies are injected into the images, and the recovery efficiency is measured as a function of total magnitude and central surface brightness.}
\label{fig:completeness_idv}
\end{figure*}

\begin{figure*}
\ContinuedFloat
\includegraphics[width=\textwidth]{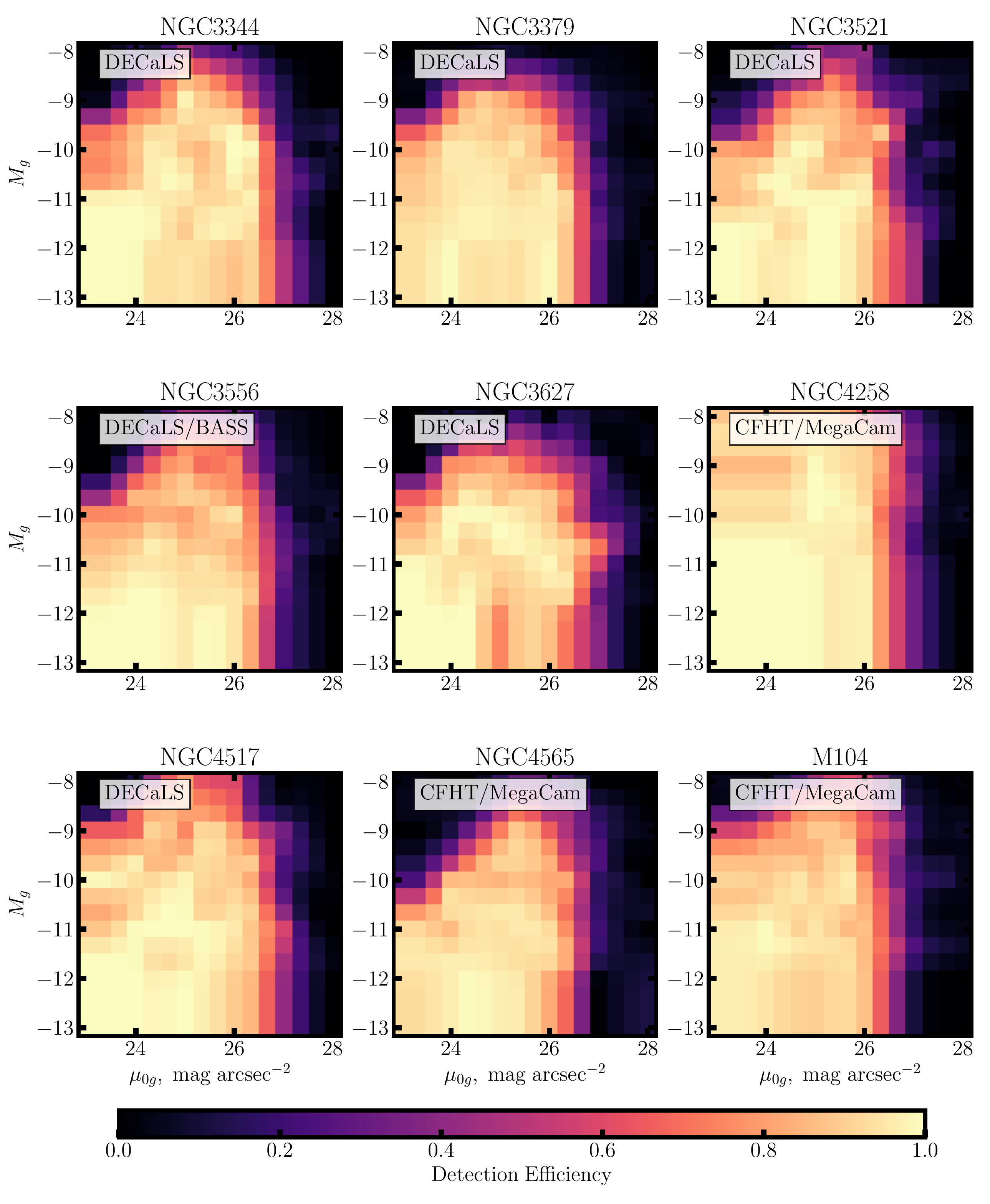}
\caption{}
\label{fig:completeness_idv}
\end{figure*}

\begin{figure*}
\ContinuedFloat
\includegraphics[width=\textwidth]{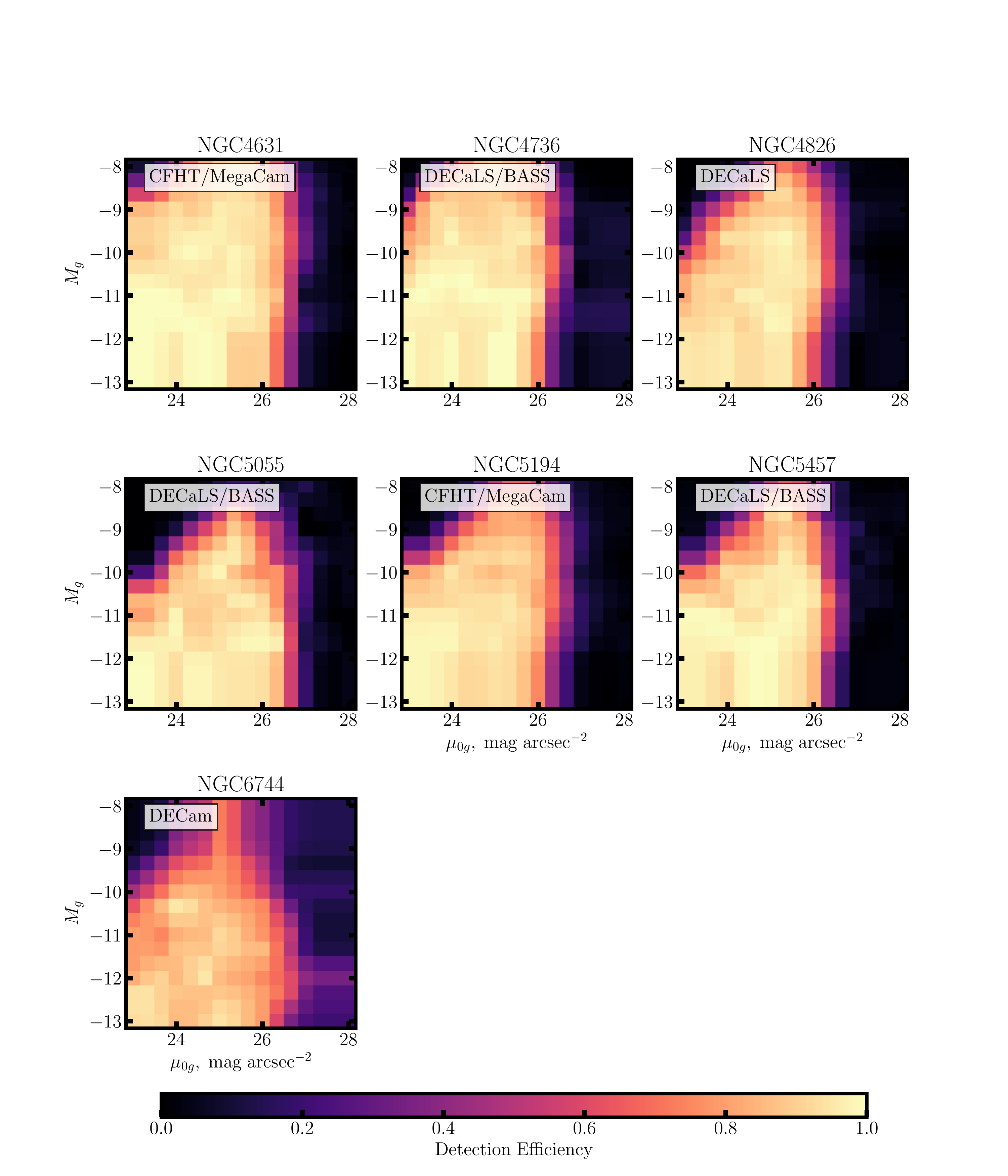}
\caption{}
\label{fig:completeness_idv}
\end{figure*}

\begin{deluxetable*}{ccccc}
\tablecaption{Comparison with the UNGC.\label{tab:ungc}}
\tablehead{
\colhead{Name}  & \colhead{Host} & \colhead{RA} & \colhead{DEC}  & \colhead{Reason}   \\ 
\colhead{}   & \colhead{}  & \colhead{(deg)}  & \colhead{(deg)} & \colhead{}  }  
\startdata
ScuSR  &  NGC253  &  8.4658  &  -27.84  &  viz reject; tidal stream of a bkgd. gal  \\ 
NGC628dwTBG  &  NGC628  &  23.2783  &  16.2514  &  viz reject; cirrus  \\ 
NGC628dwB  &  NGC628  &  24.0967  &  15.9647  &  below ELVES luminosity limit  \\ 
N1291-DW11  &  NGC1291  &  48.1842  &  -42.7983  &  not in DECaLS tract footprints  \\ 
N1291-DW15  &  NGC1291  &  49.9071  &  -40.4214  &  not detected but bkgd. galaxy from bulge+disk morph.  \\ 
N1291-DW14  &  NGC1291  &  51.2617  &  -40.3589  &  viz reject; bkgd. galaxy with spiral arms  \\ 
dw1044+11  &  NGC3379  &  161.1362  &  11.2694  &  viz reject; tidal stream of a bkgd. gal  \\ 
BST1047+1156  &  NGC3379  &  161.9325  &  11.9336  &  Below ELVES SB completeness  \\ 
NGC3521dwTBG  &  NGC3521  &  166.8046  &  0.1875  &  Does not appear to be real; nothing seen in deep HSC imaging  \\ 
LVJ1215+4732  &  NGC4258  &  183.9629  &  47.5489  &  In chip gap of CFHT/Megacam survey and known bkgd  \\ 
M106edgeN4217  &  NGC4258  &  184.0521  &  47.1344  &  Tidal stream/fluff of bkgd. galaxy  \\ 
NGC4288  &  NGC4258  &  185.1588  &  46.2917  &  Outside of CFHT/Megacam footprint; clear bkgd. morphology  \\ 
CVnHI  &  NGC4258  &  185.1808  &  46.2092  &  Not seen in deep optical images  \\ 
SUCD1  &  M104  &  190.0129  &  -11.6678  &  ELVES is not sensitive to UCDs  \\ 
KKSG33  &  M104  &  190.0371  &  -12.3647  &  Outside of CFHT/Megacam footprint  \\ 
NGC4656UV  &  NGC4631  &  191.0654  &  32.2833  &  TDG; not a distinct galaxy  \\ 
NGC4485  &  NGC4736  &  187.63  &  41.7  &  rejected because TRGB  \\ 
NGC4490  &  NGC4736  &  187.6517  &  41.6436  &  rejected because TRGB  \\ 
UGC07678  &  NGC4736  &  188.0017  &  39.8319  &  rejected because redshift  \\ 
PGC041749  &  NGC4736  &  188.4696  &  39.6258  &  rejected because redshift  \\ 
UGC7719  &  NGC4736  &  188.5025  &  39.0194  &  rejected because redshift  \\ 
UGC7751  &  NGC4736  &  188.7992  &  41.0608  &  rejected because redshift  \\ 
NGC4618  &  NGC4736  &  190.3867  &  41.1508  &  rejected because TRGB  \\ 
NGC4625  &  NGC4736  &  190.4696  &  41.2739  &  rejected because TRGB  \\ 
dw1305+41  &  NGC4736  &  196.3708  &  41.89  &  Falls in star-mask but clear bkgd. morphology  \\ 
dw1305+41  &  NGC5055  &  196.3708  &  41.89  &  Falls in star-mask but clear bkgd. morphology  \\ 
NGC5055dwTBG2  &  NGC5055  &  198.7129  &  41.725  &  Viz reject; tidal stream of a bkgd. gal  \\ 
GBT1355+5439  &  NGC5457  &  208.7108  &  54.6472  &  Nothing seen in deep optical images  \\ 
PGC2448110  &  NGC5457  &  211.2408  &  53.6914  &  Just an HII region  \\ 
PGC725719  &  NGC5236  &  200.7367  &  -29.7386  &  Distant edge-on disk galaxy?  \\ 
dw1326-29  &  NGC5236  &  201.5167  &  -29.4044  &  SBF reject  \\ 
dw1328-29  &  NGC5236  &  202.05  &  -29.4792  &  SBF reject  \\ 
dw1329-32  &  NGC5236  &  202.4917  &  -32.4961  &  SBF reject  \\ 
dw1330-32  &  NGC5236  &  202.725  &  -32.3058  &  SBF reject  \\ 
dw1336-32  &  NGC5236  &  204.1375  &  -32.3014  &  Cirrus  \\ 
dw1337-26  &  NGC5236  &  204.3042  &  -26.8028  &  Deep imaging shows just scattered light  \\ 
NGC5253  &  NGC5236  &  204.9825  &  -31.64  &  TRGB indicates foreground  \\ 
\enddata
\tablecomments{Reasons why the UNGC entries that fall within the coverage radii of the ELVES hosts but are not in the ELVES catalogs are missed. The only cases where legitimate, likely satellite candidates are missed is where they fell just outside the survey tracts (while still being within the coverage radius). In many cases ELVES uses deeper imaging for targeting which revealed previously noted candidates to be clear background galaxies.}
\end{deluxetable*}

\section{Comparison to Previous LV Satellite Searches}
\label{app:prev_work}
In this section, we compare the ELVES satellite searches with previous searches of the same host in the literature. We only focus on the hosts new in this work here. \citet{carlsten2020a} provided an exhaustive comparison with previous searches for the hosts included there (NGC 1023, NGC 4258, NGC 4631, NGC4565, NGC 5194, M104), and we refer the reader to that discussion for comparison of those satellite surveys.

\textbf{UNGC \citep{karachentsev2013}}: By far the current best reference for LV dwarfs is the Updated Nearby Galaxy Catalog of \citet{karachentsev2013}. This catalog is kept up-to-date with new dwarf searches and, thus, includes the objects from many of the individual host surveys we discuss below. We use a version of the UNGC from March, 2021. For each of the 25 hosts surveyed in ELVES, we go through the UNGC and find all objects that fall within the survey coverage radius of the host ($r_\mathrm{cover}$ from Table \ref{tab:hosts}). All UNGC entries that fall within the covered radius are matched with ELVES candidate satellites, and those that do not have a match are listed in Table \ref{tab:ungc}. Overall there are 245 UNGC objects that fall within the ELVES survyed regions, with 37 not having a matching ELVES candidate\footnote{One UNGC galaxy appears twice, once as a ``candidate satellite'' of NGC 4736 and once of NGC 5055}. For each of these, we provide a brief reason as to why it is not included in the ELVES satellite catalogs. These reasons are listed in Table \ref{tab:ungc}. About $\sim1/2$ of the missing UNGC galaxies were detected in ELVES but rejected based on their morphology (none of these UNGC entries have an actual distance measurement). ELVES uses deeper targeting imaging than what was likely available in the original detection of these objects and reveals clear indicators of being in the background. These include bulge+disk morphology, spiral arms present in a small ($r_e<5$\arcs) galaxy, or a tidal bridge connecting it to a nearby massive background galaxy. The only cases where a legitimate candidate satellite is missed is where the candidate is below the completeness limits of ELVES or it falls outside of the survey footprint (while still being within the coverage radius given in Table \ref{tab:hosts}).

\textbf{NASA-Sloan Atlas \citep[NSA;][]{blanton2011}}: We additionally look for any galaxies with known redshifts in the NSA in the regions surrounding the ELVES hosts. No additional valid candidates are found that are not already included in the ELVES satellite lists.

\textbf{NGC 1291 \citep{byun2020}}: We recover all the satellite candidates reported by \citet{byun2020} that fall within 300 kpc projected of NGC 1291 except for three: Dw11, Dw15, and Dw14. The first falls just outside the footprints of the DECaLS tracts used in our search. The latter two had morphologies strongly indicating they are background. One is a distant edge-on disk galaxy with a clear bulge, and the other has a bar and spiral arms yet is quite small ($r_e<5$\arcs). We note that our DECaLS/DES search recovers 6 candidates not included by \citet{byun2020}, including two we have rejected as background contaminants, one we have confirmed via SBF, and three that remain without distance confirmation.

\textbf{NGC 5457 \citep{javanmardi_m101, muller101, bennet2017}}: We recover all candidates detected around NGC 5457 by \citet{javanmardi_m101}, \citet{muller101}, and \citet{bennet2017} that are $M_V<-9$ mag (i.e. within the ELVES completeness limit) except for three: dwC, dw26, and dw31. The first was detected by our algorithm but rejected in the visual inspection stage. \citet{bennet2019} showed this galaxy is, indeed, a background contaminant. Dw26 was not detected by our algorithm as it is a relatively high-surface brightness background spiral \citep{bennet2017, sbf_m101}. Dw31 was covered by a star-mask in our DECaLS search but was shown by \citet{sbf_m101} to be a background contaminant. Our DECaLS search produced many additional candidates not noted by these previous works. Admittedly, most are dwarfs in the background NGC 5485 group, but two have been confirmed as NGC 5457 satellites via SBF (one is $M_V>-9$ mag and thus not in the ELVES list).

Finally, we mention that our DECaLS searches of NGC 4631 and NGC 4258 recover all confirmed satellites of those hosts found in the deeper CFHT/Megacam surveys of \citet{carlsten2020a} with the single exception of dw1242p3237 around NGC 4631. However, this satellite is below the typical $\mu_{0,V}\sim26.5$ mag arcsec$^{-2}$ sensitivity of the DECaLS searches.

\section{Candidate Satellite Distance Constraints}
\label{app:distances}
In Table \ref{tab:dists_confirmed}, we present the distance constraints for the confirmed satellites. Candidate name, coordinates, host, and host distance are given along with the distance information used in the confirmation. SBF distance is given along with $\pm1,2\sigma$ distance uncertainties, the S/N of the measurement, and the source of the data used (e.g. CFHT/MegaCam, HSC, etc.). Known redshifts or TRGB distances for dwarfs that have them are also listed, along with their source. Redshifts without a reference come from SIMBAD. Dwarfs marked by a $^\dagger$ had some problem with the SBF measurement and are explained in more detail below.

Table \ref{tab:dists_rejected} similarly presents the distance constraints for the rejected candidates. In this case, only the $2\sigma$ lower bound to the SBF distance is given. 

Table \ref{tab:dists_maybes} lists the candidates who did not have a conclusive distance constraint. The source, if any, of an attempted SBF measurement is listed.

In the remaining part of this section, we discuss the few cases that are exceptions to the usual criteria for confirming or rejecting candidate satellites. Some of these had problematic SBF measurements while some have distance information coming from sources other than the most common three methods used (redshift, TRGB, and SBF). 

\begin{itemize}
    \item dw1242p3231 from the NGC 4631 system has archival HST imaging in which it is unresolved, indicating that it is not at $D=7.4$ Mpc.
    \item dw1330p4708 from the NGC 5194 system similarly has archival HST coverage showing it is unresolved and, hence, background.
    \item dw1908m6343 from the NGC 6744 system had a high S/N SBF measurement but yielded far too small of a distance. This is a blue, very irregular dwarf which can bias the fluctuation too high since a S\'{e}rsc model is assumed in the SBF measurement. Considering the fact that the SBF amount is visually about the same as other, more regular blue satellites in this system and dw1908m6343 is quite near NGC 6744 in projection, this is very likely a satellite. 
    \item dw0031m2246 from the NGC 253 system barely makes the $\Delta cz <275$ km/s cut but has a Tully-Fisher distance in NED indicating it is background.
    \item dw0933p2030 from the NGC 2903 system has an SDSS redshift with $\Delta cz >275$ km/s which appears to be extremely low S/N and erroneous upon visual inspection, and thus we trust the SBF distance result.
    \item dw1240p3800 and UGC7690 from the NGC 4736 system make the $\Delta cz <275$ km/s cut but both have Tully-Fisher distances in NED indicating background.
    \item dw1300p1843 from the NGC 4826 system has a SBF distance consistent with that of the host but only with S/N$\sim3$. However, it has visually noticeable SBF that appears to be the same level has other satellites in this system. Additionally, its large $r_e\sim30$\arcs~size indicates it is unlikely to be background (it is already an outlier to the mass-size relation). Thus, we consider it confirmed.
    \item dw1046p1259 from the NGC 3379 system has a very strong SBF signal but yields a slightly lower distance than its host. This dwarf is in a halo of scattered light from a nearby star which often biases the fluctuation measurement high (and thus distance low) \citep{carlsten2020b}. We consider this a confirmed satellite.
    \item dw1119p1404 from the NGC 3627 system has an SDSS redshift with $\Delta cz = -303$ km/s which appears to be extremely low S/N and just noise upon visual inspection, and thus we trust the SBF distance result.
    \item dw1120p1337 from the NGC 3627 system is an actively disrupting nucleated dwarf \citep[see, e.g.,][]{jennings2015} and no distance measurement is required to confirm its satellite status.
    \item dw0316m4244 and dw0323m4105 from the NGC 1291 system were imaged with deeper Magellan imaging in order to measure SBF. However, the deeper imaging clearly revealed them to be tidal features of massive background galaxies, and we reject them without an SBF measurement.
    \item dw0852p3347 from the NGC 2683 system is tidally distorted by NGC 2683 and thus is a confirmed satellite without need for a distance measurement.
    \item dw1105p0006 from the NGC 3521 system is also tidally distorted by its host, removing any need for a distance measurement.
    \item UGCA337 (dw1312p4147) from the NGC 5055 system has an extremely strong SBF signal but the $2\sigma$ distance upper bound is slightly closer than NGC 5055. It is slightly irregular which is biasing the SBF a little high. It has a redshift that makes the $\Delta cz <275$ km/s cut, and is almost certainly a satellite.
    \item dw1240m1140 from the M104 system shows a strong SBF signal that puts it in the foreground. However, this dwarf is located very close to M104 in projection, and the halo of M104 appears to be adding fluctuation power to the SBF measurement. Considering its dSph morphology, proximity to M104, and SBF, we include it in the `confirmed' category.
    \item dw1241p3251 from the NGC 4631 system appears slightly foreground from the SBF measurement. However, this galaxy is not well fit by a S\'{e}rsic profile, and thus the distance is likely underestimated. This galaxy has a redshift consistent with NGC 4631 ($\Delta cz\sim60$ km/s) and is considered `confirmed'.
    \item dw1106p5644 from the NGC 3556 is not seen in deeper CFHT imaging, indicating it was just an artifact in DECaLS. It is included in the `rejected' category.
    \item dw1235p2606 from the NGC 4565 system is located directly in the middle of an H\textsc{I} warp on the northwest edge of the disk. \citet{radburn2014} used HST observations to show that this candidate consists of young ($\sim600$ Myr) stars that likely formed in-situ in the warp. Thus, this is not a dwarf satellite and is `rejected'.

\end{itemize}

\startlongtable
\begin{longrotatetable}
\movetabledown=10mm
\begin{deluxetable}{cccccccccccc}
\tablecaption{Confirmed dwarf satellite distance results\label{tab:dists_confirmed}}
\tablehead{
\colhead{Name} & \colhead{RA} & \colhead{DEC}  & \colhead{Host}  & \colhead{$D_\mathrm{host}$}  & \colhead{SBF Distance}  & \colhead{SBF S/N} & \colhead{SBF Source} & \colhead{$v_\mathrm{rec}$}  & \colhead{$v_\mathrm{rec}$ Source} & \colhead{$D_\mathrm{TRGB}$} & \colhead{$D_\mathrm{TRGB}$ Source} \\ 
\colhead{} & \colhead{(deg)}  & \colhead{(deg)}   & \colhead{}  & \colhead{(Mpc)}  & \colhead{(Mpc)}  & \colhead{} & \colhead{}  & \colhead{(km/s)} & \colhead{}  & \colhead{(Mpc)} & \colhead{} }  
\startdata
NGC247 & 11.783 & -20.757 & NGC253 & 3.56 & -- & -- & -- & 159 & -- & 3.72 & a\\
dw0047m2623 & 11.894 & -26.390 & NGC253 & 3.56 & -- & -- & -- & -- & -- & 3.90 & b\\
dw0049m2100 & 12.454 & -21.017 & NGC253 & 3.56 & -- & -- & -- & 295 & -- & 3.44 & a\\
dw0050m2444 & 12.575 & -24.737 & NGC253 & 3.56 & -- & -- & -- & -- & -- & 3.12 & c\\
dw0055m2309 & 13.754 & -23.169 & NGC253 & 3.56 & -- & -- & -- & 250 & a & -- & --\\
dw0132p1422 & 23.249 & 14.374 & NGC628 & 9.77 & -- & -- & -- & 669 & -- & -- & --\\
dw0133p1543 & 23.484 & 15.731 & NGC628 & 9.77 & $9.18^{+0.98,2.04}_{-0.94,1.9}$ & 8.29 & GEMINI & -- & -- & -- & --\\
dw0134p1544 & 23.554 & 15.746 & NGC628 & 9.77 & $10.5^{+0.92,1.86}_{-0.86,1.74}$ & 9.63 & GEMINI & -- & -- & -- & --\\
dw0134p1438 & 23.674 & 14.644 & NGC628 & 9.77 & -- & -- & -- & 731 & -- & -- & --\\
dw0136p1628 & 24.084 & 16.470 & NGC628 & 9.77 & $9.45^{+0.72,1.42}_{-0.74,1.57}$ & 22.76 & GEMINI & -- & -- & -- & --\\
dw0137p1537 & 24.324 & 15.633 & NGC628 & 9.77 & $10.12^{+1.87,3.97}_{-1.67,3.08}$ & 8.05 & CFHT & -- & -- & -- & --\\
dw0137p1607 & 24.416 & 16.132 & NGC628 & 9.77 & $9.05^{+1.26,2.79}_{-1.14,2.22}$ & 5.90 & GEMINI & -- & -- & -- & --\\
dw0138p1458 & 24.502 & 14.982 & NGC628 & 9.77 & $8.61^{+0.87,1.91}_{-0.74,1.36}$ & 6.34 & GEMINI & 738 & -- & -- & --\\
UGC1171 & 24.937 & 15.899 & NGC628 & 9.77 & $10.13^{+1.28,2.92}_{-0.96,1.7}$ & 5.41 & GEMINI & 696 & -- & -- & --\\
dw0139p1433 & 24.961 & 14.556 & NGC628 & 9.77 & $10.82^{+0.96,2.05}_{-0.8,1.5}$ & 8.04 & GEMINI & -- & -- & -- & --\\
dw0140p1556 & 25.040 & 15.939 & NGC628 & 9.77 & $11.78^{+1.49,3.31}_{-1.18,2.11}$ & 6.10 & GEMINI & -- & -- & -- & --\\
UGC1176 & 25.043 & 15.905 & NGC628 & 9.77 & -- & -- & -- & 632 & -- & -- & --\\
dw0143p1541 & 25.901 & 15.694 & NGC628 & 9.77 & $8.64^{+0.71,1.48}_{-0.61,1.22}$ & 8.57 & GEMINI & 789 & -- & -- & --\\
dw0221p4221 & 35.301 & 42.364 & NGC891 & 9.12 & $10.48^{+1.46,3.01}_{-1.38,2.67}$ & 10.57 & CFHT & -- & -- & 10.28 & d\\
UGC1807 & 35.306 & 42.763 & NGC891 & 9.12 & -- & -- & -- & 630 & -- & -- & --\\
dw0222p4242 & 35.730 & 42.712 & NGC891 & 9.12 & $10.28^{+1.57,3.46}_{-1.38,2.6}$ & 6.29 & CFHT & -- & -- & -- & --\\
\enddata
\tablecomments{The names, coordinates, hosts, and distance results for the confirmed satellites. Dwarfs marked with a $^\dagger$ are exceptions to the usual confirmation criteria. The full version of the table will be published in the online journal or will be provided upon request to the authors. SIMBAD is the source for $v_\mathrm{rec}$ unless otherwise noted. Other sources listed are: (a) \citet{karachentsev2013}(b) \citet{sand2014}, (c) \citet{toloba2016}, (d) \citet{muller2019_n891}, (e) \citet{hipass2004}, (f) \citet{irwin2009}, (g) \citet{cohen2018}, (h) \citet{sabbi2018}, (i) \citet{tully2016}, (j) \citet{haynes2018}, (k) \citet{karunakaran_smudges}, (l) \citet{cairns}, (m) \citet{smercina2018}, (n) \citet{t15}, (o) \citet{danieli2017}, (p) \citet{bennet2019}, .}
\end{deluxetable}
\end{longrotatetable}
\startlongtable
\begin{longrotatetable}
\movetabledown=10mm
\begin{deluxetable}{ccccccccccc}
\tablecaption{Rejected background contaminants distance results\label{tab:dists_rejected}}
\tablehead{
\colhead{Name} & \colhead{RA} & \colhead{DEC}  & \colhead{Host}  & \colhead{$D_\mathrm{host}$}  & \colhead{SBF Distance lower bound}  & \colhead{SBF Source} & \colhead{$v_\mathrm{rec}$}  & \colhead{$v_\mathrm{rec}$ Source} & \colhead{$D_\mathrm{TRGB}$} & \colhead{$D_\mathrm{TRGB}$ Source} \\ 
\colhead{} & \colhead{(deg)}  & \colhead{(deg)}   & \colhead{}  & \colhead{(Mpc)}  & \colhead{(Mpc)} & \colhead{}  & \colhead{(km/s)} & \colhead{}  & \colhead{(Mpc)} & \colhead{} }  
\startdata
dw0031m2246 & 7.843 & -22.766 & NGC253 & 3.56 & -- & -- & 530 & -- & -- & --\\
IC1574 & 10.766 & -22.247 & NGC253 & 3.56 & -- & -- & 360 & -- & 4.92 & a\\
dw0137p1439 & 24.355 & 14.661 & NGC628 & 9.77 & 10.97 & GEMINI & -- & -- & -- & --\\
dw0139p1505 & 24.943 & 15.086 & NGC628 & 9.77 & 10.32 & GEMINI & -- & -- & -- & --\\
dw0140p1459 & 25.221 & 14.999 & NGC628 & 9.77 & 22.45 & GEMINI & -- & -- & -- & --\\
dw0222p4206 & 35.728 & 42.104 & NGC891 & 9.12 & 9.4 & CFHT & -- & -- & -- & --\\
dw0236p3925 & 39.129 & 39.422 & NGC1023 & 10.40 & 14.05 & CFHT & -- & -- & -- & --\\
dw0237p3903 & 39.289 & 39.061 & NGC1023 & 10.40 & 13.14 & CFHT & -- & -- & -- & --\\
dw0240p3844 & 40.031 & 38.749 & NGC1023 & 10.40 & 16.66 & CFHT & -- & -- & -- & --\\
dw0240p3829 & 40.123 & 38.493 & NGC1023 & 10.40 & 21.61 & CFHT & -- & -- & -- & --\\
dw0241p3934 & 40.435 & 39.582 & NGC1023 & 10.40 & 13.46 & CFHT & -- & -- & -- & --\\
dw0316m4244 & 49.032 & -42.738 & NGC1291 & 9.08 & -- & -- & -- & -- & -- & --\\
dw0319m4003 & 49.858 & -40.063 & NGC1291 & 9.08 & 9.67 & MAGELLAN & -- & -- & -- & --\\
dw0323m4105 & 50.977 & -41.093 & NGC1291 & 9.08 & -- & -- & -- & -- & -- & --\\
dw0501m3811 & 75.468 & -38.196 & NGC1808 & 9.29 & 20.67 & GEMINI & -- & -- & -- & --\\
dw0502m3837 & 75.611 & -38.625 & NGC1808 & 9.29 & 24.67 & GEMINI & -- & -- & -- & --\\
dw0505m3620 & 76.422 & -36.335 & NGC1808 & 9.29 & 13.13 & MAGELLAN & -- & -- & -- & --\\
dw0515m3703 & 78.846 & -37.059 & NGC1808 & 9.29 & 15.81 & MAGELLAN & -- & -- & -- & --\\
dw0850p3307 & 132.651 & 33.130 & NGC2683 & 9.40 & 10.64 & HSC & -- & -- & -- & --\\
dw0850p3310 & 132.662 & 33.168 & NGC2683 & 9.40 & 15.04 & HSC & -- & -- & -- & --\\
dw0925p2207 & 141.258 & 22.122 & NGC2903 & 9.00 & 17.16 & GEMINI & -- & -- & -- & --\\
\enddata
\tablecomments{The names, coordinates, hosts, and distance results for the rejected background contaminants. Objects marked with a $^\dagger$ are exceptions to the usual rejection criteria. The full version of the table will be published in the online journal or will be provided upon request to the authors. SIMBAD is the source for $v_\mathrm{rec}$ unless otherwise noted. Other sources listed are: (a) \citet{karachentsev2013}, (b) \citet{cohen2018}, (c) \citet{dalcanton2009}, (d) \citet{dalcanton1997}.}
\end{deluxetable}
\end{longrotatetable}
\begin{deluxetable*}{cccccc}
\tablecaption{Unconfirmed/possible candidate satellites.\label{tab:dists_maybes}}
\tablehead{
\colhead{Name} & \colhead{RA} & \colhead{DEC}  & \colhead{Host}  & \colhead{$D_\mathrm{host}$}  & \colhead{SBF attempt}   \\ 
\colhead{} & \colhead{(deg)}  & \colhead{(deg)}   & \colhead{}  & \colhead{(Mpc)} & \colhead{}   }  
\startdata
dw0036m2828 & 9.127 & -28.470 & NGC253 & 3.56 & NONE\\
dw0141p1651 & 25.417 & 16.872 & NGC628 & 9.77 & GEMINI\\
dw0223p4314 & 35.892 & 43.240 & NGC891 & 9.12 & NONE\\
dw0224p4158 & 36.097 & 41.975 & NGC891 & 9.12 & NONE\\
dw0225p4153 & 36.319 & 41.892 & NGC891 & 9.12 & CFHT\\
dw0226p4206 & 36.669 & 42.111 & NGC891 & 9.12 & NONE\\
dw0238p3805 & 39.670 & 38.085 & NGC1023 & 10.40 & NONE\\
dw0241p3829 & 40.476 & 38.498 & NGC1023 & 10.40 & CFHT\\
dw0243p3915 & 40.979 & 39.256 & NGC1023 & 10.40 & CFHT\\
dw0312m3955 & 48.038 & -39.918 & NGC1291 & 9.08 & MAGELLAN\\
dw0315m4112 & 48.942 & -41.202 & NGC1291 & 9.08 & MAGELLAN\\
dw0316m4207 & 49.182 & -42.128 & NGC1291 & 9.08 & MAGELLAN\\
dw0321m4149 & 50.485 & -41.818 & NGC1291 & 9.08 & MAGELLAN\\
dw0502m3845 & 75.639 & -38.765 & NGC1808 & 9.29 & GEMINI\\
dw0506m3800 & 76.535 & -38.010 & NGC1808 & 9.29 & GEMINI\\
dw0507m3733 & 76.872 & -37.554 & NGC1808 & 9.29 & NONE\\
dw0510m3717 & 77.710 & -37.286 & NGC1808 & 9.29 & GEMINI\\
dw0846p3300 & 131.559 & 32.997 & NGC2683 & 9.40 & NONE\\
dw0846p3348 & 131.630 & 33.811 & NGC2683 & 9.40 & NONE\\
dw0852p3249 & 133.199 & 32.829 & NGC2683 & 9.40 & NONE\\
dw0854p3314 & 133.585 & 33.249 & NGC2683 & 9.40 & NONE\\
\enddata
\tablecomments{The names, coordinates, hosts, and source for attempted SBF measurement for the unconfirmed/possible dwarf satellites. The full version of the table will be published in the online journal or will be provided upon request to the authors.}
\end{deluxetable*}

\begin{figure*}
\includegraphics[width=\textwidth]{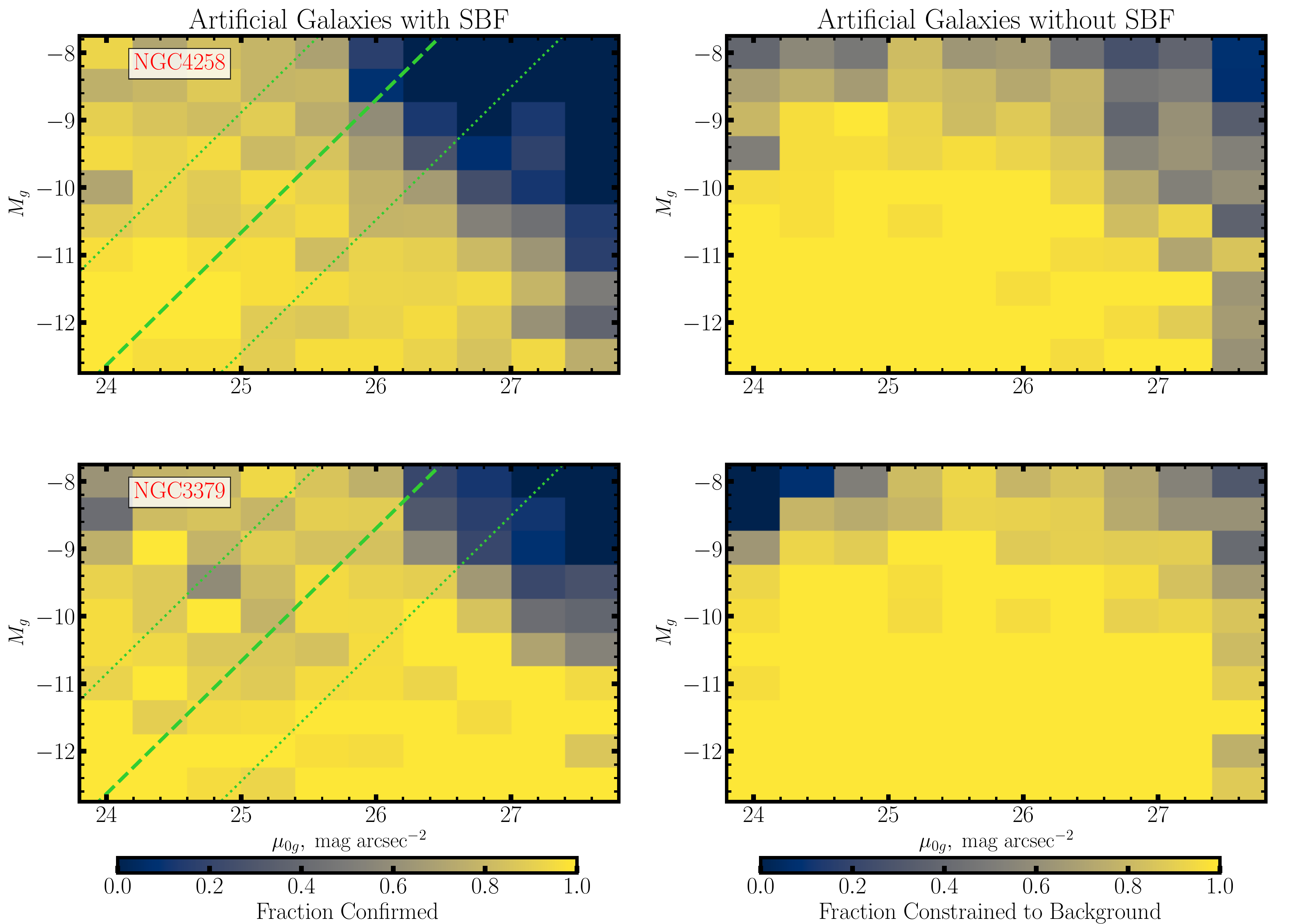}
\caption{Results of performing the SBF measurement on suites of simulated dwarfs around two example hosts (top and bottom). \textit{Right}: Dwarfs are simulated with SBF at the level appropriate to being real satellites of their host. The plot shows the fraction of simulated dwarfs that are successfully confirmed as satellites at different luminosities and surface brightness. The dwarfs that are not confirmed have too low S/N in the SBF measurement and stay as `unconfirmed/candidate' satellites. The green dashed line shows the mass-size relation of \citet{carlsten2021a}. \textit{Left}: Dwarfs are simulated without SBF, as if they are in the distant background. The plot shows the fraction of simulated dwarfs that are successfully rejected as background contaminants due to their lack of SBF. Again, the dwarfs that are not rejected have too low S/N in the SBF measurement and stay as `unconfirmed/candidate' satellites. }
\label{fig:sbf_sims}
\end{figure*}

\section{SBF Image Simulations}
\label{app:sbf_sims}
In this section, we describe simple image simulations with SBF that we performed to investigate the confirmation and rejection criteria outlined in \S\ref{sec:sbf_criteria}. For these simulations, we take tracts of very deep imaging data from two hosts: $g/r$ CFHT/Megacam imaging for NGC 4258 and $g/i$ Subaru/HSC imaging for NGC 3379. For each host, we simulate dwarfs both with and without SBF, representing actual satellites and background contaminates, respectively. We then attempt an SBF measurement and quantify how successful the measurement is in confirming or rejecting candidate satellites as a function of dwarf luminosity and surface brightness. 

In particular, we simulate dwarfs with exponential profiles and colors of $g-r=0.45$ and $g-i=0.7$. For dwarfs that are simulated with SBF, we give them the appropriate level of SBF given their color and host distance. Once the simulated dwarfs are injected into the imaging frames, we perform an SBF measurement in the usual way starting with fitting the dwarf with a S\'{e}rsic profile. We then record if the SBF measurement successfully produced a distance constraint for that dwarf. For dwarfs with SBF, this means an SBF measurement with $S/N>5$ and distance consistent with that of the host. For dwarfs without SBF, this means having a $2\sigma$ distance lower bound beyond the host's distance. 

Figure \ref{fig:sbf_sims} shows the results of these simulations. Both sides show the success rate of the SBF measurement as a function of luminosity (assuming the host's distance) and surface brightness. The right panels are for the dwarfs simulated with SBF and shows the fraction of them at each luminosity and surface brightness which yield SBF measurements with $S/N>5$ and distances consistent with the host. The left panels are for the dwarfs without SBF and show the fraction that are successfully constrained to be in the background. Note that in both cases the dwarfs with `unsuccessful' SBF measurements are just those with too low S/N, and the dwarf would remain in the `unconfirmed/candidate' category. The few cases where an artificial galaxy with SBF gets erroneously rejected to be background or vice versa are due to problems in the S\'{e}rsic fitting and/or masking that would be caught in the SBF process on real dwarfs. 

Both confirming or rejected a dwarf via SBF are more successful at higher luminosity and surface brightness. Following the mass-size relation of \citet{carlsten2021a}, both measurements start to fail fainter than $M_g\sim-9$ mag and $\mu_{0,g}\sim 26$ mag arcsec$^{-2}$. It appears that a background dwarf in this borderline parameter region is more likely to be successfully rejected than a real dwarf is to be confirmed. In other words, with imaging data to apply SBF of a given depth, it is possible to reject background contaminants using our criterion for dwarfs that are $\sim0.5-1$ mag fainter than what is possible for confirming a dwarf satellite. This difference stems from the specific threshold (S/N$>5$) we require in the SBF measurement. A lower threshold would make it easier to confirm a satellite while a higher threshold would exacerbate this difference. We discussed the rationale for this threshold in \S\ref{sec:sbf_criteria} and \citet{carlsten2020b}.

\startlongtable
\begin{longrotatetable}
\movetabledown=10mm
\begin{deluxetable}{cccccccccccccc}
\tablecaption{Satellite Photometry\label{tab:sat_photometry}}
\tablehead{
\colhead{Name} & \colhead{RA} & \colhead{DEC}  & \colhead{Host}  & \colhead{$m_g$}  & \colhead{$m_{r/i}$}     & \colhead{$M_V$}   & \colhead{$M_\star$}   & \colhead{$\mu_{0,V}$}   & \colhead{$r_e$}   & \colhead{ETG}   & \colhead{Conf.}   & \colhead{$P_\mathrm{sat}$}   & \colhead{Filt.}       \\ 
\colhead{} & \colhead{(deg)}  & \colhead{(deg)}   & \colhead{}   & \colhead{(mag)}   & \colhead{(mag)}     & \colhead{(mag)}   & \colhead{}   & \colhead{(mag arcsec$^{-2}$)}   & \colhead{(pc)}   & \colhead{}   & \colhead{}   & \colhead{}   & \colhead{}   }  
\startdata
NGC247 & 11.7834 & -20.7570 & NGC253 & $9.44\pm 0.08$ & $8.96\pm 0.09$ & $-18.60\pm 0.09$ & $9.41\pm 0.13$ & $21.27\pm 0.11$ & $4278.3\pm 199.3$ & False & True & -- & gr\\
dw0047m2623 & 11.8939 & -26.3898 & NGC253 & $18.89\pm 0.11$ & $18.33\pm 0.11$ & $-9.19\pm 0.11$ & $5.77\pm 0.14$ & $26.59\pm 0.15$ & $453.9\pm 39.8$ & True & True & -- & gr\\
dw0049m2100 & 12.4537 & -21.0167 & NGC253 & $15.07\pm 0.09$ & $14.89\pm 0.09$ & $-12.79\pm 0.09$ & $6.62\pm 0.13$ & $23.20\pm 0.11$ & $627.1\pm 31.8$ & False & True & -- & gr\\
dw0050m2444 & 12.5749 & -24.7367 & NGC253 & $18.22\pm 0.10$ & $17.95\pm 0.10$ & $-9.70\pm 0.10$ & $5.53\pm 0.13$ & $27.91\pm 0.13$ & $1539.2\pm 115.0$ & True & True & -- & gr\\
dw0055m2309 & 13.7538 & -23.1688 & NGC253 & $17.49\pm 0.09$ & $17.14\pm 0.10$ & $-10.47\pm 0.10$ & $5.96\pm 0.13$ & $23.75\pm 0.12$ & $279.1\pm 18.1$ & False & True & -- & gr\\
dw0036m2828 & 9.1273 & -28.4701 & NGC253 & $18.67\pm 0.11$ & $18.47\pm 0.11$ & $-9.21\pm 0.11$ & $5.23\pm 0.14$ & $24.39\pm 0.14$ & $311.7\pm 25.8$ & False & False & 0.05 & gr\\
dw0132p1422 & 23.2486 & 14.3736 & NGC628 & $16.05\pm 0.09$ & $15.85\pm 0.09$ & $-14.02\pm 0.09$ & $7.14\pm 0.13$ & $22.02\pm 0.11$ & $715.4\pm 38.7$ & False & True & -- & gr\\
dw0133p1543 & 23.4842 & 15.7314 & NGC628 & $19.94\pm 0.14$ & $19.38\pm 0.13$ & $-10.34\pm 0.14$ & $6.22\pm 0.15$ & $23.93\pm 0.20$ & $214.4\pm 25.6$ & True & True & -- & gr\\
dw0134p1544 & 23.5537 & 15.7463 & NGC628 & $17.57\pm 0.10$ & $17.07\pm 0.10$ & $-12.68\pm 0.10$ & $7.07\pm 0.13$ & $24.00\pm 0.12$ & $1078.1\pm 65.9$ & True & True & -- & gr\\
dw0134p1438 & 23.6741 & 14.6444 & NGC628 & $17.06\pm 0.09$ & $16.86\pm 0.10$ & $-13.01\pm 0.09$ & $6.74\pm 0.13$ & $21.73\pm 0.12$ & $497.9\pm 30.2$ & False & True & -- & gr\\
dw0136p1628 & 24.0842 & 16.4701 & NGC628 & $17.75\pm 0.10$ & $17.19\pm 0.10$ & $-12.52\pm 0.10$ & $7.10\pm 0.13$ & $22.50\pm 0.13$ & $336.7\pm 22.6$ & True & True & -- & gr\\
dw0137p1537 & 24.3236 & 15.6329 & NGC628 & $17.86\pm 0.10$ & $17.24\pm 0.10$ & $-12.45\pm 0.10$ & $7.17\pm 0.13$ & $22.59\pm 0.13$ & $380.2\pm 26.4$ & True & True & -- & gr\\
dw0137p1607 & 24.4156 & 16.1320 & NGC628 & $20.20\pm 0.14$ & $19.57\pm 0.13$ & $-10.12\pm 0.15$ & $6.24\pm 0.17$ & $25.96\pm 0.21$ & $489.2\pm 64.2$ & True & True & -- & gr\\
dw0138p1458 & 24.5024 & 14.9825 & NGC628 & $16.98\pm 0.09$ & $16.61\pm 0.09$ & $-13.19\pm 0.09$ & $7.07\pm 0.13$ & $22.42\pm 0.12$ & $349.5\pm 21.0$ & False & True & -- & gr\\
UGC1171 & 24.9366 & 15.8993 & NGC628 & $15.71\pm 0.09$ & $15.45\pm 0.09$ & $-14.40\pm 0.09$ & $7.38\pm 0.13$ & $22.36\pm 0.11$ & $840.5\pm 44.3$ & False & True & -- & gr\\
dw0139p1433 & 24.9611 & 14.5563 & NGC628 & $17.89\pm 0.10$ & $17.59\pm 0.10$ & $-12.24\pm 0.10$ & $6.59\pm 0.13$ & $22.25\pm 0.13$ & $258.9\pm 18.1$ & False & True & -- & gr\\
dw0140p1556 & 25.0395 & 15.9394 & NGC628 & $18.64\pm 0.11$ & $18.47\pm 0.11$ & $-11.41\pm 0.11$ & $6.05\pm 0.13$ & $24.03\pm 0.14$ & $449.5\pm 36.9$ & False & True & -- & gr\\
UGC1176 & 25.0430 & 15.9054 & NGC628 & $14.24\pm 0.09$ & $13.86\pm 0.09$ & $-15.93\pm 0.09$ & $8.19\pm 0.13$ & $23.44\pm 0.11$ & $2350.7\pm 115.3$ & False & True & -- & gr\\
dw0143p1541 & 25.9012 & 15.6937 & NGC628 & $15.75\pm 0.09$ & $15.35\pm 0.09$ & $-14.44\pm 0.09$ & $7.63\pm 0.13$ & $24.46\pm 0.11$ & $1437.6\pm 76.0$ & False & True & -- & gr\\
dw0141p1651 & 25.4173 & 16.8721 & NGC628 & $18.14\pm 0.10$ & $17.68\pm 0.10$ & $-12.08\pm 0.10$ & $6.78\pm 0.13$ & $23.06\pm 0.13$ & $473.4\pm 34.7$ & False & False & 0.30 & gr\\
dw0221p4221 & 35.3010 & 42.3640 & NGC891 & $17.28\pm 0.09$ & $16.79\pm 0.10$ & $-12.80\pm 0.10$ & $7.10\pm 0.13$ & $22.58\pm 0.12$ & $513.0\pm 31.9$ & True & True & -- & gr\\
\enddata
\tablecomments{The main photometric results for the ELVES Survey. Includes both satellites with conclusive distance confirmation and those without distinguished by the Confirmed column. $P_\mathrm{sat}$ estimates the likelihood of the unconfirmed satellites being real based on the candidates luminosity and surface brightness, see \S\ref{sec:cont_mod}. Photometry references are the same as given in Table \ref{tab:hosts}. Dwarfs marked with a $^\dagger$ had some issue with the photometry and the results might be biased. The full table will be published online or will be provided upon request from the authors.}
\end{deluxetable}
\end{longrotatetable}
\begin{deluxetable*}{ccccccc}
\tablecaption{GALEX Photometry for confirmed satellites.\label{tab:galex}}
\tablehead{
\colhead{Name} & \colhead{RA} & \colhead{DEC}  & \colhead{Host}  & \colhead{$m_\mathrm{NUV}$}   & \colhead{$m_\mathrm{FUV}$}   & \colhead{GALEX Frame}  \\ 
\colhead{} & \colhead{(deg)}  & \colhead{(deg)}   & \colhead{}  & \colhead{(mag)} & \colhead{(mag)}    & \colhead{} }  
\startdata
dw0047m2623 & NGC253 & 11.8939 & -26.3898 & $>22.08$ & $>21.15$ & MIS2DFSGP\_30532\_0143\\
dw0049m2100 & NGC253 & 12.4537 & -21.0167 & $16.74\pm 0.05$ & $16.97\pm 0.05$ & GI1\_009002\_UGCA015\\
dw0055m2309 & NGC253 & 13.7538 & -23.1688 & $18.93\pm 0.07$ & $19.17\pm 0.11$ & AIS\_276\_1\_59\\
dw0133p1543 & NGC628 & 23.4842 & 15.7314 & $>22.45$ & $>22.38$ & AIS\_183\_1\_66\\
dw0134p1544 & NGC628 & 23.5537 & 15.7463 & $20.65\pm 0.16$ & $>22.37$ & GI3\_050001\_NGC628\\
dw0136p1628 & NGC628 & 24.0842 & 16.4701 & $21.00\pm 0.55$ & $>21.68$ & AIS\_183\_1\_65\\
dw0137p1537 & NGC628 & 24.3236 & 15.6329 & $21.50\pm 0.17$ & $>22.97$ & NGA\_NGC0628\\
dw0137p1607 & NGC628 & 24.4156 & 16.1320 & $>22.58$ & $>22.84$ & NGA\_NGC0628\\
UGC1171 & NGC628 & 24.9366 & 15.8993 & $17.94\pm 0.05$ & $18.39\pm 0.05$ & GI1\_047008\_UGC01176\\
dw0139p1433 & NGC628 & 24.9611 & 14.5563 & $19.35\pm 0.05$ & $19.47\pm 0.05$ & MISDR2\_17173\_0426\\
dw0140p1556 & NGC628 & 25.0395 & 15.9394 & $20.44\pm 0.11$ & $20.49\pm 0.11$ & GI1\_047008\_UGC01176\\
UGC1176 & NGC628 & 25.0430 & 15.9054 & $15.78\pm 0.05$ & $16.07\pm 0.05$ & GI1\_047008\_UGC01176\\
dw0143p1541 & NGC628 & 25.9012 & 15.6937 & $17.73\pm 0.11$ & $17.87\pm 0.13$ & AIS\_183\_1\_83\\
dw0221p4221 & NGC891 & 35.3010 & 42.3640 & $20.71\pm 0.08$ & $22.34\pm 0.42$ & GI2\_019004\_3C66B\\
UGC1807 & NGC891 & 35.3058 & 42.7629 & $15.93\pm 0.05$ & $16.17\pm 0.05$ & GI2\_019004\_3C66B\\
dw0222p4242 & NGC891 & 35.7298 & 42.7123 & $19.78\pm 0.09$ & $>22.04$ & GI2\_019004\_3C66B\\
dw0239p3926 & NGC1023 & 39.8325 & 39.4339 & $>20.05$ & $>19.98$ & AIS\_59\_1\_64\\
dw0239p3903 & NGC1023 & 39.8432 & 39.0554 & $>21.43$ & $>21.03$ & AIS\_59\_1\_64\\
dw0239p3902 & NGC1023 & 39.9460 & 39.0473 & $>23.24$ & $>22.78$ & NGA\_NGC1023\\
UGC2157 & NGC1023 & 40.1044 & 38.5630 & $17.13\pm 0.05$ & $17.72\pm 0.09$ & AIS\_59\_1\_74\\
dw0240p3854 & NGC1023 & 40.1374 & 38.9004 & $18.82\pm 0.05$ & $19.18\pm 0.05$ & NGA\_NGC1023\\
\enddata
\tablecomments{The GALEX photometry for the distance-confirmed satellite that have archival coverage. For dwarfs with a detection with $S/N<2$, a $2\sigma$ lower limit to the UV magnitude is given. The full version of the table will be published in the online journal or will be provided upon request to the authors.}
\end{deluxetable*}

\section{Satellite System Results}
\label{app:sat_lists}
In Table \ref{tab:sat_photometry}, we present the main optical photometry for the confirmed and possible satellites. Note that only satellites and candidates with $M_V<-9$ mag are included.

Note that these lists are slightly different than the dwarf lists in the ELVES paper on dwarf structure \citep{carlsten2021a} and the ELVES paper on globular clusters and nuclear star clusters \citep{carlsten2021b}. The former was restricted to small survey coverage radii, a sub-sample of the full ELVES host sample, and only dwarfs in the mass range $10^{5.5}<M_\star<10^{8.5}$\msun. The photometry given here is essentially identical to that given there with a few exceptions where different imaging data or slightly different S\'{e}rsic fits were used. The photometry is also often slightly different than that given in \citet{carlsten2020b} for overlapping dwarfs since in that work we were not using an iterative masking process in the S\'{e}rsic fitting \citep[see][for an in-depth description of this]{carlsten2021a}.

On the other hand, \citet{carlsten2021b} considered confirmed early-type satellites at all radii (even as far out as 500 kpc projected) and, thus, included some satellites not included in the current catalogs.

Finally, we note that three candidate satellites (dw1305p4206, dw1308p4054, and dw1311p4051) are repeated twice each in the list, once as satellite candidates of NGC 5055 and once as candidates of NGC 4736 (NGC 5055 and NGC 4736 have overlapping projected virial regions on the sky). The distance independent quantities for the copies are essentially the same (even though the S\'{e}rsic fits are independent), but the luminosities assume the host's distance which will be different.

Table \ref{tab:galex} presents the GALEX photometry for the confirmed satellites which have archival coverage. In addition to the satellites without archival coverage, this list does not include several of the most massive ($M_\star>10^9$\msun) satellites for which we did not trust the effective radii from the optical photometry.

\end{document}